%% file: main.tex
\documentclass[preprint,12pt]{elsarticle}

\usepackage{graphicx}
\usepackage{gensymb}
\usepackage{tabularx}
\usepackage{float}
\usepackage{pgfplots}\pgfplotsset{width=10cm,compat=1.9}
\usepackage{amssymb}
\usepackage[version=4]{mhchem}
\usepackage{color}
\usepackage{esvect}
\usepackage{arydshln} 
\usepackage{graphics}
\usepackage{url}
\usepackage{ulem} 

\newcommand{\trm}{\textrm}

\newcommand{\ta}{\overline} 
\newcommand{\etah}{\eta_\trm{H}} 

\usepackage{lineno}

\biboptions{sort&compress}

\journal{International Journal of Hydrogen Energy}

\begin{document}

\begin{frontmatter}


\title{The effect of hydrogen enrichment on the forced response of \ce{CH4}/\ce{H2}/Air laminar flames}



\author{Zhengli Lim$^{[*1]}$}
\author{Jingxuan Li$^{[*2]}$}
\author{Aimee S. Morgans$^{[*1]}$}

\cortext[]{Author for correspondence: Zhengli Lim -- zhengli.lim15@imperial.ac.uk}

\address{$^{[*1]}$Department of Mechanical Engineering, Imperial College London \\ 
            $^{[*2]}$School of Astronautics, Beihang University, Beijing 100083, China}

\begin{abstract}
Hydrogen-enrichment of conventional natural gas mixtures is an actively-explored strategy for reducing pollutant emissions from combustion. This study investigates the effect of hydrogen enrichment on the unsteady flame response to perturbations, with a view to understanding the implications for thermoacoustic stability. The Level Set Approach for kinematically tracking the flame front was applied to a laminar conical premixed methane / hydrogen / air flame subjected to 2D incompressible velocity perturbations. For hydrogen enrichment levels ranging from 0\% to 80\% by volume, the resulting unsteady heat release rate of the flame was used to generate the Flame Describing Functions (FDFs). This was performed across a range of perturbation frequencies and levels at ambient pressure. The mean heat release rate of the flame was fixed at $\ta{\dot{Q}} = 2.69\ \trm{kW}$ and the equivalence ratio was set to $\varphi=1.08$ for all hydrogen enrichment levels. Hydrogen-enrichment was found to shift the FDF gain drop-off to higher frequencies, which will increase propensity to thermoacoustic instability. It also reduced the effective flame time delay. Sensitivity analyses at $\varphi = 0.8$ revealed that the changes in FDF were driven predominantly by the flame burning speed, and were insensitive to changes in Markstein length.
\end{abstract}


\end{frontmatter}


\section{Introduction}
\label{sec: Introduction}

The combustion of hydrogen-enriched fuel forms an area of increasingly active research interest as gas turbine technology progresses toward fuel decarbonisation \cite{RefWorks:doc:6012d9628f086f8f125cdd53,RefWorks:doc:6012d92b8f08b1b18ddcd92a,RefWorks:doc:6012d8e88f0859159dc40de0,RefWorks:doc:6012d81d8f08fd59bf0f8c9b}. The enrichment of conventional hydrocarbon fuels with hydrogen causes significant changes to their base properties, such as diffusivity, heating value and laminar flame burning speed \cite{RefWorks:doc:6012e0938f088cef04500998,RefWorks:doc:6012e2e48f08f067fad6d72e}, which lead to alterations in the more complex behaviour associated with the combustion of the fuel \cite{RefWorks:doc:6012e0ad8f08f63d0e7b95a6,RefWorks:doc:6012e0938f088cef04500998,RefWorks:doc:6012e07c8f08fd59bf0f8de4,RefWorks:doc:608035608f085340501f6183}. These include the susceptibility to thermoacoustic instabilities and the conditions under which they occur. Thermoacoustic instabilities, also known as combustion instabilities, have been found to be specifically sensitive to the presence of hydrogen, which burns faster and hotter than conventional natural gas fuel \cite{RefWorks:doc:608035c98f087588f882c267,RefWorks:doc:608035608f085340501f6183,RefWorks:doc:608035e18f087588f882c26d}.

Thermoacoustic instabilities comprise a major challenge faced by modern gas turbines as they produce unwanted effects ranging from reduced performance to component damage \cite{RefWorks:doc:6006caac8f08273396b01144,RefWorks:doc:6022cb8a8f08c31852dfb51f,RefWorks:doc:6022cba48f085eb58f1810ed}. The instabilities are an especially apparent side effect for lean premixed combustion \cite{RefWorks:doc:6006caac8f08273396b01144}, which is becoming the preferred route over non-premixed combustion due to tighter regulation of $\trm{NO}_x$ emissions, which are more easily controlled via the former route \cite{RefWorks:doc:6012de548f08d3139a17782d,RefWorks:doc:6012de218f088cef04500905,RefWorks:doc:6012ddf78f08d37a8bbb5143}. Thermoacoustic instabilities arise due to the coupling between an unsteady heat source and acoustic waves propagating within a geometry. They occur when the rate of acoustic energy added to the system from the heat source exceeds the rate of acoustic energy lost via transfer across the system boundaries \cite{RefWorks:doc:5efdf855e4b0b28b98c87dc1,RefWorks:doc:6006cc0d8f08323a86bc3e4e}.

Multi-scale modelling of thermoacoustic instability is often performed for both efficient computational prediction and to offer insights into instability and how to mitigate it. It comprises two main parts: modelling the flame's response to velocity perturbations \cite{RefWorks:doc:5f075518e4b09ff9ed7fcac5,RefWorks:doc:5f063bcae4b079caaf76862a}, and modelling the acoustic wave propagation within the specified combustor geometry \cite{RefWorks:doc:5f04aef1e4b045fc706a6927}. In the context of understanding thermoacoustic stability changes due to hydrogen enrichment, the multi-scale modelling part more susceptible to changes associated with hydrogen enrichment will be the flame's response. Furthermore, understanding the dynamics of laminar premixed flames is an important precursor to understanding the dynamics of the turbulent premixed flames that are typically of practical interest  \cite{RefWorks:doc:608034ec8f08664842fd093a}.

The present work performs a systematic study of the response of a laminar \ce{CH4}/\ce{H2}/Air flame to velocity perturbations, under different levels of hydrogen enrichment. It achieves this by combining a kinematic flame model with an incompressible flow field model. This facilitates computations which are extremely fast compared to turbulent flame studies which employ Large Eddy Simulation (LES) \cite{RefWorks:doc:60828c968f08e320036f1927,RefWorks:doc:60828c738f08f07243c8a4b4,RefWorks:doc:60828c4e8f08d9f5b584371a,RefWorks:doc:60828c2b8f087abb2f9a3d3f,RefWorks:doc:60828c0a8f08f07243c8a4ad,RefWorks:doc:60828bd58f081ad12666ac20,RefWorks:doc:60828bae8f084199ad49ce52,RefWorks:doc:60828b8b8f084199ad49ce4f,RefWorks:doc:60828b5a8f08e25c2cbf3860,RefWorks:doc:60828b2f8f084ea8c21d75f9}. It further facilitates fundamental insights into the parameters governing changes to the flame response, such as laminar flame speed, Markstein length (which directly influences flame burning speed) and bulk velocity, the latter varying in order to maintain constant flame thermal power under varying fuel density and heating value.

\section{Background Theory}
\label{sec: Background}
The definition used for equivalence ratio, $\varphi$, is,
\begin{equation}
\label{eqn: Equivalence Ratio Def}
    \varphi \equiv \frac{\trm{FAR}}{\trm{FAR}_\trm{st}} = \frac{X_\trm{fuel}/X_\trm{air}}{(X_\trm{fuel}/X_\trm{air})_\trm{st}}
\end{equation}
where $\trm{FAR} = X_\trm{fuel}/X_\trm{air}$ is fuel-to-air ratio ($X$ represents mole fraction), and the subscript `$\trm{st}$' denotes stoichiometric conditions. Note that the `fuel' here is regarded as the combined \ce{CH4}/\ce{H2} mixture, such that $X_\trm{fuel} = X_{\trm{CH}_4} + X_{\trm{H}_2}$. The hydrogen enrichment level, $\eta_\trm{H}$, is defined as,
\begin{equation}
\label{eqn: Hydrogen Enrichment Def}
    \eta_\trm{H} \equiv \frac{X_{\trm{H}_2}}{X_{\trm{CH}_4} + X_{\trm{H}_2}}
\end{equation}



The flame response may be characterised by a `Flame Describing Function' (FDF), also known as a nonlinear Flame Transfer Function (FTF). This takes the normalised velocity perturbation upstream of the flame, characterised by both its frequency and its amplitude, as the input, and the resulting heat release rate fluctuation as the output. The FDF approach assumes that the dominant frequency of the flame response matches that of the upstream velocity forcing, and that the steady-state response is equivalent to the input forcing modified by a gain and phase lag. This may be expressed mathematically as \cite{RefWorks:doc:5f063b54e4b07c3b7e23f165,RefWorks:doc:5f063bcae4b079caaf76862a},
\begin{equation}\label{eqn: FDF Definition}
    \mathcal{F}(\omega,\beta) \equiv \frac{\hat{\dot{Q}}'(\omega,\beta)/\ta{\dot{Q}}}{\hat{u}_1'(\omega)/\ta{u}_1}\ \ \ \ \ ,\ \beta\equiv |\hat{u}'_1/\ta{u}_1|
\end{equation}
where $\hat{u}_1'(s)$ and $\hat{\dot{Q}}'(s,\beta)$ are the upstream velocity perturbation and heat release rate fluctuation at the forcing frequency $\omega$ respectively, both of which have been transformed into the frequency domain from their time domain variables $u'_1(t)$ and $\dot{Q}'(t)$. The overbar notation denotes time-averaged quantities. $\mathcal{F}(s,\beta)$ itself is generally complex-valued and contains information regarding the gain and phase lag imparted by the process. 

In the case of laminar flames, turbulence modelling is not required. Several approaches have been devised to solve for the flame kinematics, with one commonly used kinematic model being the `G-Equation' \cite{RefWorks:doc:5efca7dae4b04862538548f4,RefWorks:doc:5f073ac4e4b09ff9ed7fc6a0,RefWorks:doc:5f075535e4b002a825213c2e,RefWorks:doc:60198eb78f08e3dee16e0ebc}. This assumes that the flame is an extremely thin interface, such that a scalar variable, $G$, can be used to track the flame surface location. The isocontour $G=0$ defines the flame front, which separates fresh reacting flow (denoted by $G<0$) from the burned gases ($G>0$):
\begin{equation}\label{eqn: G-Equation}
    \frac{\partial G}{\partial t} + \vv{u}\cdot\nabla G = s_\trm{L}|\nabla G|
\end{equation}
where $s_\trm{L}$ is the flame burning speed. The resulting flame surface may be integrated to obtain the heat release rate $\dot{Q}$, which is assumed to scale proportionately with surface area $A$, such that $\dot{Q}(t)/\ta{\dot{Q}} = A(t)/\ta{A}$. Recently, when combined with an appropriate model for the incompressible velocity field, this was shown to predict the FDF of a laminar conical methane/air flame \cite{RefWorks:doc:5f0859aae4b071e9d867ead9}, in strong agreement with experimental results \cite{RefWorks:doc:6019905e8f088cef04526f9d}. 

The flame burning speed $s_\trm{L}$ may be treated as the base laminar flame speed, modified by the Markstein parameter which accounts for local flame curvature \cite{RefWorks:doc:5f0857fce4b071e9d867ead3,RefWorks:doc:5f0a5188e4b0e655406b2383},
\begin{equation}\label{eqn: sL Model}
    s_\trm{L} = s_{\trm{L}0}(1-\mathcal{L}\kappa)
\end{equation}
where $s_{\trm{L}0}(T,P,\varphi,\eta_\trm{H})$ is the plane laminar flame speed, $\kappa = \nabla\cdot\nabla G/|\nabla G|$ is the local flame curvature and $\mathcal{L}(T,P,\varphi,\eta_\trm{H})$ is the `Markstein length', which is a length scale describing the thermal thickness of the flame. Both $s_{\trm{L}0}$ and $\mathcal{L}$ are functions of both the thermal conditions and fuel composition.

\section{Methodology}
\label{sec: Methodology}

\subsection{Principal Test Case}
\begin{figure}[H]
    \centering
    \includegraphics[width=0.7\hsize]{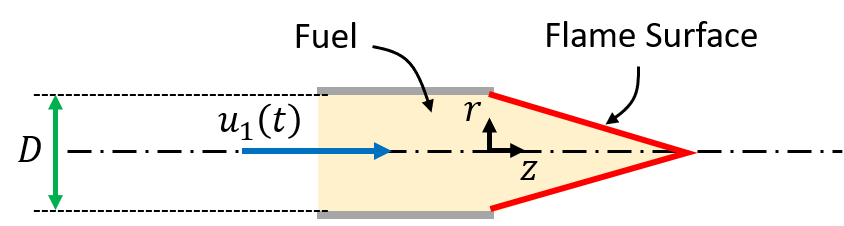}
    \caption{Illustration of test case.}
    \label{fig: HLSA_1}
\end{figure}

We consider a conical flame with its base rooted to a cylindrical burner of diameter $D$. Premixed \ce{CH4}/\ce{H2}/Air is fed into the burner of composition defined by $(\varphi,\etah)$ at an upstream velocity of $u_1(t) = \ta{u}_1 + u_1'(t)$, where $\ta{u}_1$ and $u_1'(t)$ are the mean and fluctuating components respectively. This is illustrated in Figure \ref{fig: HLSA_1}.

The kinematic response of the flamefront to the upstream velocity fluctuations is simulated using the G-Equation (equation~\eqref{eqn: G-Equation}), solved using the Level Set Approach (LSA) \cite{RefWorks:doc:5f0857fce4b071e9d867ead3}. 

Downstream of the burner exit, a two-dimensional flow field is considered, such that $\vv{u}(z,r,t) = [u_z,u_r]$, where $u_z$ is the major axial component and $u_r$ is the minor radial component. It has been established that an incompressible flow field assumption is reasonable for modelling the flame response \cite{RefWorks:doc:5f075518e4b09ff9ed7fcac5,RefWorks:doc:5f075535e4b002a825213c2e}. In addition to this, the coupling between the flame kinematics and the flow field itself may be neglected, such that independent forcing functions for $\vv{u}$ may be applied \cite{RefWorks:doc:5f09d383e4b0c32800f938fd, RefWorks:doc:5f0859aae4b071e9d867ead9},
\begin{equation}\label{eqn: u_z model}
    u_z(z,t) = \ta{u}_1 + u_1' = \ta{u}_1 + \hat{u}_1\cos(kz - \omega t)
\end{equation}
\begin{equation}\label{eqn: u_r model}
    u_r(z,r,t) = u_r' = k\frac{r}{2}\hat{u}_1\sin(kz - \omega t)
\end{equation}
where $\hat{u}_1$ is the perturbation amplitude, which is varied as a parameter once normalised into $\beta = \hat{u}_1/\ta{u}_1$. $\omega$ is the perturbation angular frequency and $k = \omega/\ta{u}_1$ is the wavenumber. The model has been validated by Direct Numerical Simulation \cite{RefWorks:doc:5f09d474e4b07c3b7e24c2bf}.

\subsection{Selection of Variables and Constraints}
The burner diameter was set at $D=22\ \trm{mm}$ for all simulations. As the main focus of this work is the effect of $\etah$ on $\mathcal{F}$, the equivalence ratio was fixed $\varphi = 1.08$, and the hydrogen enrichment level was tested across five hydrogen enrichment levels, $\etah = [0:0.2:0.8]$, for a total of 5 FDFs. The mean upstream velocity $\ta{u}_1$ was varied based on fuel composition to satisfy the constraint of a constant mean heat release rate of $\ta{\dot{Q}} = 2.69\ \trm{kW}$. Inlet thermal conditions were fixed at a temperature and pressure of $T_\trm{in} = 300\ \trm{K}$ and $P_\trm{in} = 1\ \trm{atm}$. We note that the chosen test burner geometry, equivalence ratio and flame power, along with the incompressible velocity model of Equations~\eqref{eqn: u_z model} and~\eqref{eqn: u_r model}, correspond to those for which the G-equation recently yielded excellent agreement with the experimentally measured FDF for a methane/air flame \cite{RefWorks:doc:5f0859aae4b071e9d867ead9}. This provides confidence in the main assumptions underpinning the approach.

For identification of the FDF, harmonically forced simulations were performed by implementing Equations \eqref{eqn: u_z model} and \eqref{eqn: u_r model} over $16$ forcing frequencies, $f_\trm{p} = \omega/2\pi = [10:10:160]\ \trm{Hz}$ and four amplitude perturbation levels, $\beta = [0.036,0.071,0.143,0.214]$. This implies a total of $4\times16 = 64$ data points per FDF.

At this point, we make some comments on hydrogen's exceptionally low density. While the effect of hydrogen's density is fully accounted for in terms of the reactant density, velocity and flow rate and hence on the burner power, the G-equation neglects density effects on the kinematic response of the flame. This means that in terms of the flame's forced response, the density is assumed constant throughout space and equal to that of the reactant mixture. This assumption has been shown sufficiently accurate for methane/air flames ($\etah=0$) \cite{RefWorks:doc:5f0859aae4b071e9d867ead9, RefWorks:doc:6019905e8f088cef04526f9d}. For the hydrogen-enriched compositions tested in this study, the reactant mixture density remains within $10\%$ of the $\etah=0$ case up to $\etah=0.6$, with the $\etah=0.8$ case showing a $16\%$ deviation. This was assumed sufficiently small for the assumption to remain reasonable. We note that the G-equation is not able to account for Darrius-Landau or thermodiffusive instabilities \cite{RefWorks:doc:601bf9028f08f547dbdd10d7,RefWorks:doc:601bf8f28f08fe29727c66c8} which, although often of relevance to hydrogen-rich flames, are beyond the scope of the current work.

The Markstein length was fixed at $\mathcal{L} = 1\ \trm{mm}$ for the main study. Although $\mathcal{L}$ is known to vary according to the mixture state, it is shown in Section \ref{subsec: Markstein Sensitivity} that its variations do not significantly affect the overall results.

Table \ref{tab: Main Simulation Inputs} summarises the simulation parameters examined in the main study. Note that $\ta{u}_1$ and $s_{L0}$ are not independent variables here but are computed based on $(\varphi,\etah)$ and the heat release rate constraint set for this study.

\begin{table}[H]
\centering
\caption{Summary of simulation parameters for the main study.}
\label{tab: Main Simulation Inputs}
\begin{tabular}{c|c|c|c|c|c|c}
 $\varphi$ & $\etah$ & $\beta$ & $f_\trm{p}\ [\trm{Hz}]$ & $\ta{u}_1\ [\trm{m s}^{-1}]$ & ${s_\trm{L0}}^{[*2]}\ [\trm{m s}^{-1}]$ & $\mathcal{L}\ [\trm{mm}]$\\ \hline
 1.08 & 0.0 & $[*1]$  & 10:10:160 & 2.12 & 0.3859 & 1.00 \\ 
 1.08 & 0.2 & 0.036 & 10:10:160 & 2.13 & 0.4494 & 1.00 \\
 1.08 & 0.4 & 0.071 & 10:10:160 & 2.15 & 0.5538 & 1.00 \\
 1.08 & 0.6 & 0.143 & 10:10:160 & 2.18 & 0.7523 & 1.00 \\
 1.08 & 0.8 & 0.214 & 10:10:160 & 2.22 & 1.2175 & 1.00 \\ \hline
\end{tabular}
\begin{raggedright}
\\ \small{$^{[*1]}$ Note that all four $\beta$ values listed were tested for each $\etah$ value.}
\\ \small{$^{[*2]}$ Evaluated by Cantera, see Section \ref{sec: Numerical Solution Method} for details.} \\
\end{raggedright}
\end{table}

\subsection{Numerical Solution Method}\label{sec: Numerical Solution Method}
\begin{figure}[H]
\centering
\input{figures/s_L0}
\caption{Comparison of laminar flame speeds determined by Cantera and experimentally.}
\label{fig: Cantera Flame Speed}
\end{figure}
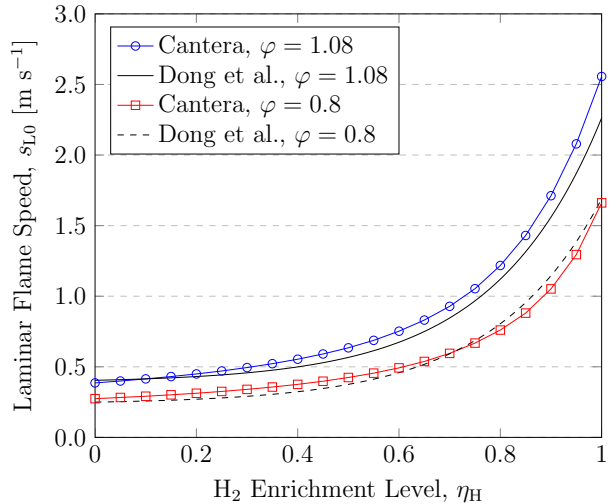

The plane laminar flame speed $s_{L0}$ was evaluated for each unique fuel state using the open-source chemical kinetics code `Cantera' \cite{cantera}, coupled with the GRI-Mech 3.0 mechanism, utilised predominantly for natural gas combustion \cite{RefWorks:doc:5fc1080ae4b0c6c777b3b1f9}. Figure \ref{fig: Cantera Flame Speed} shows the computed $s_\trm{L0}$ values produced by Cantera for the domain $\etah = [0,1]$, both at the main test case equivalence ratio of $\varphi = 1.08$, and at the equivalence ratio considered in Section~\ref{subsec: Markstein Sensitivity} of  $\varphi = 0.8$. The results are corroborated by those of Dong et al. \cite{RefWorks:doc:5f0864bae4b09ff9ed7ffee9}, whose experimental data best-fit curves are shown alongside the predictions by Cantera. Noting that a significant margin of error was expected in the experimental data, this study used the predictions by Cantera, which were deemed sufficiently accurate for the test cases under consideration.

The G-Equation model (Equation \ref{eqn: G-Equation}) was solved with a global spatial step of $\Delta x = 0.5\ \trm{mm}$ and temporal step of $\Delta t = \Delta x/40 = 1.25\times 10^{-5}\ \trm{s}$ to ensure stability for the range of convective velocities considered. $G$ was solved spatially using a ``fifth-order Weighted Essentially Non-Oscillatory (WENO)" scheme \cite{RefWorks:doc:5f09eda2e4b031c64073ffb3} and temporally using a ``third-order Total Variation Diminishing (TVD) Runge-Kutta" scheme \cite{RefWorks:doc:5f09edd0e4b0f485f3af5991}. 

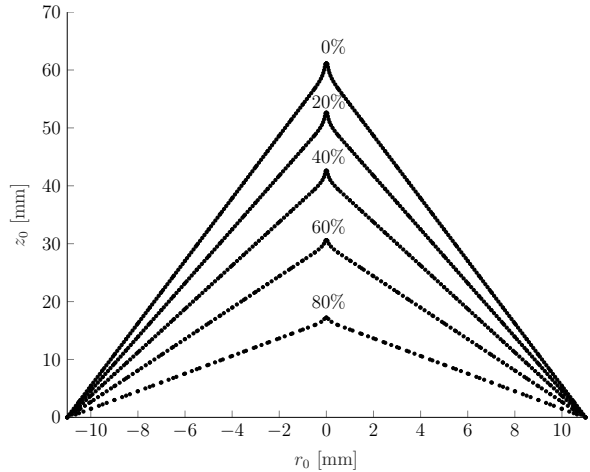
\begin{figure}[H]
    \centering
    \input{figures/FF0_Length}
    \caption{Unforced flame fronts at $\varphi = 1.08$ as $\etah$ is varied from $0-80\%$. The flame becomes shorter as $\etah$ increases.}
    \label{fig: Unforced Flames}
\end{figure}

The unforced flames were first generated by setting $\beta = 0$ across $\etah = [0:0.2:0.8]$. These plots are shown in Figure \ref{fig: Unforced Flames}, and show that the axial flame length decreases as $\etah$ is increased, supporting numerous experimental observations \cite{RefWorks:doc:6080351d8f08cc798956c9ab}.

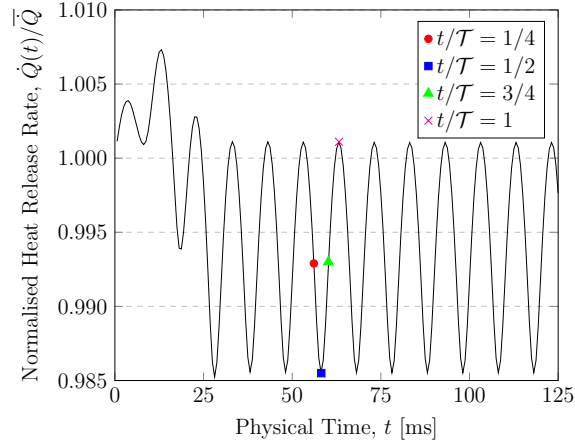
\begin{figure}[H]
\centering
\input{figures/QR_Example}
\caption{Example of a flame response function for $(\varphi,\etah,\beta,f_\trm{p}) = (1.08,0,0.036,100)$ with data points corresponding to flame front instances as shown in Figure \ref{fig: FF Example}.}
\label{fig: QR Example}
\end{figure}

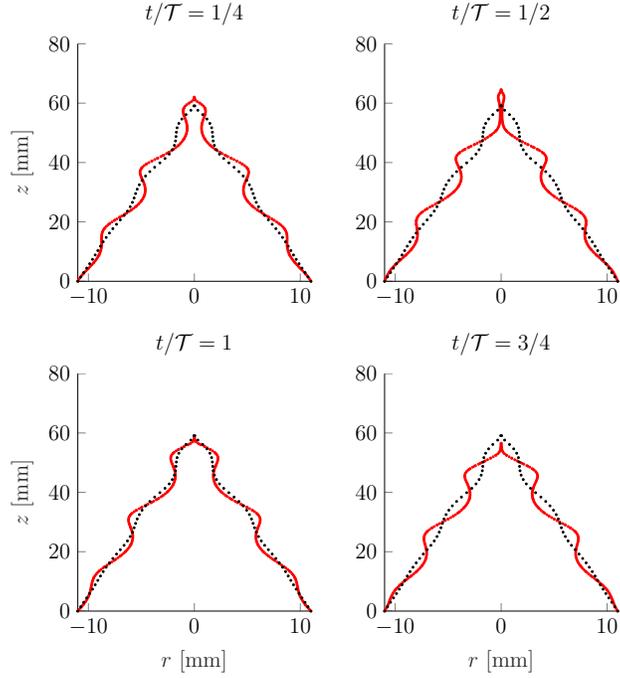
\begin{figure}[H]
    \centering
    \input{figures/FF_Example}
    \caption{Example of flame front evolution per forcing cycle for $(\varphi,\etah,\beta,f_\trm{p}) = (1.08,0,0.036,100)$, corresponding to heat release rate points as shown in Figure \ref{fig: QR Example}. Plots in solid red and black dotted lines correspond to the instantaneous and time-mean flame fronts respectively.}
    \label{fig: FF Example}
\end{figure}

Subsequently, forcing was introduced. Each simulation was time-marched until the flame response was observed to settle from its initial transients into periodic oscillatory behaviour. Figure \ref{fig: QR Example} shows an example of the flame response function produced from an arbitrarily chosen simulation. Note that the normalised heat release rate as shown in Figure \ref{fig: QR Example} is computed as the ratio of the instantaneous heat release rate $\dot{Q}(t)$ to that of the unforced case $\dot{Q}_0$, the latter of which is assumed approximately equal to the mean heat release rate, such that $\ta{\dot{Q}} \approx \dot{Q}_0$. Figure \ref{fig: FF Example} shows examples of the flame front evolution corresponding to the periodic heat release rate in Figure \ref{fig: QR Example}.

The FDF gain $|\mathcal{F}|$ was determined as the ratio of the steady-state amplitude of $\dot{Q}(t)/\ta{\dot{Q}}$ to the perturbation level $\beta$. The phase shift $\angle\mathcal{F}$ relative to the major axial perturbation was determined by cross-correlation of the two functions.

\section{Results and Discussion}
\label{sec: Results}
\subsection{FDF Results}\label{subsec: FDF Results}

\begin{figure}[H]
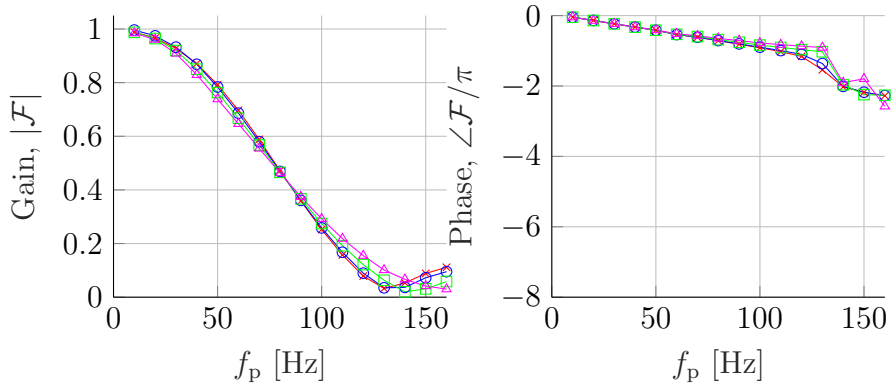

    \centering
    \caption{Flame Describing Functions for varying hydrogen enrichment, $\etah$. Each row of plots corresponds to a different value of $\etah$, with the top row corresponding to $\etah=0$ and $\etah$ increasing for subsequent rows by $0.2$, up to a final value of $0.8$.}
    \label{fig: FDF Results}
\end{figure}

The FDF gain and phase are shown in Figure \ref{fig: FDF Results}, with each row of plots corresponding to a given value of $\etah$, with $\etah$ increasing in increments of $0.2$; $\etah = [0:0.2:0.8]$. Each of the curves on the FDF plot corresponds to one of the four $\beta$ values tested. The FDF for $\etah=0$ corroborates that found by Li \& Morgans \cite{RefWorks:doc:5f0859aae4b071e9d867ead9}.

Three main observations regarding the effect of hydrogen enrichment on the flame response can be drawn from the results. An increase in hydrogen enrichment causes: (i) an increase in cut-off frequency, (ii) a decrease in phase lag and, (iii) a reduction in non-linearity.

\vspace{0.5cm}
\includegraphics[width=0.2\hsize]{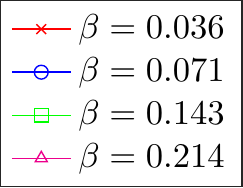}

\begin{center}
    (a),(b) -- $0\%$ Hydrogen Enrichment
    \input{figures/0.0H_FDF}
    \vspace{0.5cm}
    
    (c),(d) -- $20\%$ Hydrogen Enrichment
    \input{figures/0.2H_FDF}
    \vspace{0.5cm}
    
    (e),(f) -- $40\%$ Hydrogen Enrichment
    \input{figures/0.4H_FDF}
    \vspace{0.5cm}
    
    (g),(h) -- $60\%$ Hydrogen Enrichment
    \input{figures/0.6H_FDF}
    \vspace{0.5cm}
    
    (i),(j) -- $80\%$ Hydrogen Enrichment
    \input{figures/0.8H_FDF}
\end{center}
\vspace{-1cm}

From the gain plots, hydrogen enrichment is seen to increase the threshold frequency at which gain drop-off occurs. The consequence of this is that systems involving hydrogen-enriched fuel are likely to exhibit increased propensity to thermoacoustic instability -- a finding that is corroborated by other numerical studies on hydrogen-enriched fuels \cite{RefWorks:doc:5f085e13e4b0ff46bb659685,RefWorks:doc:5f085e84e4b09ff9ed7ffe3d,RefWorks:doc:5f085ec7e4b00f138978178d,RefWorks:doc:5f085ee8e4b0e0332d64941c}.

From the phase plots, hydrogen enrichment is seen to decrease the phase lag between the perturbation and the response. This is consistent with the fact that increased hydrogen content produces shorter flames (Figure \ref{fig: Unforced Flames}), leading to lower convection times associated with the response, in corroboration with the findings of \cite{RefWorks:doc:608035a98f08012ce2025fd4}. The thermoacoustic stability of a combustor depends on strongly the phase of the FDF, but in a more complicated manner, with the Rayleigh criterion source term depending on the phase between the heat release rate and acoustic pressure perturbation at the flame \cite{RefWorks:doc:5efdf855e4b0b28b98c87dc1}. The latter depends on the acoustic modeshape of the relevant mode within the combustor, and so cannot be predicted from the flame response alone. Thus while it can be said that the reduced phase will affect the thermoacoustic stability, the direction of this effect will depend on the geometry of the combustor within which the flame is placed.

Finally, both plots show that hydrogen enrichment suppresses the dependence of $\mathcal{F}$ on the perturbation level $\beta$, as the four curves on each plot converge toward a single curve as hydrogen content increases. However, it was noted that all plots show that the curves begin converging at low perturbation frequencies, and diverge upon crossing a threshold frequency, this threshold increasing with hydrogen enrichment. Hence, it is very likely that non-linearity is not wholly suppressed but instead shifted to higher frequencies.

\begin{figure}[H]
    \centering
    \input{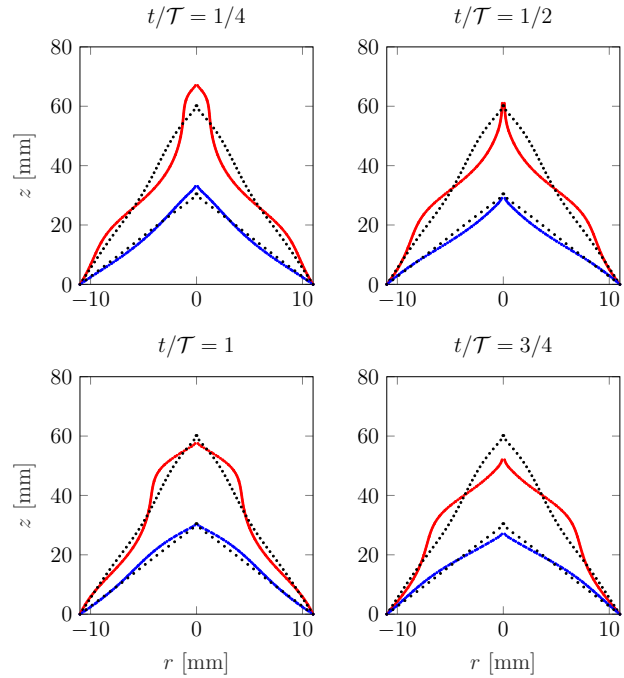}
    \caption{Comparison of $\etah=0$ (red) and $\etah=0.6$ (blue) flame front evolution for one forcing cycle at $(\varphi,\beta,f_\trm{p}) = (1.08,0.071,40)$. Axial flame length decreases with increasing $\etah$. Plots in solid and dotted lines correspond to the instantaneous and time-mean flame fronts respectively.}
    \label{fig: FF Comparison}
\end{figure}

The increase in the gain drop-off frequency for hydrogen-enriched flames was determined a trait of interest to analyse further. Hence, flame front evolution tracking was performed for the two test cases $\etah = 0$ and $\etah = 0.6$, both held at $(\varphi,\beta,f_\trm{p}) = (1.08,0.071,40)$. The $40\ \trm{Hz}$ frequency was chosen because the gain of the $\etah = 0$ case was relatively low at that frequency, being beyond the drop-off frequency, while the gain of the $\etah = 0.6$ case remained relatively high. The resulting flame front snapshots are shown in Figure \ref{fig: FF Comparison}, with the $\etah=0$ case plotted in solid red and the $\etah=0.6$ case plotted in solid blue. Note that the snapshots $t/\mathcal{T} = 1/2$ and $t/\mathcal{T} = 1$ correspond to the minimum and maximum normalised heat release rate states respectively.

First, the similarities between the two cases were noted. Namely, the flame surface area (and consequently the heat release rate) depends on two features of the flame front: 1. the overall tip length, and 2. the concavity of the region leading up to the flame tip. A greater flame tip length allows more area to be covered over the axial span. Additionally, a flame shape that maintains a relatively large radius along its axial length has a larger surface area since $\delta A \propto 2\pi r \delta z$. The result is that the final flame surface area is governed by the combination of these two factors.

The $t/\mathcal{T} = 1/4$ and $t/\mathcal{T} = 3/4$ snapshots represent times when the gain is at its mean value, and it was noted that the two mechanisms mentioned are competing at these times. At, $t/\mathcal{T} = 1/4$, the flame tip length is at its highest, but this is undermined by the concavity of the flame which decays to lower radii relatively quickly. The opposite is true at $t/\mathcal{T} = 3/4$. In contrast, the maximum and minimum gains occur when both mechanisms act constructively.

As for the differences, it was noted that the $\etah=0$ flame exhibited a greater distortion of curvature from the mean shape at all times, while the $\etah=0.6$ flame retained a profile more closely resembling its unforced shape. It is shown in Section \ref{sec: General Sensitivity Analysis} that the higher laminar flame speed of hydrogen-enriched fuel is by far the dominant driver behind such differences. Hence, it can be postulated that the higher flame burning speed of the $\etah=0.6$ case acts to stabilise the flame against distortion by a given perturbation.

\begin{figure}[H]
    \centering
    \input{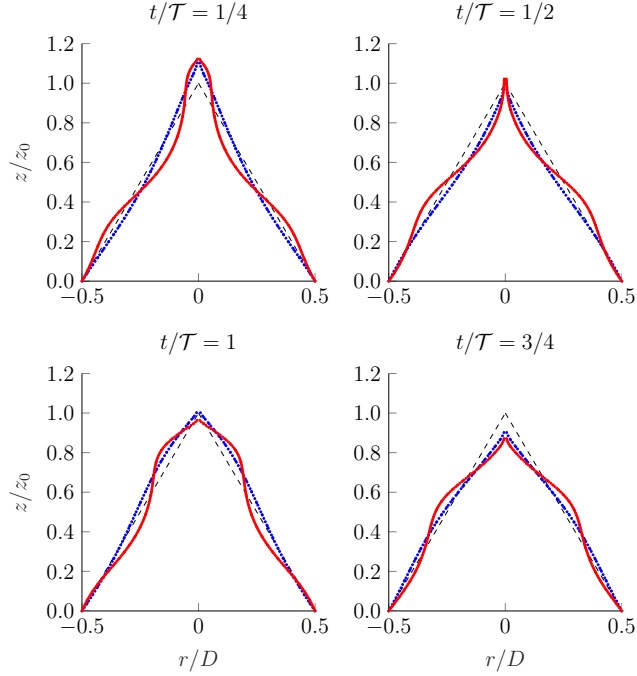}
    \caption{Dimensionless plots of $\etah=0$ (red) and $\etah=0.6$ (blue) flame fronts for one forcing cycle at $(\varphi,\beta,f_\trm{p}) = (1.08,0.071,40)$. Plots in solid and dotted lines correspond to the instantaneous and time-mean flame fronts respectively.}
    \label{fig: Pi Comparison}
\end{figure}

It may appear counter-intuitive that the $\etah=0.6$ case is the one with higher gain when comparing the cases on an absolute scale. A fairer comparison was achieved by creating dimensionless plots of the flame fronts, with $z/z_0$ against $r/D$, where $z_0$ is the axial distance from the base to the unforced flame tip. This is shown in Figure \ref{fig: Pi Comparison}.

Within the dimensionless plots, the flame tip lengths were observed to approximately match across both cases at all times. While greater curvature does lead to a greater surface area in absolute terms, it is important to note that the gain is effectively determined by the difference between the maximum and minimum areas obtained in a cycle. Although the $\etah=0$ flame displays significantly more curvature distortion than the $\etah=0.6$ case, this distortion increases both its maximum and minimum areas in absolute terms, but may not necessarily translate to a greater difference between the two. It was observed in this case that the curvature-distortion of the flame actually dampened the driving mechanisms of area fluctuation, resulting in a lower gain.

\begin{figure}[H]
    \centering
    \input{figures/dA_Compare}
    \caption{Dimensionless plots of surface area contribution as a function of the axial coordinate, $z$, for $\etah = 0$ (red) and $\etah = 0.6$ (blue) during a forcing cycle. Dotted lines represent the mean values.}
    \label{fig: dA Compare}
\end{figure}
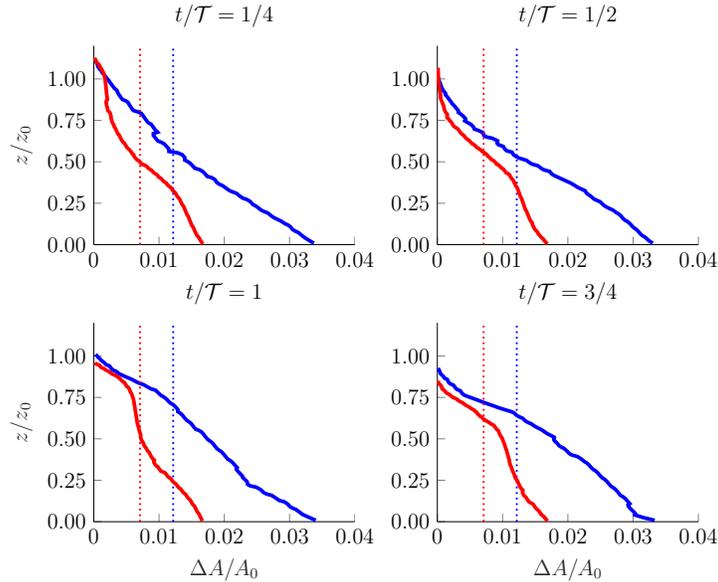

Figure \ref{fig: dA Compare} shows the dimensionless plots of $\Delta A/A_0$ against $z/z_0$ for both flames at various points in the forcing cycle, where $A_0$ is the unforced flame surface area. The dimensionless quantity $\Delta A/A_0$ denotes the relative contribution of a flame section (at a given axial location) to the total flame surface area. Since the total area $A$ of the flame is calculated by numerically integrating across the grid points of spatial step $\Delta x$, the area contribution between any two such points is calculated as $\Delta A$. This means that a higher value of $\Delta A/A_0$ at a given $z/z_0$ signifies a greater contribution to the normalised flame surface area at that location. Note that $A_0 = 210\ \trm{mm}^2$ and $110\ \trm{mm}^2$ for the $\etah = 0$ and $0.6$ flames respectively. The plots in Figure \ref{fig: dA Compare} are consistent with the suggestion that the greatest contribution to the flame surface area occurs at the base.

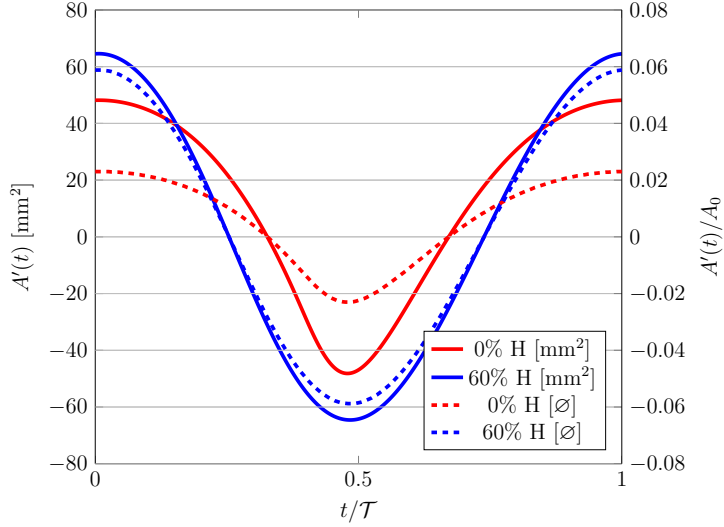
\begin{figure}[H]
    \centering
    \input{figures/A_Transient}
    \caption{Absolute and dimensionless flame surface area fluctuations across one period at $f_\trm{p} = 40\ \trm{Hz}$.}
    \label{fig: A Transient}
\end{figure}

The absolute and dimensionless area fluctuations, $A'(t)$ and $A'(t)/A_0$, are shown over one forcing period in Figure \ref{fig: A Transient}. It can be seen that both the absolute and dimensionless fluctuations for the $\etah = 0.6$ case exceed those of the $\etah = 0$ case, supporting the idea that the higher local curvature of the latter (as observed in Figures \ref{fig: FF Comparison} and \ref{fig: Pi Comparison}) does not necessarily imply greater deviations in the total flame surface area.

\subsection{Sensitivity of Results to Markstein Length}\label{subsec: Markstein Sensitivity}

The FDF results obtained in Figure \ref{fig: FDF Results} were based on the assumption of a constant Markstein length of $\mathcal{L}_\trm{ct} = 1\ \trm{mm}$. This assumption is now investigated by conducting a separate analysis at an equivalence ratio of $0.8$, this corresponding to one where the actual Markstein lengths $\mathcal{L}_\trm{act}$ are known \cite{RefWorks:doc:5fc4176ae4b03ac8d70e6d3b}. A low forcing amplitude level of $\beta = 0.071$ is used, enabling the flame transfer function (FTF) rather than the more complicated FDF to be compared. The FTFs were calculated across the same frequency range, $f_\trm{p} = 10:10:160\ \trm{Hz}$, using both the actual and assumed Markstein lengths, with the hydrogen content varying from $\etah = [0:0.2:0.6]$, as summarised in Table \ref{tab: Markstein Sensitivity Inputs}.

\begin{table}[H]
\centering
\caption{Summary of simulation parameters for the Markstein length sensitivity analysis.}
\label{tab: Markstein Sensitivity Inputs}
\begin{tabular}{c|c|c|c|c|c}
 $\varphi$ & $\etah$ & $\beta$ & $f_\trm{p}\ [\trm{Hz}]$ & $\mathcal{L}_\trm{act}\ [\trm{mm}]$ & $\mathcal{L}_\trm{ct}\ [\trm{mm}]$\\ \hline
 0.8 & 0.0 & 0.071 & 10:10:160 & 0.56 & 1.00 \\ 
 0.8 & 0.2 & 0.071 & 10:10:160 & 0.34 & 1.00 \\
 0.8 & 0.4 & 0.071 & 10:10:160 & 0.12 & 1.00 \\
 0.8 & 0.6 & 0.071 & 10:10:160 & -0.11 & 1.00 \\ \hline
\end{tabular}
\end{table}

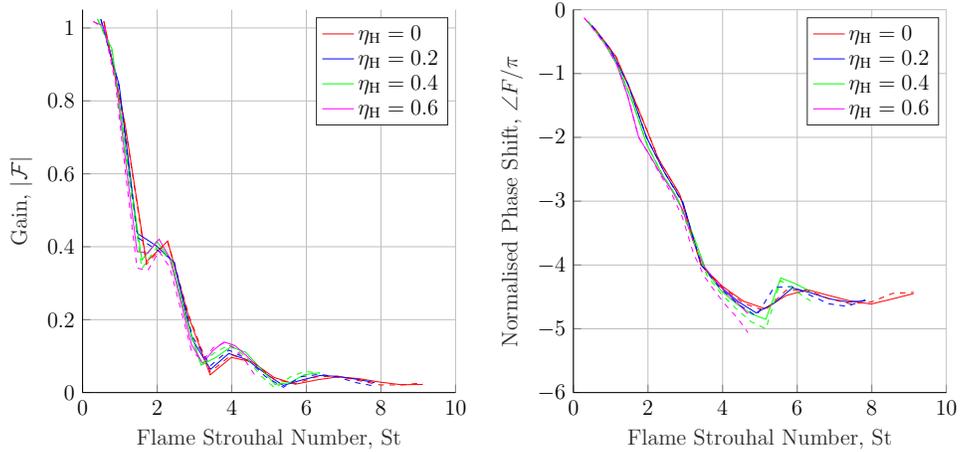
\begin{figure}[H]
    \centering
    \input{figures/L_Compare}
    \caption{Comparison of Flame Transfer Functions with actual (solid lines) against assumed (dashed lines) Markstein lengths, across varying hydrogen enrichment levels.}
    \label{fig: L Comparison}
\end{figure}

The resulting FTFs are shown in Figure \ref{fig: L Comparison}. Note that solid line plots represent the FTFs produced with $\mathcal{L}_\trm{act}$, while dashed line plots represent the FTFs produced with $\mathcal{L}_\trm{ct}$. The forcing frequency is expressed in its dimensionless form, the flame Strouhal number, $\trm{St} \equiv f_\trm{p}D/s_\trm{L0}$.

The results show that the flame response exhibited very little sensitivity to $\mathcal{L}$, considering the range of $\mathcal{L}$ achievable in practical combustion systems. We note that this present study is restricted to laminar flames; $\mathcal{L}$ variations may play a larger role in the response of turbulent flames, where flame wrinkling is substantial.

\subsection{General Sensitivity Analysis}\label{sec: General Sensitivity Analysis}

Insights into the effect of hydrogen enrichment on the flame response can be obtained by performing an analysis of the sensitivity of $\mathcal{F}$ to the properties $s_\trm{L0}$, $\mathcal{L}$ and $\ta{u}_1$. The analysis was conducted at $(\varphi,\beta) = (0.8,0.071)$ for two cases: $\etah = 0$ and $\etah = 0.6$, in order to obtain two FTFs upon which comparisons may be made. The Markstein lengths used were the actual values in literature \cite{RefWorks:doc:5fc4176ae4b03ac8d70e6d3b}.

To attempt to isolate the effects of different parameters, the $\etah = 0$ case was then artificially modified in order to perform three separate simulations, where the variables $s_\trm{L0}$, $\mathcal{L}$ and $\ta{u}_1$ were changed in turn to match those of the $\etah = 0.6$ case. The other variables were left unchanged. This was done in order to observe the extent to which the artificially modified FTFs shifted from the $\etah = 0$ reference case toward the $\etah = 0.6$ reference case, and hence determine the relative importance of each variable in governing the flame response.

Table \ref{tab: Sensitivity Analysis Inputs} summarises the simulations performed for the sensitivity analysis. Results from $16\times 5 = 80$ simulations were used for this. $\ta{\dot{Q}}$ was found to vary as a result of artificially modifying the base parameters, but this variation was not significant.

\begin{table}[H]
\centering
\caption{Summary of simulation parameters for the sensitivity analysis.}
\label{tab: Sensitivity Analysis Inputs}
\resizebox{\columnwidth}{!}{
\begin{tabular}{c|c|c|c|c|c|c|c}
 $\varphi$ & $\etah$ & $\beta$ & $f_\trm{p}\ [\trm{Hz}]$ & $\ta{u}_1\ [\trm{m s}^{-1}]$ & ${s_\trm{L0}}^{[*2]}\ [\trm{m s}^{-1}]$ & $\mathcal{L}\ [\trm{mm}]$ & $\ta{\dot{Q}}\ [\trm{kW}]$\\ \hline
 0.8 & 0.0    & 0.071 & 10:10:160 & 2.7878 & 0.2742 & 0.56 & 2.69\\ 
 0.8 & 0.6    & 0.071 & 10:10:160 & 2.8069 & 0.4922 & -0.11 & 2.69\\ \hdashline
 0.8 & $[*1]$ & 0.071 & 10:10:160 & 2.8069 & 0.2742 & 0.56 & 2.71\\
 0.8 & $[*1]$ & 0.071 & 10:10:160 & 2.7878 & 0.4922 & 0.56 & 2.67\\
 0.8 & $[*1]$ & 0.071 & 10:10:160 & 2.7878 & 0.2742 & -0.11 & 2.69\\ \hline
\end{tabular}
}
\begin{raggedright}
\\ \small{$^{[*1]}$ Simulations with artificially modified parameters.}
\\ \small{$^{[*2]}$ Evaluated by Cantera, see Section \ref{sec: Numerical Solution Method} for details.} \\
\end{raggedright}
\end{table}

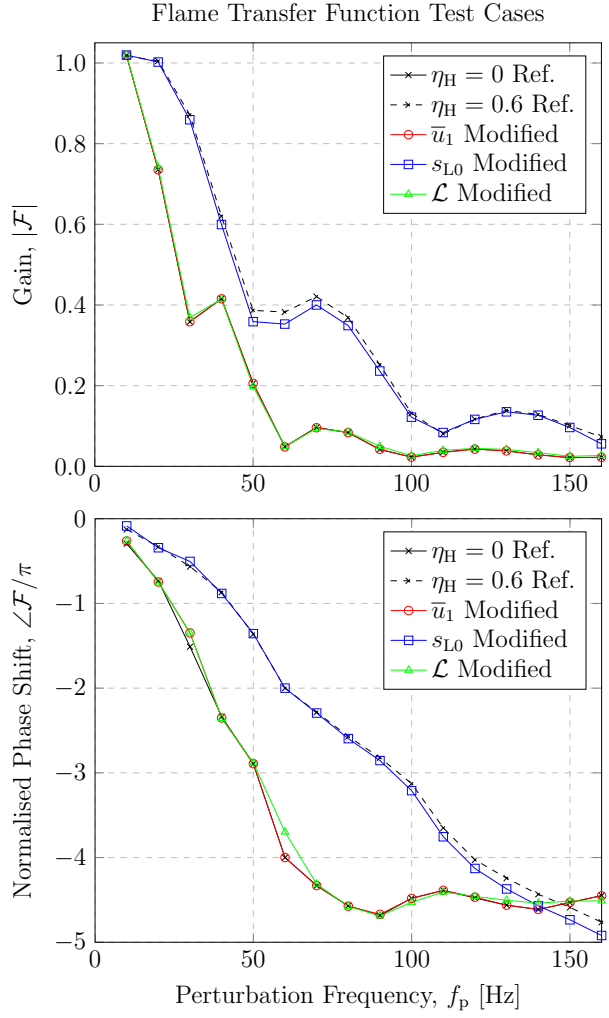
\begin{figure}[H]
\centering
\input{figures/GainSensitivity}
\input{figures/PhaseSensitivity}
\caption{Flame speed is the dominant driver of hydrogen-enriched flame response.}
\label{fig: Sensitivity Analysis}
\end{figure}

Figure \ref{fig: Sensitivity Analysis} shows the results of the sensitivity analysis. It is clear that the laminar flame speed, $s_\trm{L0}$, is the dominant cause of the FTF/FDF shift at elevated hydrogen enrichment, while the effects of $\ta{u}_1$ and $\mathcal{L}$ are minimal. Note that the $\etah=0$ reference curve overlaps the modified $\ta{u}_1$ and $\mathcal{L}$ curves to the point where the three curves are indistinguishable across the majority of the frequency range. The same is true for the $\etah=0$ reference and the modified $s_\trm{L0}$ curves.

These insensitivity of $\mathcal{F}$ to $\ta{u}_1$ is an expected result for these test cases, since both of these base variables do not change significantly in absolute terms across the $\etah=0$ and $\etah=0.6$ cases. While $\ta{u}_1$ is calculated based on the fuel mixture density and heating value in order to satisfy a constant power constraint, it changes negligibly with $\etah$ since the dominant mass component in the fuel mixture is air. This implies that $\mathcal{F}$ may change significantly with $\ta{u}_1$ if differences in their absolute values become significantly large, but for lean premixed gaseous fuels, this is unlikely to be the case. 

The insensitivity of $\mathcal{F}$ to $\mathcal{L}$ is supported by the study conducted in Section \ref{subsec: Markstein Sensitivity}, which suggested that flame wrinkling in laminar flames was not substantial enough for variations in $\mathcal{L}$ to drive significant changes in the flame response.

\newpage

\section{Conclusion}
The enrichment of conventional methane fuel with hydrogen causes significant changes to several of its base properties, the most significant being laminar flame speed which increases from $0.386\ \trm{m s}^{-1}$ at zero hydrogen enrichment to $1.218\ \trm{m s}^{-1}$ at $80\%$ hydrogen enrichment (at equivalence ratio $1.08$). This study has investigated the effect of hydrogen enrichment on the forced response of the flame in the context of thermoacoustic stability. Across hydrogen enrichment levels ranging from 0\% to 80\% and with the flame's thermal power maintained constant, it was found that with increasing hydrogen enrichment, the frequency fall off of the flame's gain was pushed to higher frequencies, as was the frequency at which nonlinear effects became important. The phase lag in the flame's response reduced with hydrogen enrichment level, in accordance with the shortening of the mean flame shape, since the convection time between the flame base and the flame front is reduced.

Further investigations showed that changes to the base laminar flame speed $s_{\trm{L}0}$ were by far the dominant factor underpinning the changes in the forced flame response. Variations in Markstein length $\mathcal{L}$ were found to have no practical effect on the results. This is important because the Markstein lengths of mixtures containing multiple fuel components of different molecular diffusivities are difficult to predict with theory, and few studies have attempted to do this specifically for $\trm{CH}_4/\trm{H}_2/\trm{Air}$ blends. This work shows that a precise prediction of the Markstein length may not be critical within the scope of practical engineering application.

Analysis of the flame dynamics suggested that the higher frequency fall off of the flame's gain with hydrogen enrichment was linked to a higher resistance to distortion due to the higher flame speed. The increased resistance to flame distortion suggests that the turbulent flames may behave in a similar manner, provided the hydrogen-enrichment level is high enough.

The higher gain fall-off frequency of hydrogen-enriched flames suggests a generally higher propensity to thermoacoustic instability. In practice, changes in the phase response of the flame will also play a role, and these will depend on the geometry of the combustor within which the flame is placed; this is consistent with the observation that hydrogen enrichment can inhibit thermoacoustic oscillations in some combustors \cite{RefWorks:doc:6080351d8f08cc798956c9ab,RefWorks:doc:608035e18f087588f882c26d}.


\hrulefill
\vspace{0.5cm}
\begin{center}
    \textbf{Acknowledgements}
\end{center}
This work is supported by the Beit Fellowship Trustees of Imperial College London and the European Research Council (ERC) Consolidator Grant AFIRMATIVE (2018-23).

\newpage






\bibliographystyle{elsarticle-num-names}
\bibliography{export.bib}







\end{document}

%% file: figures/s_L0.tex
\begin{tikzpicture}[scale = 0.8]
\begin{axis}[
    xlabel={\ce{H2} Enrichment Level, $\etah$},
    ylabel={Laminar Flame Speed, $s_\trm{L0}$ [$\trm{m}\ \trm{s}^{-1}$]},
    xmin=0, xmax=1,
    ymin=0, ymax=3,
    xtick={0,0.2,0.4,0.6,0.8,1.0},
    ytick={0,0.5,1.0,1.5,2.0,2.5,3.0},
    y tick label style={
        /pgf/number format/.cd,
        fixed,
        fixed zerofill,
        precision=1,
        /tikz/.cd},
    legend pos=north west,
    legend cell align={left},
    ymajorgrids=true,
    grid style=dashed,
]

\addplot[
    color=blue,
    mark=o,
    ]
    coordinates {
    (0,0.3859)(0.05,0.3993)(0.1,0.4144)(0.15,0.4308)(0.2,0.4494)(0.25,0.4704)(0.3,0.4947)(0.35,0.5221)
    (0.4,0.5538)(0.45,0.5909)(0.5,0.6344)(0.55,0.6874)(0.6,0.7523)(0.65,0.8311)(0.7,0.9283)(0.75,1.0533)
    (0.8,1.2175)(0.85,1.4302)(0.9,1.7119)(0.95,2.0791)(1,2.5563)
    };
    \addlegendentry{Cantera, $\varphi = 1.08$}

\addplot [
    domain=0:1, 
    samples=100, 
    color=black,
]
{0.3829 + 1.90837*(0.00221 + 0.009*exp(100*x/21.30807))};
\addlegendentry{Dong et al., $\varphi = 1.08$}

\addplot[
    color=red,
    mark=square,
    ]
    coordinates {
    (0,0.2742)(0.05,0.2827)(0.1,0.2915)(0.15,0.3015)(0.2,0.3127)(0.25,0.3255)(0.3,0.3401)(0.35,0.3567)
    (0.4,0.3756)(0.45,0.3978)(0.5,0.4242)(0.55,0.4551)(0.6,0.4923)(0.65,0.5379)(0.7,0.5952)(0.75,0.6679)
    (0.8,0.7604)(0.85,0.8804)(0.9,1.0518)(0.95,1.2949)(1,1.6614)
    };
    \addlegendentry{Cantera, $\varphi = 0.8$}
    
\addplot [
    domain=0:1, 
    samples=100, 
    color=black,
    dashed,
]
{0.23315 + 1.4791*(0.00221 + 0.009*exp(100*x/21.30807))};
\addlegendentry{Dong et al., $\varphi = 0.8$}
    
\end{axis}
\end{tikzpicture}

%% file: figures/FF0_Length.tex
%
%
\begin{tikzpicture}[scale=0.6]

\begin{axis}[%
width=4.521in,
height=3.538in,
at={(0.758in,0.509in)},
scale only axis,
mark size = 1,
xmin=-11,
xmax=11,
xlabel style={font=\color{white!15!black}},
xlabel={$r_0\text{ [mm]}$},
ymin=0,
ymax=70,
ylabel style={font=\color{white!15!black}},
ylabel={$z_0\text{ [mm]}$},
axis background/.style={fill=white},
axis x line*=bottom,
axis y line*=left
]
\addplot [color=black, only marks, mark=*, mark options={solid, black}, forget plot]
  table[row sep=crcr]{%
-11	0\\
-10.9101526843356	0.5\\
-10.8181803604254	1\\
-10.7259710088181	1.5\\
-10.6335841844507	2\\
-10.5411044592407	2.5\\
-10.5	2.72220800253636\\
-10.4485938219191	3\\
-10.3560803953703	3.5\\
-10.2635682617576	4\\
-10.1710559888912	4.5\\
-10.0785434533989	5\\
-10	5.42450105337296\\
-9.98603073380236	5.5\\
-9.8935178187079	6\\
-9.80100492847181	6.5\\
-9.70849189953164	7\\
-9.61597873150776	7.5\\
-9.52346552000788	8\\
-9.5	8.12682259177835\\
-9.43095235518812	8.5\\
-9.33843924940748	9\\
-9.24592616234336	9.5\\
-9.1534130705312	10\\
-9.06089999070761	10.5\\
-9	10.8291427744556\\
-8.96838696148285	11\\
-8.87587399894684	11.5\\
-8.78336110876167	12\\
-8.69084827402378	12.5\\
-8.59833546485851	13\\
-8.5058226488634	13.5\\
-8.5	13.5314694120954\\
-8.41330981876945	14\\
-8.32079697098725	14.5\\
-8.22828410550845	15\\
-8.13577122556577	15.5\\
-8.04325833607497	16\\
-8	16.2337962417338\\
-7.95074543675041	16.5\\
-7.8582325205109	17\\
-7.76571958479581	17.5\\
-7.6732066263089	18\\
-7.5806936414375	18.5\\
-7.5	18.9361204860281\\
-7.48818062707017	19\\
-7.39566758858162	19.5\\
-7.30315453025492	20\\
-7.21064145806181	20.5\\
-7.11812837975598	21\\
-7.02561530424834	21.5\\
-7	21.6384415473742\\
-6.93310223741835	22\\
-6.84058918338392	22.5\\
-6.74807614608125	23\\
-6.65556312774784	23.5\\
-6.56305012907317	24\\
-6.5	24.3407636934823\\
-6.47053714997504	24.5\\
-6.37802419131115	25\\
-6.28551125179512	25.5\\
-6.19299832956306	26\\
-6.10048542220395	26.5\\
-6.00797252690593	27\\
-6	27.0430887365349\\
-5.91545963900042	27.5\\
-5.82294675372147	28\\
-5.73043386655726	28.5\\
-5.63792097315028	29\\
-5.54540806951528	29.5\\
-5.5	29.745414754875\\
-5.45289515263681	30\\
-5.36038222083851	30.5\\
-5.26786927307232	31\\
-5.17535630918531	31.5\\
-5.08284332978317	32\\
-5	32.4477388887419\\
-4.99033033624273	32.5\\
-4.89781732837402	33\\
-4.80530430979009	33.5\\
-4.7127912751353	34\\
-4.62027813898002	34.5\\
-4.52776498641766	35\\
-4.5	35.1500600279717\\
-4.43525213789448	35.5\\
-4.34273935231104	36\\
-4.25022647517678	36.5\\
-4.15771355104256	37\\
-4.06520060558698	37.5\\
-4	37.8523862609796\\
-3.97268761654176	38\\
-3.8801745634909	38.5\\
-3.78766148534078	39\\
-3.69514840869509	39.5\\
-3.60263536773731	40\\
-3.51012239166696	40.5\\
-3.5	40.5547080219523\\
-3.4176095675198	41\\
-3.32509682922676	41.5\\
-3.23258412747904	42\\
-3.14007140209636	42.5\\
-3.04755859106128	43\\
-3	43.2570374206078\\
-2.95504557609143	43.5\\
-2.86253235236119	44\\
-2.77001902256031	44.5\\
-2.67750567634586	45\\
-2.58499242135689	45.5\\
-2.5	45.9593535953638\\
-2.4924793664928	46\\
-2.39996673543333	46.5\\
-2.30745430722827	47\\
-2.21494195311381	47.5\\
-2.12242950198256	48\\
-2.02991679125307	48.5\\
-2	48.661689308859\\
-1.93740342380218	49\\
-1.84488955905843	49.5\\
-1.75237547536814	50\\
-1.65986144654323	50.5\\
-1.56734776460582	51\\
-1.5	51.3639905235362\\
-1.47483479929702	51.5\\
-1.38232280314112	52\\
-1.28981124583069	52.5\\
-1.19730025216602	53\\
-1.10478790110103	53.5\\
-1.01227823332606	54\\
-1	54.066356329043\\
-0.919759738923879	54.5\\
-0.827252090609783	55\\
-0.734724528870491	55.5\\
-0.642224392311322	56\\
-0.549674670673622	56.5\\
-0.5	56.7706931458726\\
-0.460384663429613	57\\
-0.37756030716621	57.5\\
-0.301001742088043	58\\
-0.232997448230456	58.5\\
-0.175784394706142	59\\
-0.130459763698503	59.5\\
-0.0950128426007118	60\\
-0.0682292065580554	60.5\\
-0.0471685227627621	61\\
0	61.1422996033018\\
0	61.1422996033018\\
11	0\\
10.9101526843356	0.5\\
10.8181803604254	1\\
10.7259710088181	1.5\\
10.6335841844507	2\\
10.5411044592407	2.5\\
10.5	2.72220800253636\\
10.4485938219191	3\\
10.3560803953703	3.5\\
10.2635682617576	4\\
10.1710559888912	4.5\\
10.0785434533989	5\\
10	5.42450105337296\\
9.98603073380236	5.5\\
9.8935178187079	6\\
9.80100492847181	6.5\\
9.70849189953164	7\\
9.61597873150776	7.5\\
9.52346552000788	8\\
9.5	8.12682259177835\\
9.43095235518812	8.5\\
9.33843924940748	9\\
9.24592616234336	9.5\\
9.1534130705312	10\\
9.06089999070761	10.5\\
9	10.8291427744556\\
8.96838696148285	11\\
8.87587399894684	11.5\\
8.78336110876167	12\\
8.69084827402378	12.5\\
8.59833546485851	13\\
8.5058226488634	13.5\\
8.5	13.5314694120954\\
8.41330981876945	14\\
8.32079697098725	14.5\\
8.22828410550845	15\\
8.13577122556577	15.5\\
8.04325833607497	16\\
8	16.2337962417338\\
7.95074543675041	16.5\\
7.8582325205109	17\\
7.76571958479581	17.5\\
7.6732066263089	18\\
7.5806936414375	18.5\\
7.5	18.9361204860281\\
7.48818062707017	19\\
7.39566758858162	19.5\\
7.30315453025492	20\\
7.21064145806181	20.5\\
7.11812837975598	21\\
7.02561530424834	21.5\\
7	21.6384415473742\\
6.93310223741835	22\\
6.84058918338392	22.5\\
6.74807614608125	23\\
6.65556312774784	23.5\\
6.56305012907317	24\\
6.5	24.3407636934823\\
6.47053714997504	24.5\\
6.37802419131115	25\\
6.28551125179512	25.5\\
6.19299832956306	26\\
6.10048542220395	26.5\\
6.00797252690593	27\\
6	27.0430887365349\\
5.91545963900042	27.5\\
5.82294675372147	28\\
5.73043386655726	28.5\\
5.63792097315028	29\\
5.54540806951528	29.5\\
5.5	29.745414754875\\
5.45289515263681	30\\
5.36038222083851	30.5\\
5.26786927307232	31\\
5.17535630918531	31.5\\
5.08284332978317	32\\
5	32.4477388887419\\
4.99033033624273	32.5\\
4.89781732837402	33\\
4.80530430979009	33.5\\
4.7127912751353	34\\
4.62027813898002	34.5\\
4.52776498641766	35\\
4.5	35.1500600279717\\
4.43525213789448	35.5\\
4.34273935231104	36\\
4.25022647517678	36.5\\
4.15771355104256	37\\
4.06520060558698	37.5\\
4	37.8523862609796\\
3.97268761654176	38\\
3.8801745634909	38.5\\
3.78766148534078	39\\
3.69514840869509	39.5\\
3.60263536773731	40\\
3.51012239166696	40.5\\
3.5	40.5547080219523\\
3.4176095675198	41\\
3.32509682922676	41.5\\
3.23258412747904	42\\
3.14007140209636	42.5\\
3.04755859106128	43\\
3	43.2570374206078\\
2.95504557609143	43.5\\
2.86253235236119	44\\
2.77001902256031	44.5\\
2.67750567634586	45\\
2.58499242135689	45.5\\
2.5	45.9593535953638\\
2.4924793664928	46\\
2.39996673543333	46.5\\
2.30745430722827	47\\
2.21494195311381	47.5\\
2.12242950198256	48\\
2.02991679125307	48.5\\
2	48.661689308859\\
1.93740342380218	49\\
1.84488955905843	49.5\\
1.75237547536814	50\\
1.65986144654323	50.5\\
1.56734776460582	51\\
1.5	51.3639905235362\\
1.47483479929702	51.5\\
1.38232280314112	52\\
1.28981124583069	52.5\\
1.19730025216602	53\\
1.10478790110103	53.5\\
1.01227823332606	54\\
1	54.066356329043\\
0.919759738923879	54.5\\
0.827252090609783	55\\
0.734724528870491	55.5\\
0.642224392311322	56\\
0.549674670673622	56.5\\
0.5	56.7706931458726\\
0.460384663429613	57\\
0.37756030716621	57.5\\
0.301001742088043	58\\
0.232997448230456	58.5\\
0.175784394706142	59\\
0.130459763698503	59.5\\
0.0950128426007118	60\\
0.0682292065580554	60.5\\
0.0471685227627621	61\\
-0	61.1422996033018\\
-0	61.1422996033018\\
};
\node[right, align=left]
at (axis cs:-0.5,64) {0\%};
\addplot [color=black, only marks, mark=*, mark options={solid, black}, forget plot]
  table[row sep=crcr]{%
-11	0\\
-10.8896647291961	0.5\\
-10.7807701336116	1\\
-10.6731333929757	1.5\\
-10.5654998404671	2\\
-10.5	2.30415614440744\\
-10.4577707658725	2.5\\
-10.3500188594516	3\\
-10.242274103091	3.5\\
-10.134530688755	4\\
-10.0267859512217	4.5\\
-10	4.62430160928566\\
-9.91903969837042	5\\
-9.81129352140395	5.5\\
-9.70354768529677	6\\
-9.59580193230454	6.5\\
-9.5	6.94457382658864\\
-9.48805612310779	7\\
-9.38031026010429	7.5\\
-9.27256439662799	8\\
-9.16481855747197	8.5\\
-9.05707274970005	9\\
-9	9.26484912150622\\
-8.94932697153297	9.5\\
-8.84158123083291	10\\
-8.73383552698272	10.5\\
-8.62608986476215	11\\
-8.5183442528666	11.5\\
-8.5	11.5851276470827\\
-8.41059868920412	12\\
-8.30285316396447	12.5\\
-8.19510766485072	13\\
-8.0873621760463	13.5\\
-8	13.9054098834431\\
-7.97961668218874	14\\
-7.87187117768211	14.5\\
-7.76412565810952	15\\
-7.6563801220596	15.5\\
-7.54863457077633	16\\
-7.5	16.2256917506518\\
-7.4408890044711	16.5\\
-7.33314342033401	17\\
-7.22539781669634	17.5\\
-7.11765219107488	18\\
-7.00990654055515	18.5\\
-7	18.5459718712408\\
-6.90216086504932	19\\
-6.79441516572368	19.5\\
-6.68666944429181	20\\
-6.57892370374788	20.5\\
-6.5	20.8662497106913\\
-6.47117794782897	21\\
-6.36343217950148	21.5\\
-6.25568640279664	22\\
-6.14794062180674	22.5\\
-6.04019484063986	23\\
-6	23.1865262927769\\
-5.93244906400729	23.5\\
-5.82470329709041	24\\
-5.7169575443579	24.5\\
-5.60921180977042	25\\
-5.50146609653679	25.5\\
-5.5	25.5068035042518\\
-5.39372040643341	26\\
-5.28597473975574	26.5\\
-5.17822909581748	27\\
-5.07048347294091	27.5\\
-5	27.8270828182957\\
-4.96273786701524	28\\
-4.85499227230941	28.5\\
-4.74724665683171	29\\
-4.6395009389206	29.5\\
-4.5317552349934	30\\
-4.5	30.1473622903531\\
-4.42400987392908	30.5\\
-4.316264540901	31\\
-4.20851908162886	31.5\\
-4.10077354599006	32\\
-4	32.4676458153739\\
-3.99302795639386	32.5\\
-3.88528228430078	33\\
-3.77753657138975	33.5\\
-3.66979084185002	34\\
-3.56204512527095	34.5\\
-3.5	34.7879239881916\\
-3.45429947417953	35\\
-3.34655389481002	35.5\\
-3.23880833304822	36\\
-3.13106274315215	36.5\\
-3.02331707716792	37\\
-3	37.1082041090476\\
-2.91557122347426	37.5\\
-2.80782526053265	38\\
-2.70007927246806	38.5\\
-2.59233335051626	39\\
-2.5	39.428477867949\\
-2.48458760585427	39.5\\
-2.37684217943166	40\\
-2.26909688279789	40.5\\
-2.16135158919213	41\\
-2.05360614938103	41.5\\
-2	41.7487622127055\\
-1.94586028000477	42\\
-1.83811401664167	42.5\\
-1.73036763999932	43\\
-1.62262138828418	43.5\\
-1.51487549685827	44\\
-1.5	44.0690308174777\\
-1.40713039283465	44.5\\
-1.29938582352056	45\\
-1.19164084876637	45.5\\
-1.08389805834946	46\\
-1	46.3893183326616\\
-0.976147855098347	46.5\\
-0.868410048392932	47\\
-0.760648920398691	47.5\\
-0.652918052163405	48\\
-0.545151758293404	48.5\\
-0.5	48.7121114114148\\
-0.442860431317453	49\\
-0.348896141219309	49.5\\
-0.264156879683897	50\\
-0.19186338028884	50.5\\
-0.135174399791263	51\\
-0.0928899761344624	51.5\\
-0.0627822498984209	52\\
-0.0406845848202985	52.5\\
0	52.6146756900666\\
0	52.6146756900666\\
11	0\\
10.8896647291961	0.5\\
10.7807701336116	1\\
10.6731333929757	1.5\\
10.5654998404671	2\\
10.5	2.30415614440744\\
10.4577707658725	2.5\\
10.3500188594516	3\\
10.242274103091	3.5\\
10.134530688755	4\\
10.0267859512217	4.5\\
10	4.62430160928566\\
9.91903969837042	5\\
9.81129352140395	5.5\\
9.70354768529677	6\\
9.59580193230454	6.5\\
9.5	6.94457382658864\\
9.48805612310779	7\\
9.38031026010429	7.5\\
9.27256439662799	8\\
9.16481855747197	8.5\\
9.05707274970005	9\\
9	9.26484912150622\\
8.94932697153297	9.5\\
8.84158123083291	10\\
8.73383552698272	10.5\\
8.62608986476215	11\\
8.5183442528666	11.5\\
8.5	11.5851276470827\\
8.41059868920412	12\\
8.30285316396447	12.5\\
8.19510766485072	13\\
8.0873621760463	13.5\\
8	13.9054098834431\\
7.97961668218874	14\\
7.87187117768211	14.5\\
7.76412565810952	15\\
7.6563801220596	15.5\\
7.54863457077633	16\\
7.5	16.2256917506518\\
7.4408890044711	16.5\\
7.33314342033401	17\\
7.22539781669634	17.5\\
7.11765219107488	18\\
7.00990654055515	18.5\\
7	18.5459718712408\\
6.90216086504932	19\\
6.79441516572368	19.5\\
6.68666944429181	20\\
6.57892370374788	20.5\\
6.5	20.8662497106913\\
6.47117794782897	21\\
6.36343217950148	21.5\\
6.25568640279664	22\\
6.14794062180674	22.5\\
6.04019484063986	23\\
6	23.1865262927769\\
5.93244906400729	23.5\\
5.82470329709041	24\\
5.7169575443579	24.5\\
5.60921180977042	25\\
5.50146609653679	25.5\\
5.5	25.5068035042518\\
5.39372040643341	26\\
5.28597473975574	26.5\\
5.17822909581748	27\\
5.07048347294091	27.5\\
5	27.8270828182957\\
4.96273786701524	28\\
4.85499227230941	28.5\\
4.74724665683171	29\\
4.6395009389206	29.5\\
4.5317552349934	30\\
4.5	30.1473622903531\\
4.42400987392908	30.5\\
4.316264540901	31\\
4.20851908162886	31.5\\
4.10077354599006	32\\
4	32.4676458153739\\
3.99302795639386	32.5\\
3.88528228430078	33\\
3.77753657138975	33.5\\
3.66979084185002	34\\
3.56204512527095	34.5\\
3.5	34.7879239881916\\
3.45429947417953	35\\
3.34655389481002	35.5\\
3.23880833304822	36\\
3.13106274315215	36.5\\
3.02331707716792	37\\
3	37.1082041090476\\
2.91557122347426	37.5\\
2.80782526053265	38\\
2.70007927246806	38.5\\
2.59233335051626	39\\
2.5	39.428477867949\\
2.48458760585427	39.5\\
2.37684217943166	40\\
2.26909688279789	40.5\\
2.16135158919213	41\\
2.05360614938103	41.5\\
2	41.7487622127055\\
1.94586028000477	42\\
1.83811401664167	42.5\\
1.73036763999932	43\\
1.62262138828418	43.5\\
1.51487549685827	44\\
1.5	44.0690308174777\\
1.40713039283465	44.5\\
1.29938582352056	45\\
1.19164084876637	45.5\\
1.08389805834946	46\\
1	46.3893183326616\\
0.976147855098347	46.5\\
0.868410048392932	47\\
0.760648920398691	47.5\\
0.652918052163405	48\\
0.545151758293404	48.5\\
0.5	48.7121114114148\\
0.442860431317453	49\\
0.348896141219309	49.5\\
0.264156879683897	50\\
0.19186338028884	50.5\\
0.135174399791263	51\\
0.0928899761344624	51.5\\
0.0627822498984209	52\\
0.0406845848202985	52.5\\
-0	52.6146756900666\\
-0	52.6146756900666\\
};
\node[right, align=left]
at (axis cs:-0.9,54.5) {20\%};
\addplot [color=black, only marks, mark=*, mark options={solid, black}, forget plot]
  table[row sep=crcr]{%
-11	0\\
-10.8602326829428	0.5\\
-10.7258329311372	1\\
-10.5929314103431	1.5\\
-10.5	1.84967608575561\\
-10.4598432571793	2\\
-10.3265593532218	2.5\\
-10.1932976637763	3\\
-10.0600388697573	3.5\\
-10	3.72527370799649\\
-9.92677708839088	4\\
-9.79351438793718	4.5\\
-9.66025199512036	5\\
-9.52698954272934	5.5\\
-9.5	5.60126452752915\\
-9.39372685071439	6\\
-9.26046419709031	6.5\\
-9.12720156953587	7\\
-9	7.47725897080802\\
-8.99393894159975	7.5\\
-8.86067631834965	8\\
-8.7274137081956	8.5\\
-8.59415110775412	9\\
-8.5	9.35325406551833\\
-8.46088851159804	9.5\\
-8.32762591585683	10\\
-8.19436331802712	10.5\\
-8.06110071663728	11\\
-8	11.2292492944332\\
-7.92783811076196	11.5\\
-7.79457550043473	12\\
-7.66131288572052	12.5\\
-7.52805026677451	13\\
-7.5	13.1052443140503\\
-7.39478764310273	13.5\\
-7.26152501370312	14\\
-7.12826237755309	14.5\\
-7	14.9812390541605\\
-6.99499973347697	15\\
-6.86173708067921	15.5\\
-6.72847441879993	16\\
-6.59521174748733	16.5\\
-6.5	16.8572333488606\\
-6.46194906646896	17\\
-6.32868637527196	17.5\\
-6.19542367368112	18\\
-6.06216096143249	18.5\\
-6	18.7332271178985\\
-5.92889823836868	19\\
-5.79563550485287	19.5\\
-5.66237276153987	20\\
-5.52911000928439	20.5\\
-5.5	20.6092203453662\\
-5.39584724913124	21\\
-5.26258448256631	21.5\\
-5.12932170878034	22\\
-5	22.485213129371\\
-4.9960589206732	22.5\\
-4.86279606983972	23\\
-4.7295330732225	23.5\\
-4.59627006226245	24\\
-4.5	24.361203820307\\
-4.46300731831149	24.5\\
-4.32974474909094	25\\
-4.19648208360814	25.5\\
-4.06321937310997	26\\
-4	26.2371982544539\\
-3.92995663050923	26.5\\
-3.79669387414993	27\\
-3.66343113202048	27.5\\
-3.53016842477945	28\\
-3.5	28.1131915975294\\
-3.39690579348038	28.5\\
-3.26364320220656	29\\
-3.13038061180188	29.5\\
-3	29.989186708052\\
-2.99711798341436	30\\
-2.86385524372469	30.5\\
-2.73059248446843	31\\
-2.59732976535861	31.5\\
-2.5	31.8651803141625\\
-2.46406717741395	32\\
-2.33080475885481	32.5\\
-2.19754236435853	33\\
-2.0642798847283	33.5\\
-2	33.7411773562743\\
-1.93101712335911	34\\
-1.79775414783445	34.5\\
-1.66449118295506	35\\
-1.5312284065354	35.5\\
-1.5	35.6171688098548\\
-1.39796609967135	36\\
-1.26470415701167	36.5\\
-1.13144104668437	37\\
-1	37.4931812041193\\
-0.998182672932108	37.5\\
-0.864912264407995	38\\
-0.731660300417125	38.5\\
-0.598370854830698	39\\
-0.5	39.3713149437289\\
-0.468557238008618	39.5\\
-0.352646814567745	40\\
-0.249495150361903	40.5\\
-0.164894180474873	41\\
-0.102961682814652	41.5\\
-0.0622464940850883	42\\
-0.0360457360004166	42.5\\
0	42.5999613710245\\
0	42.5999613710245\\
11	0\\
10.8602326829428	0.5\\
10.7258329311372	1\\
10.5929314103431	1.5\\
10.5	1.84967608575561\\
10.4598432571793	2\\
10.3265593532218	2.5\\
10.1932976637763	3\\
10.0600388697573	3.5\\
10	3.72527370799649\\
9.92677708839088	4\\
9.79351438793718	4.5\\
9.66025199512036	5\\
9.52698954272934	5.5\\
9.5	5.60126452752915\\
9.39372685071439	6\\
9.26046419709031	6.5\\
9.12720156953587	7\\
9	7.47725897080802\\
8.99393894159975	7.5\\
8.86067631834965	8\\
8.7274137081956	8.5\\
8.59415110775412	9\\
8.5	9.35325406551833\\
8.46088851159804	9.5\\
8.32762591585683	10\\
8.19436331802712	10.5\\
8.06110071663728	11\\
8	11.2292492944332\\
7.92783811076196	11.5\\
7.79457550043473	12\\
7.66131288572052	12.5\\
7.52805026677451	13\\
7.5	13.1052443140503\\
7.39478764310273	13.5\\
7.26152501370312	14\\
7.12826237755309	14.5\\
7	14.9812390541605\\
6.99499973347697	15\\
6.86173708067921	15.5\\
6.72847441879993	16\\
6.59521174748733	16.5\\
6.5	16.8572333488606\\
6.46194906646896	17\\
6.32868637527196	17.5\\
6.19542367368112	18\\
6.06216096143249	18.5\\
6	18.7332271178985\\
5.92889823836868	19\\
5.79563550485287	19.5\\
5.66237276153987	20\\
5.52911000928439	20.5\\
5.5	20.6092203453662\\
5.39584724913124	21\\
5.26258448256631	21.5\\
5.12932170878034	22\\
5	22.485213129371\\
4.9960589206732	22.5\\
4.86279606983972	23\\
4.7295330732225	23.5\\
4.59627006226245	24\\
4.5	24.361203820307\\
4.46300731831149	24.5\\
4.32974474909094	25\\
4.19648208360814	25.5\\
4.06321937310997	26\\
4	26.2371982544539\\
3.92995663050923	26.5\\
3.79669387414993	27\\
3.66343113202048	27.5\\
3.53016842477945	28\\
3.5	28.1131915975294\\
3.39690579348038	28.5\\
3.26364320220656	29\\
3.13038061180188	29.5\\
3	29.989186708052\\
2.99711798341436	30\\
2.86385524372469	30.5\\
2.73059248446843	31\\
2.59732976535861	31.5\\
2.5	31.8651803141625\\
2.46406717741395	32\\
2.33080475885481	32.5\\
2.19754236435853	33\\
2.0642798847283	33.5\\
2	33.7411773562743\\
1.93101712335911	34\\
1.79775414783445	34.5\\
1.66449118295506	35\\
1.5312284065354	35.5\\
1.5	35.6171688098548\\
1.39796609967135	36\\
1.26470415701167	36.5\\
1.13144104668437	37\\
1	37.4931812041193\\
0.998182672932108	37.5\\
0.864912264407995	38\\
0.731660300417125	38.5\\
0.598370854830698	39\\
0.5	39.3713149437289\\
0.468557238008618	39.5\\
0.352646814567745	40\\
0.249495150361903	40.5\\
0.164894180474873	41\\
0.102961682814652	41.5\\
0.0622464940850883	42\\
0.0360457360004166	42.5\\
-0	42.5999613710245\\
-0	42.5999613710245\\
};
\node[right, align=left]
at (axis cs:-0.9,45) {40\%};
\addplot [color=black, only marks, mark=*, mark options={solid, black}, forget plot]
  table[row sep=crcr]{%
-11	0\\
-10.8002576824338	0.5\\
-10.6159591411316	1\\
-10.5	1.31873484005051\\
-10.4327429468349	1.5\\
-10.2482771140159	2\\
-10.0639524332992	2.5\\
-10	2.67360030116494\\
-9.87966025671281	3\\
-9.69540439323524	3.5\\
-9.51116245374161	4\\
-9.5	4.03029211229807\\
-9.32690814925814	4.5\\
-9.14264819781265	5\\
-9	5.38708378485466\\
-8.95838800440748	5.5\\
-8.77412929807922	6\\
-8.58987142096044	6.5\\
-8.5	6.74387390814143\\
-8.40561342291785	7\\
-8.22135514205586	7.5\\
-8.03709676946976	8\\
-8	8.10066509817266\\
-7.85283842544141	8.5\\
-7.66858011462918	9\\
-7.5	9.45745593810531\\
-7.48432180613897	9.5\\
-7.30006348581436	10\\
-7.11580515770409	10.5\\
-7	10.8142467366356\\
-6.93154682786946	11\\
-6.74728849886016	11.5\\
-6.5630301706686	12\\
-6.5	12.1710375088122\\
-6.37877184311489	12.5\\
-6.19451351650244	13\\
-6.01025519132377	13.5\\
-6	13.5278282986741\\
-5.82599686797855	14\\
-5.64173854694679	14.5\\
-5.5	14.8846191281658\\
-5.45748022956265	15\\
-5.27322191816255	15.5\\
-5.08896359782992	16\\
-5	16.2414098769591\\
-4.90470515514756	16.5\\
-4.72044661147867	17\\
-4.53618818002565	17.5\\
-4.5	17.5981995930819\\
-4.3519299662693	18\\
-4.16767170119715	18.5\\
-4	18.9549909130954\\
-3.98341340168568	19\\
-3.79915508507431	19.5\\
-3.61489676997304	20\\
-3.5	20.311781783027\\
-3.43063845937083	20.5\\
-3.24638015169248	21\\
-3.06212183987016	21.5\\
-3	21.6685726864517\\
-2.87786351638853	22\\
-2.69360519041872	22.5\\
-2.50934687113886	23\\
-2.5	23.025363499709\\
-2.32508856871929	23.5\\
-2.14083026339568	24\\
-2	24.3821544371401\\
-1.95657194885139	24.5\\
-1.77231363158278	25\\
-1.58805531171683	25.5\\
-1.5	25.738945238648\\
-1.40379694675228	26\\
-1.2195384773432	26.5\\
-1.03528080920877	27\\
-1	27.0957358683643\\
-0.851019683933238	27.5\\
-0.666750497234466	28\\
-0.5	28.4536299219845\\
-0.484605420663932	28.5\\
-0.328798918364408	29\\
-0.197277409887609	29.5\\
-0.101967794640622	30\\
-0.0467650678636185	30.5\\
0	30.6246253566983\\
0	30.6246253566983\\
11	0\\
10.8002576824338	0.5\\
10.6159591411316	1\\
10.5	1.31873484005051\\
10.4327429468349	1.5\\
10.2482771140159	2\\
10.0639524332992	2.5\\
10	2.67360030116494\\
9.87966025671281	3\\
9.69540439323524	3.5\\
9.51116245374161	4\\
9.5	4.03029211229807\\
9.32690814925814	4.5\\
9.14264819781265	5\\
9	5.38708378485466\\
8.95838800440748	5.5\\
8.77412929807922	6\\
8.58987142096044	6.5\\
8.5	6.74387390814143\\
8.40561342291785	7\\
8.22135514205586	7.5\\
8.03709676946976	8\\
8	8.10066509817266\\
7.85283842544141	8.5\\
7.66858011462918	9\\
7.5	9.45745593810531\\
7.48432180613897	9.5\\
7.30006348581436	10\\
7.11580515770409	10.5\\
7	10.8142467366356\\
6.93154682786946	11\\
6.74728849886016	11.5\\
6.5630301706686	12\\
6.5	12.1710375088122\\
6.37877184311489	12.5\\
6.19451351650244	13\\
6.01025519132377	13.5\\
6	13.5278282986741\\
5.82599686797855	14\\
5.64173854694679	14.5\\
5.5	14.8846191281658\\
5.45748022956265	15\\
5.27322191816255	15.5\\
5.08896359782992	16\\
5	16.2414098769591\\
4.90470515514756	16.5\\
4.72044661147867	17\\
4.53618818002565	17.5\\
4.5	17.5981995930819\\
4.3519299662693	18\\
4.16767170119715	18.5\\
4	18.9549909130954\\
3.98341340168568	19\\
3.79915508507431	19.5\\
3.61489676997304	20\\
3.5	20.311781783027\\
3.43063845937083	20.5\\
3.24638015169248	21\\
3.06212183987016	21.5\\
3	21.6685726864517\\
2.87786351638853	22\\
2.69360519041872	22.5\\
2.50934687113886	23\\
2.5	23.025363499709\\
2.32508856871929	23.5\\
2.14083026339568	24\\
2	24.3821544371401\\
1.95657194885139	24.5\\
1.77231363158278	25\\
1.58805531171683	25.5\\
1.5	25.738945238648\\
1.40379694675228	26\\
1.2195384773432	26.5\\
1.03528080920877	27\\
1	27.0957358683643\\
0.851019683933238	27.5\\
0.666750497234466	28\\
0.5	28.4536299219845\\
0.484605420663932	28.5\\
0.328798918364408	29\\
0.197277409887609	29.5\\
0.101967794640622	30\\
0.0467650678636185	30.5\\
-0	30.6246253566983\\
-0	30.6246253566983\\
};
\node[right, align=left]
at (axis cs:-0.9,33) {60\%};
\addplot [color=black, only marks, mark=*, mark options={solid, black}, forget plot]
  table[row sep=crcr]{%
-11	0\\
-10.6396128127771	0.5\\
-10.5	0.72002371616831\\
-10.3117232203203	1\\
-10	1.47531708063312\\
-9.98364798554827	1.5\\
-9.65462905868893	2\\
-9.5	2.23564784052401\\
-9.32614094149603	2.5\\
-9	2.99633471588397\\
-8.99759022069897	3\\
-8.66899019283085	3.5\\
-8.5	3.75716077187651\\
-8.34040547933225	4\\
-8.01181776052517	4.5\\
-8	4.51798289648139\\
-7.68322913062069	5\\
-7.5	5.27881336511994\\
-7.35464115084147	5.5\\
-7.02605302413998	6\\
-7	6.03964391310157\\
-6.69746485031114	6.5\\
-6.5	6.80047471345941\\
-6.36887669584678	7\\
-6.04028853825637	7.5\\
-6	7.56130552619914\\
-5.71170038636496	8\\
-5.5	8.32213637572253\\
-5.38311224566527	8.5\\
-5.0545241074067	9\\
-5	9.08296724656385\\
-4.72593595136789	9.5\\
-4.5	9.84379809208062\\
-4.39734781929796	10\\
-4.06875969729007	10.5\\
-4	10.604629007725\\
-3.74017158113146	11\\
-3.5	11.365459924343\\
-3.4115834512389	11.5\\
-3.08299534080386	12\\
-3	12.1262908769005\\
-2.75440736393297	12.5\\
-2.5	12.8871221301986\\
-2.42581933751596	13\\
-2.09723153228697	13.5\\
-2	13.647954024004\\
-1.76864495731378	14\\
-1.5	14.4087888949568\\
-1.44005821553328	14.5\\
-1.1114743625696	15\\
-1	15.1696188087739\\
-0.782907909204289	15.5\\
-0.5	15.9326297240533\\
-0.462295447913783	16\\
-0.218819989401057	16.5\\
-0.0846489018360257	17\\
0	17.2120095983087\\
0	17.2120095983087\\
11	0\\
10.6396128127771	0.5\\
10.5	0.72002371616831\\
10.3117232203203	1\\
10	1.47531708063312\\
9.98364798554827	1.5\\
9.65462905868893	2\\
9.5	2.23564784052401\\
9.32614094149603	2.5\\
9	2.99633471588397\\
8.99759022069897	3\\
8.66899019283085	3.5\\
8.5	3.75716077187651\\
8.34040547933225	4\\
8.01181776052517	4.5\\
8	4.51798289648139\\
7.68322913062069	5\\
7.5	5.27881336511994\\
7.35464115084147	5.5\\
7.02605302413998	6\\
7	6.03964391310157\\
6.69746485031114	6.5\\
6.5	6.80047471345941\\
6.36887669584678	7\\
6.04028853825637	7.5\\
6	7.56130552619914\\
5.71170038636496	8\\
5.5	8.32213637572253\\
5.38311224566527	8.5\\
5.0545241074067	9\\
5	9.08296724656385\\
4.72593595136789	9.5\\
4.5	9.84379809208062\\
4.39734781929796	10\\
4.06875969729007	10.5\\
4	10.604629007725\\
3.74017158113146	11\\
3.5	11.365459924343\\
3.4115834512389	11.5\\
3.08299534080386	12\\
3	12.1262908769005\\
2.75440736393297	12.5\\
2.5	12.8871221301986\\
2.42581933751596	13\\
2.09723153228697	13.5\\
2	13.647954024004\\
1.76864495731378	14\\
1.5	14.4087888949568\\
1.44005821553328	14.5\\
1.1114743625696	15\\
1	15.1696188087739\\
0.782907909204289	15.5\\
0.5	15.9326297240533\\
0.462295447913783	16\\
0.218819989401057	16.5\\
0.0846489018360257	17\\
-0	17.2120095983087\\
-0	17.2120095983087\\
};
\node[right, align=left]
at (axis cs:-0.9,20) {80\%};
\end{axis}

\begin{axis}[%
width=5.833in,
height=4.375in,
at={(0in,0in)},
scale only axis,
xmin=0,
xmax=1,
ymin=0,
ymax=1,
axis line style={draw=none},
ticks=none,
axis x line*=bottom,
axis y line*=left
]
\end{axis}
\end{tikzpicture}%

%% file: figures/QR_Example.tex
\begin{tikzpicture}[scale=0.7]
\begin{axis}[
    xlabel={Physical Time, $t\ [\trm{ms}]$},
    ylabel={Normalised Heat Release Rate, $\dot{Q}(t)/\ta{\dot{Q}}$},
    xmin=0, xmax=125,
    ymin=0.985, ymax=1.01,
    xtick={0,25,50,75,100,125},
    ytick={0.985,0.990,0.995,1.000,1.005,1.01},
    y tick label style={
        /pgf/number format/.cd,
        fixed,
        fixed zerofill,
        precision=3,
        /tikz/.cd
    },
    legend pos=north east,
    legend cell align={left},
    ymajorgrids=true,
    grid style=dashed,
]

\addplot[
    color=black,
    forget plot]
    table {
    0.625000 1.001134
1.250000 1.002099
1.875000 1.002908
2.500000 1.003476
3.125000 1.003804
3.750000 1.003881
4.375000 1.003689
5.000000 1.003246
5.625000 1.002670
6.250000 1.002095
6.875000 1.001490
7.500000 1.001089
8.125000 1.000920
8.750000 1.001135
9.375000 1.001800
10.000000 1.002862
10.625000 1.004143
11.250000 1.005418
11.875000 1.006483
12.500000 1.007159
13.125000 1.007318
13.750000 1.006910
14.375000 1.005946
15.000000 1.004460
15.625000 1.002547
16.250000 1.000255
16.875000 0.997760
17.500000 0.995349
18.125000 0.993921
18.750000 0.993875
19.375000 0.995061
20.000000 0.996931
20.625000 0.998962
21.250000 1.000765
21.875000 1.002095
22.500000 1.002788
23.125000 1.002767
23.750000 1.002006
24.375000 1.000502
25.000000 0.998304
25.625000 0.995546
26.250000 0.992294
26.875000 0.988867
27.500000 0.986234
28.125000 0.985243
28.750000 0.986202
29.375000 0.988640
30.000000 0.991724
30.625000 0.994783
31.250000 0.997408
31.875000 0.999410
32.500000 1.000660
33.125000 1.001107
33.750000 1.000745
34.375000 0.999580
35.000000 0.997680
35.625000 0.995205
36.250000 0.992216
36.875000 0.989028
37.500000 0.986550
38.125000 0.985581
38.750000 0.986393
39.375000 0.988708
40.000000 0.991739
40.625000 0.994784
41.250000 0.997404
41.875000 0.999403
42.500000 1.000652
43.125000 1.001098
43.750000 1.000736
44.375000 0.999567
45.000000 0.997668
45.625000 0.995189
46.250000 0.992201
46.875000 0.989015
47.500000 0.986530
48.125000 0.985558
48.750000 0.986370
49.375000 0.988686
50.000000 0.991718
50.625000 0.994763
51.250000 0.997382
51.875000 0.999382
52.500000 1.000632
53.125000 1.001079
53.750000 1.000717
54.375000 0.999547
55.000000 0.997649
55.625000 0.995170
56.250000 0.992182
56.875000 0.988996
57.500000 0.986511
58.125000 0.985541
58.750000 0.986352
59.375000 0.988667
60.000000 0.991701
60.625000 0.994746
61.250000 0.997366
61.875000 0.999366
62.500000 1.000616
63.125000 1.001063
63.750000 1.000702
64.375000 0.999532
65.000000 0.997634
65.625000 0.995156
66.250000 0.992169
66.875000 0.988983
67.500000 0.986499
68.125000 0.985530
68.750000 0.986345
69.375000 0.988661
70.000000 0.991694
70.625000 0.994739
71.250000 0.997359
71.875000 0.999359
72.500000 1.000609
73.125000 1.001057
73.750000 1.000696
74.375000 0.999527
75.000000 0.997628
75.625000 0.995151
76.250000 0.992163
76.875000 0.988978
77.500000 0.986495
78.125000 0.985526
78.750000 0.986341
79.375000 0.988659
80.000000 0.991693
80.625000 0.994738
81.250000 0.997358
81.875000 0.999358
82.500000 1.000609
83.125000 1.001056
83.750000 1.000695
84.375000 0.999526
85.000000 0.997628
85.625000 0.995150
86.250000 0.992163
86.875000 0.988978
87.500000 0.986495
88.125000 0.985527
88.750000 0.986341
89.375000 0.988657
90.000000 0.991693
90.625000 0.994738
91.250000 0.997358
91.875000 0.999358
92.500000 1.000609
93.125000 1.001056
93.750000 1.000696
94.375000 0.999526
95.000000 0.997628
95.625000 0.995150
96.250000 0.992163
96.875000 0.988978
97.500000 0.986495
98.125000 0.985527
98.750000 0.986343
99.375000 0.988659
100.000000 0.991693
100.625000 0.994738
101.250000 0.997358
101.875000 0.999358
102.500000 1.000609
103.125000 1.001056
103.750000 1.000696
104.375000 0.999526
105.000000 0.997628
105.625000 0.995150
106.250000 0.992163
106.875000 0.988978
107.500000 0.986495
108.125000 0.985526
108.750000 0.986341
109.375000 0.988659
110.000000 0.991693
110.625000 0.994738
111.250000 0.997358
111.875000 0.999358
112.500000 1.000609
113.125000 1.001056
113.750000 1.000695
114.375000 0.999526
115.000000 0.997628
115.625000 0.995150
116.250000 0.992163
116.875000 0.988978
117.500000 0.986495
118.125000 0.985527
118.750000 0.986341
119.375000 0.988657
120.000000 0.991692
120.625000 0.994738
121.250000 0.997358
121.875000 0.999358
122.500000 1.000609
123.125000 1.001056
123.750000 1.000695
124.375000 0.999526
125.000000 0.997628
};

\addplot[only marks, mark=*, mark options={solid, red}, mark size=2pt]
coordinates{(56.1,0.9929)};
\addlegendentry{$t/\mathcal{T} = 1/4$}

\addplot[only marks, mark=square*, mark options={solid, blue}, mark size=2pt]
coordinates{(58.2,0.9855)};
\addlegendentry{$t/\mathcal{T} = 1/2$}

\addplot[only marks, mark=triangle*, mark options={solid, green}, mark size=3pt]
coordinates{(60.2,0.9930)};
\addlegendentry{$t/\mathcal{T} = 3/4$}

\addplot[only marks, mark=x, mark options={solid, magenta}, mark size=3pt]
coordinates{(63.2,1.0011)};
\addlegendentry{$t/\mathcal{T} = 1$}

\end{axis}
\end{tikzpicture}

%% file: figures/FF_Example.tex
%
%
\begin{tikzpicture}[scale=0.7]

\begin{axis}[%
width=1.743in,
height=1.777in,
at={(0.677in,3.041in)},
scale only axis,
xmin=-11,
xmax=11,
xtick={-10,   0,  10},
ymin=0,
ymax=80,
ytick={ 0, 20, 40, 60, 80},
ylabel style={font=\color{white!15!black}},
ylabel={$z$ [mm]},
axis background/.style={fill=white},
title={$t/\mathcal{T} = 1/4$},
axis x line*=bottom,
axis y line*=left
]

\addplot [color=red, only marks, mark=*, mark options={solid, red}, mark size=0.5pt, forget plot]
  table[row sep=crcr]{%
-11	0\\
-10.9249520899817	0.5\\
-10.8294934109004	1\\
-10.7245174965417	1.5\\
-10.6106269243	2\\
-10.5	2.45419571468371\\
-10.4887951321466	2.5\\
-10.3605399881615	3\\
-10.2277886777402	3.5\\
-10.0924390094735	4\\
-10	4.33973087627795\\
-9.95640197118805	4.5\\
-9.82153244785092	5\\
-9.68966082111244	5.5\\
-9.56257789686622	6\\
-9.5	6.25896689204695\\
-9.44197405054831	6.5\\
-9.3292232616987	7\\
-9.22541440862522	7.5\\
-9.13141683844462	8\\
-9.04783713290217	8.5\\
-9	8.82819417226743\\
-8.97504263831364	9\\
-8.91314747557791	9.5\\
-8.86204918486945	10\\
-8.82146062438822	10.5\\
-8.79087450490365	11\\
-8.76954788090248	11.5\\
-8.75647937933854	12\\
-8.75038002797663	12.5\\
-8.74963899360751	13\\
-8.75228744721663	13.5\\
-8.75597546932219	14\\
-8.7579617805969	14.5\\
-8.75510039071427	15\\
-8.74389851217116	15.5\\
-8.72063688085278	16\\
-8.68155597994042	16.5\\
-8.62311273151125	17\\
-8.5422806678601	17.5\\
-8.5	17.7016813925057\\
-8.43672124556067	18\\
-8.30529758855132	18.5\\
-8.1484193868514	19\\
-8	19.4119124759721\\
-7.96771954291109	19.5\\
-7.76576846869582	20\\
-7.54746786772789	20.5\\
-7.5	20.603734401943\\
-7.31717128661607	21\\
-7.08040464942759	21.5\\
-7	21.6686028372957\\
-6.84207510179312	22\\
-6.60660557490045	22.5\\
-6.5	22.7321088608239\\
-6.3778902958993	23\\
-6.15887147787019	23.5\\
-6	23.882512639768\\
-5.95173979898502	24\\
-5.7584536793006	24.5\\
-5.57966142407837	25\\
-5.5	25.2425520187427\\
-5.41637750714721	25.5\\
-5.26887535317003	26\\
-5.13712912836617	26.5\\
-5.02117202351121	27\\
-5	27.1050586520783\\
-4.9209625114127	27.5\\
-4.83613263975368	28\\
-4.76640247868822	28.5\\
-4.71148277701625	29\\
-4.67105632487221	29.5\\
-4.64477840754965	30\\
-4.63226674594118	30.5\\
-4.63307954398334	31\\
-4.64667866342646	31.5\\
-4.6723747675152	32\\
-4.70925119854582	32.5\\
-4.75606335422924	33\\
-4.81111041244298	33.5\\
-4.8720765541569	34\\
-4.93586492700853	34.5\\
-4.99848847213127	35\\
-5	35.0133789147081\\
-5.05487757642825	35.5\\
-5.09888824758841	36\\
-5.12353426763028	36.5\\
-5.12147037804902	37\\
-5.08577123362424	37.5\\
-5.01081757746309	38\\
-5	38.0466684246058\\
-4.89252517297061	38.5\\
-4.73017086751199	39\\
-4.52676235376425	39.5\\
-4.5	39.5572149712002\\
-4.28687057330901	40\\
-4.02123415346224	40.5\\
-4	40.5380199537731\\
-3.73897588549566	41\\
-3.5	41.4149551855106\\
-3.45099309355943	41.5\\
-3.16578709168738	42\\
-3	42.2991788278281\\
-2.89009140269527	42.5\\
-2.62844449349911	43\\
-2.5	43.259889268844\\
-2.38323530186607	43.5\\
-2.15643227073841	44\\
-2	44.3755540005139\\
-1.94911654881508	44.5\\
-1.76146657608844	45\\
-1.59167978828544	45.5\\
-1.5	45.7987777626261\\
-1.4386009900869	46\\
-1.30120584212458	46.5\\
-1.17904669538334	47\\
-1.07143631677291	47.5\\
-1	47.8805576182562\\
-0.977758282302026	48\\
-0.897773346724916	48.5\\
-0.830807242090765	49\\
-0.776415190823382	49.5\\
-0.734178585468476	50\\
-0.703701936586085	50.5\\
-0.684625977445545	51\\
-0.676641338284295	51.5\\
-0.679483340327009	52\\
-0.692879067960957	52.5\\
-0.716430134216283	53\\
-0.749455510489247	53.5\\
-0.790813415918444	54\\
-0.83871815437568	54.5\\
-0.890563115977319	55\\
-0.942852634794939	55.5\\
-0.991280757102035	56\\
-1	56.1110069155138\\
-1.03035232062756	56.5\\
-1.05408021648044	57\\
-1.05595497855566	57.5\\
-1.02878929378204	58\\
-1	58.2384028538829\\
-0.96768729569766	58.5\\
-0.865402364649699	59\\
-0.716005287591114	59.5\\
-0.514768132498065	60\\
-0.5	60.031304196286\\
-0.288783932405565	60.5\\
-0.107537048193882	61\\
-0.0381456922531533	61.5\\
-0.0128091357831203	62\\
11	0\\
10.9249520899817	0.5\\
10.8294934109004	1\\
10.7245174965417	1.5\\
10.6106269243	2\\
10.5	2.45419571468371\\
10.4887951321466	2.5\\
10.3605399881615	3\\
10.2277886777402	3.5\\
10.0924390094735	4\\
10	4.33973087627795\\
9.95640197118805	4.5\\
9.82153244785092	5\\
9.68966082111244	5.5\\
9.56257789686622	6\\
9.5	6.25896689204695\\
9.44197405054831	6.5\\
9.3292232616987	7\\
9.22541440862522	7.5\\
9.13141683844462	8\\
9.04783713290217	8.5\\
9	8.82819417226743\\
8.97504263831364	9\\
8.91314747557791	9.5\\
8.86204918486945	10\\
8.82146062438822	10.5\\
8.79087450490365	11\\
8.76954788090248	11.5\\
8.75647937933854	12\\
8.75038002797663	12.5\\
8.74963899360751	13\\
8.75228744721663	13.5\\
8.75597546932219	14\\
8.7579617805969	14.5\\
8.75510039071427	15\\
8.74389851217116	15.5\\
8.72063688085278	16\\
8.68155597994042	16.5\\
8.62311273151125	17\\
8.5422806678601	17.5\\
8.5	17.7016813925057\\
8.43672124556067	18\\
8.30529758855132	18.5\\
8.1484193868514	19\\
8	19.4119124759721\\
7.96771954291109	19.5\\
7.76576846869582	20\\
7.54746786772789	20.5\\
7.5	20.603734401943\\
7.31717128661607	21\\
7.08040464942759	21.5\\
7	21.6686028372957\\
6.84207510179312	22\\
6.60660557490045	22.5\\
6.5	22.7321088608239\\
6.3778902958993	23\\
6.15887147787019	23.5\\
6	23.882512639768\\
5.95173979898502	24\\
5.7584536793006	24.5\\
5.57966142407837	25\\
5.5	25.2425520187427\\
5.41637750714721	25.5\\
5.26887535317003	26\\
5.13712912836617	26.5\\
5.02117202351121	27\\
5	27.1050586520783\\
4.9209625114127	27.5\\
4.83613263975368	28\\
4.76640247868822	28.5\\
4.71148277701625	29\\
4.67105632487221	29.5\\
4.64477840754965	30\\
4.63226674594118	30.5\\
4.63307954398334	31\\
4.64667866342646	31.5\\
4.6723747675152	32\\
4.70925119854582	32.5\\
4.75606335422924	33\\
4.81111041244298	33.5\\
4.8720765541569	34\\
4.93586492700853	34.5\\
4.99848847213127	35\\
5	35.0133789147081\\
5.05487757642825	35.5\\
5.09888824758841	36\\
5.12353426763028	36.5\\
5.12147037804902	37\\
5.08577123362424	37.5\\
5.01081757746309	38\\
5	38.0466684246058\\
4.89252517297061	38.5\\
4.73017086751199	39\\
4.52676235376425	39.5\\
4.5	39.5572149712002\\
4.28687057330901	40\\
4.02123415346224	40.5\\
4	40.5380199537731\\
3.73897588549566	41\\
3.5	41.4149551855106\\
3.45099309355943	41.5\\
3.16578709168738	42\\
3	42.2991788278281\\
2.89009140269527	42.5\\
2.62844449349911	43\\
2.5	43.259889268844\\
2.38323530186607	43.5\\
2.15643227073841	44\\
2	44.3755540005139\\
1.94911654881508	44.5\\
1.76146657608844	45\\
1.59167978828544	45.5\\
1.5	45.7987777626261\\
1.4386009900869	46\\
1.30120584212458	46.5\\
1.17904669538334	47\\
1.07143631677291	47.5\\
1	47.8805576182562\\
0.977758282302026	48\\
0.897773346724916	48.5\\
0.830807242090765	49\\
0.776415190823382	49.5\\
0.734178585468476	50\\
0.703701936586085	50.5\\
0.684625977445545	51\\
0.676641338284295	51.5\\
0.679483340327009	52\\
0.692879067960957	52.5\\
0.716430134216283	53\\
0.749455510489247	53.5\\
0.790813415918444	54\\
0.83871815437568	54.5\\
0.890563115977319	55\\
0.942852634794939	55.5\\
0.991280757102035	56\\
1	56.1110069155138\\
1.03035232062756	56.5\\
1.05408021648044	57\\
1.05595497855566	57.5\\
1.02878929378204	58\\
1	58.2384028538829\\
0.96768729569766	58.5\\
0.865402364649699	59\\
0.716005287591114	59.5\\
0.514768132498065	60\\
0.5	60.031304196286\\
0.288783932405565	60.5\\
0.107537048193882	61\\
0.0381456922531533	61.5\\
0.0128091357831203	62\\
};

\addplot [color=black, only marks, mark=*, mark options={solid, black}, mark size=0.5pt, forget plot]
  table[row sep=crcr]{%
-11	0\\
-10.7219151272044	1.47718119273835\\
-10.4417661516666	2.9543623854767\\
-10.1595518801349	4.43154357821502\\
-9.88144490951864	5.9087247709534\\
-9.6129049952455	7.38590596369171\\
-9.35621329130449	8.86308715643004\\
-9.1078328797749	10.3402683491684\\
-8.85993048595936	11.8174495419068\\
-8.60130604825603	13.2946307346451\\
-8.32084475855851	14.7718119273834\\
-8.01706293285777	16.2489931201217\\
-7.69304885242828	17.7261743128601\\
-7.34692659901218	19.2033555055986\\
-6.98813668858756	20.6805366983368\\
-6.63382113689705	22.1577178910753\\
-6.31504345158154	23.6348990838136\\
-6.05471772889797	25.1120802765522\\
-5.84441685549626	26.5892614692902\\
-5.68383815187449	28.0664426620286\\
-5.54542356428665	29.5436238547669\\
-5.39321944966144	31.0208050475052\\
-5.19652681544212	32.4979862402434\\
-4.93671098759007	33.9751674329818\\
-4.6067575414306	35.4523486257202\\
-4.20548898842378	36.9295298184587\\
-3.74813807390739	38.4067110111972\\
-3.25272984047917	39.8838922039351\\
-2.77044634645687	41.3610733966736\\
-2.37232037189377	42.838254589412\\
-2.08437928907078	44.3154357821506\\
-1.88610443212537	45.7926169748888\\
-1.77962923863829	47.2697981676272\\
-1.73441059546781	48.7469793603653\\
-1.69902475338832	50.2241605531043\\
-1.63042440964937	51.701341745842\\
-1.46448528825503	53.1785229385805\\
-1.18365711366699	54.6557041313188\\
-0.789219855512229	56.1328853240572\\
-0.328224274220253	57.6100665167953\\
-0.0449904652768976	59.0872477095337\\
11	0\\
10.7219151272044	1.47718119273835\\
10.4417661516666	2.9543623854767\\
10.1595518801349	4.43154357821502\\
9.88144490951864	5.9087247709534\\
9.6129049952455	7.38590596369171\\
9.35621329130449	8.86308715643004\\
9.1078328797749	10.3402683491684\\
8.85993048595936	11.8174495419068\\
8.60130604825603	13.2946307346451\\
8.32084475855851	14.7718119273834\\
8.01706293285777	16.2489931201217\\
7.69304885242828	17.7261743128601\\
7.34692659901218	19.2033555055986\\
6.98813668858756	20.6805366983368\\
6.63382113689705	22.1577178910753\\
6.31504345158154	23.6348990838136\\
6.05471772889797	25.1120802765522\\
5.84441685549626	26.5892614692902\\
5.68383815187449	28.0664426620286\\
5.54542356428665	29.5436238547669\\
5.39321944966144	31.0208050475052\\
5.19652681544212	32.4979862402434\\
4.93671098759007	33.9751674329818\\
4.6067575414306	35.4523486257202\\
4.20548898842378	36.9295298184587\\
3.74813807390739	38.4067110111972\\
3.25272984047917	39.8838922039351\\
2.77044634645687	41.3610733966736\\
2.37232037189377	42.838254589412\\
2.08437928907078	44.3154357821506\\
1.88610443212537	45.7926169748888\\
1.77962923863829	47.2697981676272\\
1.73441059546781	48.7469793603653\\
1.69902475338832	50.2241605531043\\
1.63042440964937	51.701341745842\\
1.46448528825503	53.1785229385805\\
1.18365711366699	54.6557041313188\\
0.789219855512229	56.1328853240572\\
0.328224274220253	57.6100665167953\\
0.0449904652768976	59.0872477095337\\
};
\end{axis}

\begin{axis}[%
width=1.743in,
height=1.777in,
at={(2.971in,3.041in)},
scale only axis,
xmin=-11,
xmax=11,
xtick={-10,   0,  10},
ymin=0,
ymax=80,
ytick={ 0, 20, 40, 60, 80},
axis background/.style={fill=white},
title={$t/\mathcal{T} = 1/2$},
axis x line*=bottom,
axis y line*=left
]

\addplot [color=red, only marks, mark=*, mark options={solid, red}, mark size=0.5pt, forget plot]
  table[row sep=crcr]{%
-11	0\\
-10.9388019150974	0.5\\
-10.8761283467628	1\\
-10.8114660666616	1.5\\
-10.7427078193905	2\\
-10.6677962441734	2.5\\
-10.5848527207584	3\\
-10.5	3.45872165016423\\
-10.4923339450987	3.5\\
-10.389144709579	4\\
-10.2749140761114	4.5\\
-10.1498522647201	5\\
-10.0147968981038	5.5\\
-10	5.55166549025936\\
-9.8710007373363	6\\
-9.72039721215372	6.5\\
-9.56516030663102	7\\
-9.5	7.20686472576741\\
-9.40760391998131	7.5\\
-9.25004851119399	8\\
-9.09472990504672	8.5\\
-9	8.8133508195754\\
-8.94377256345502	9\\
-8.79901564772984	9.5\\
-8.66197323519689	10\\
-8.53402464530415	10.5\\
-8.5	10.6440406553916\\
-8.41637691462808	11\\
-8.30978810159117	11.5\\
-8.21486917846521	12\\
-8.13206052533881	12.5\\
-8.0616038202592	13\\
-8.0035519242574	13.5\\
-8	13.5387199839069\\
-7.95778227053374	14\\
-7.92397963251562	14.5\\
-7.90167466478838	15\\
-7.89019941608781	15.5\\
-7.88865753295612	16\\
-7.89588406104783	16.5\\
-7.91039457034235	17\\
-7.93032522513054	17.5\\
-7.95337053411774	18\\
-7.97674469992416	18.5\\
-7.99714481426035	19\\
-8	19.1050486726771\\
-8.01073287232933	19.5\\
-8.01322742051948	20\\
-8.00008155867056	20.5\\
-8	20.5012276953733\\
-7.9667486979429	21\\
-7.90904058196775	21.5\\
-7.82354620162187	22\\
-7.7080743933962	22.5\\
-7.56203183260833	23\\
-7.5	23.178624366638\\
-7.38605903138237	23.5\\
-7.18339932234357	24\\
-7	24.4096891002538\\
-6.95889146146827	24.5\\
-6.71768999535614	25\\
-6.5	25.4340739179335\\
-6.46674367251425	25.5\\
-6.21183875604562	26\\
-6	26.4179886692704\\
-5.95862049493418	26.5\\
-5.7118861283327	27\\
-5.5	27.446434012165\\
-5.47484959634385	27.5\\
-5.25078395913207	28\\
-5.04082687133359	28.5\\
-5	28.6041357777839\\
-4.84681659306074	29\\
-4.66894553390526	29.5\\
-4.50747843270802	30\\
-4.5	30.0255910680275\\
-4.36281907748964	30.5\\
-4.23441752774435	31\\
-4.12204538336507	31.5\\
-4.02540370523354	32\\
-4	32.1557640463742\\
-3.94416940641949	32.5\\
-3.87787575850719	33\\
-3.82613242495225	33.5\\
-3.7885794952052	34\\
-3.76485915282952	34.5\\
-3.75460291568727	35\\
-3.75740652311089	35.5\\
-3.77279026788866	36\\
-3.80014127588962	36.5\\
-3.83863328788323	37\\
-3.88711895027178	37.5\\
-3.94398955044513	38\\
-4	38.444543220473\\
-4.00699781241174	38.5\\
-4.07305907604317	39\\
-4.13809506061036	39.5\\
-4.19687435506443	40\\
-4.24294679467148	40.5\\
-4.26890736222574	41\\
-4.26694467762913	41.5\\
-4.22970140461195	42\\
-4.15124169351972	42.5\\
-4.02793621796502	43\\
-4	43.0845174434156\\
-3.85827796547297	43.5\\
-3.64633490696383	44\\
-3.5	44.2992665773672\\
-3.39902821125772	44.5\\
-3.1263090567306	45\\
-3	45.2215112688524\\
-2.83988490408097	45.5\\
-2.55009246124667	46\\
-2.5	46.0874607844348\\
-2.26572133835557	46.5\\
-2	46.9875079126271\\
-1.99330604433983	47\\
-1.73803571927953	47.5\\
-1.5	47.9967713489595\\
-1.49846730467017	48\\
-1.27471928094638	48.5\\
-1.06674407647202	49\\
-1	49.1722391742814\\
-0.874958285710123	49.5\\
-0.700003987288032	50\\
-0.543204951711315	50.5\\
-0.5	50.6607360074584\\
-0.416023485325692	51\\
-0.319407156242031	51.5\\
-0.248340164810165	52\\
-0.19708221586803	52.5\\
-0.160098929441082	53\\
-0.133314386841588	53.5\\
-0.11406835611192	54\\
-0.100662606726209	54.5\\
-0.0917484790268893	55\\
-0.0864517606901542	55.5\\
-0.0841706390929996	56\\
-0.0841646756387386	56.5\\
-0.0865278640622778	57\\
-0.0917838996031004	57.5\\
-0.100280604923304	58\\
-0.112154751600992	58.5\\
-0.12763665200281	59\\
-0.14759572715304	59.5\\
-0.17323198482333	60\\
-0.204082744184122	60.5\\
-0.237114144996471	61\\
-0.265961212762033	61.5\\
-0.278837815117752	62\\
-0.262324029976665	62.5\\
-0.206486530864057	63\\
-0.122315273354447	63.5\\
-0.0544705922054965	64\\
-0.0198545242107487	64.5\\
11	0\\
10.9388019150974	0.5\\
10.8761283467628	1\\
10.8114660666616	1.5\\
10.7427078193905	2\\
10.6677962441734	2.5\\
10.5848527207584	3\\
10.5	3.45872165016423\\
10.4923339450987	3.5\\
10.389144709579	4\\
10.2749140761114	4.5\\
10.1498522647201	5\\
10.0147968981038	5.5\\
10	5.55166549025936\\
9.8710007373363	6\\
9.72039721215372	6.5\\
9.56516030663102	7\\
9.5	7.20686472576741\\
9.40760391998131	7.5\\
9.25004851119399	8\\
9.09472990504672	8.5\\
9	8.8133508195754\\
8.94377256345502	9\\
8.79901564772984	9.5\\
8.66197323519689	10\\
8.53402464530415	10.5\\
8.5	10.6440406553916\\
8.41637691462808	11\\
8.30978810159117	11.5\\
8.21486917846521	12\\
8.13206052533881	12.5\\
8.0616038202592	13\\
8.0035519242574	13.5\\
8	13.5387199839069\\
7.95778227053374	14\\
7.92397963251562	14.5\\
7.90167466478838	15\\
7.89019941608781	15.5\\
7.88865753295612	16\\
7.89588406104783	16.5\\
7.91039457034235	17\\
7.93032522513054	17.5\\
7.95337053411774	18\\
7.97674469992416	18.5\\
7.99714481426035	19\\
8	19.1050486726771\\
8.01073287232933	19.5\\
8.01322742051948	20\\
8.00008155867056	20.5\\
8	20.5012276953733\\
7.9667486979429	21\\
7.90904058196775	21.5\\
7.82354620162187	22\\
7.7080743933962	22.5\\
7.56203183260833	23\\
7.5	23.178624366638\\
7.38605903138237	23.5\\
7.18339932234357	24\\
7	24.4096891002538\\
6.95889146146827	24.5\\
6.71768999535614	25\\
6.5	25.4340739179335\\
6.46674367251425	25.5\\
6.21183875604562	26\\
6	26.4179886692704\\
5.95862049493418	26.5\\
5.7118861283327	27\\
5.5	27.446434012165\\
5.47484959634385	27.5\\
5.25078395913207	28\\
5.04082687133359	28.5\\
5	28.6041357777839\\
4.84681659306074	29\\
4.66894553390526	29.5\\
4.50747843270802	30\\
4.5	30.0255910680275\\
4.36281907748964	30.5\\
4.23441752774435	31\\
4.12204538336507	31.5\\
4.02540370523354	32\\
4	32.1557640463742\\
3.94416940641949	32.5\\
3.87787575850719	33\\
3.82613242495225	33.5\\
3.7885794952052	34\\
3.76485915282952	34.5\\
3.75460291568727	35\\
3.75740652311089	35.5\\
3.77279026788866	36\\
3.80014127588962	36.5\\
3.83863328788323	37\\
3.88711895027178	37.5\\
3.94398955044513	38\\
4	38.444543220473\\
4.00699781241174	38.5\\
4.07305907604317	39\\
4.13809506061036	39.5\\
4.19687435506443	40\\
4.24294679467148	40.5\\
4.26890736222574	41\\
4.26694467762913	41.5\\
4.22970140461195	42\\
4.15124169351972	42.5\\
4.02793621796502	43\\
4	43.0845174434156\\
3.85827796547297	43.5\\
3.64633490696383	44\\
3.5	44.2992665773672\\
3.39902821125772	44.5\\
3.1263090567306	45\\
3	45.2215112688524\\
2.83988490408097	45.5\\
2.55009246124667	46\\
2.5	46.0874607844348\\
2.26572133835557	46.5\\
2	46.9875079126271\\
1.99330604433983	47\\
1.73803571927953	47.5\\
1.5	47.9967713489595\\
1.49846730467017	48\\
1.27471928094638	48.5\\
1.06674407647202	49\\
1	49.1722391742814\\
0.874958285710123	49.5\\
0.700003987288032	50\\
0.543204951711315	50.5\\
0.5	50.6607360074584\\
0.416023485325692	51\\
0.319407156242031	51.5\\
0.248340164810165	52\\
0.19708221586803	52.5\\
0.160098929441082	53\\
0.133314386841588	53.5\\
0.11406835611192	54\\
0.100662606726209	54.5\\
0.0917484790268893	55\\
0.0864517606901542	55.5\\
0.0841706390929996	56\\
0.0841646756387386	56.5\\
0.0865278640622778	57\\
0.0917838996031004	57.5\\
0.100280604923304	58\\
0.112154751600992	58.5\\
0.12763665200281	59\\
0.14759572715304	59.5\\
0.17323198482333	60\\
0.204082744184122	60.5\\
0.237114144996471	61\\
0.265961212762033	61.5\\
0.278837815117752	62\\
0.262324029976665	62.5\\
0.206486530864057	63\\
0.122315273354447	63.5\\
0.0544705922054965	64\\
0.0198545242107487	64.5\\
};

\addplot [color=black, only marks, mark=*, mark options={solid, black}, mark size=0.5pt, forget plot]
  table[row sep=crcr]{%
-11	0\\
-10.7219151272044	1.47718119273835\\
-10.4417661516666	2.9543623854767\\
-10.1595518801349	4.43154357821502\\
-9.88144490951864	5.9087247709534\\
-9.6129049952455	7.38590596369171\\
-9.35621329130449	8.86308715643004\\
-9.1078328797749	10.3402683491684\\
-8.85993048595936	11.8174495419068\\
-8.60130604825603	13.2946307346451\\
-8.32084475855851	14.7718119273834\\
-8.01706293285777	16.2489931201217\\
-7.69304885242828	17.7261743128601\\
-7.34692659901218	19.2033555055986\\
-6.98813668858756	20.6805366983368\\
-6.63382113689705	22.1577178910753\\
-6.31504345158154	23.6348990838136\\
-6.05471772889797	25.1120802765522\\
-5.84441685549626	26.5892614692902\\
-5.68383815187449	28.0664426620286\\
-5.54542356428665	29.5436238547669\\
-5.39321944966144	31.0208050475052\\
-5.19652681544212	32.4979862402434\\
-4.93671098759007	33.9751674329818\\
-4.6067575414306	35.4523486257202\\
-4.20548898842378	36.9295298184587\\
-3.74813807390739	38.4067110111972\\
-3.25272984047917	39.8838922039351\\
-2.77044634645687	41.3610733966736\\
-2.37232037189377	42.838254589412\\
-2.08437928907078	44.3154357821506\\
-1.88610443212537	45.7926169748888\\
-1.77962923863829	47.2697981676272\\
-1.73441059546781	48.7469793603653\\
-1.69902475338832	50.2241605531043\\
-1.63042440964937	51.701341745842\\
-1.46448528825503	53.1785229385805\\
-1.18365711366699	54.6557041313188\\
-0.789219855512229	56.1328853240572\\
-0.328224274220253	57.6100665167953\\
-0.0449904652768976	59.0872477095337\\
11	0\\
10.7219151272044	1.47718119273835\\
10.4417661516666	2.9543623854767\\
10.1595518801349	4.43154357821502\\
9.88144490951864	5.9087247709534\\
9.6129049952455	7.38590596369171\\
9.35621329130449	8.86308715643004\\
9.1078328797749	10.3402683491684\\
8.85993048595936	11.8174495419068\\
8.60130604825603	13.2946307346451\\
8.32084475855851	14.7718119273834\\
8.01706293285777	16.2489931201217\\
7.69304885242828	17.7261743128601\\
7.34692659901218	19.2033555055986\\
6.98813668858756	20.6805366983368\\
6.63382113689705	22.1577178910753\\
6.31504345158154	23.6348990838136\\
6.05471772889797	25.1120802765522\\
5.84441685549626	26.5892614692902\\
5.68383815187449	28.0664426620286\\
5.54542356428665	29.5436238547669\\
5.39321944966144	31.0208050475052\\
5.19652681544212	32.4979862402434\\
4.93671098759007	33.9751674329818\\
4.6067575414306	35.4523486257202\\
4.20548898842378	36.9295298184587\\
3.74813807390739	38.4067110111972\\
3.25272984047917	39.8838922039351\\
2.77044634645687	41.3610733966736\\
2.37232037189377	42.838254589412\\
2.08437928907078	44.3154357821506\\
1.88610443212537	45.7926169748888\\
1.77962923863829	47.2697981676272\\
1.73441059546781	48.7469793603653\\
1.69902475338832	50.2241605531043\\
1.63042440964937	51.701341745842\\
1.46448528825503	53.1785229385805\\
1.18365711366699	54.6557041313188\\
0.789219855512229	56.1328853240572\\
0.328224274220253	57.6100665167953\\
0.0449904652768976	59.0872477095337\\
};
\end{axis}

\begin{axis}[%
width=1.743in,
height=1.777in,
at={(0.677in,0.573in)},
scale only axis,
xmin=-11,
xmax=11,
xtick={-10,   0,  10},
xlabel style={font=\color{white!15!black}},
xlabel={$r$ [mm]},
ymin=0,
ymax=80,
ytick={ 0, 20, 40, 60, 80},
ylabel style={font=\color{white!15!black}},
ylabel={$z$ [mm]},
axis background/.style={fill=white},
title={$t/\mathcal{T} = 1$},
axis x line*=bottom,
axis y line*=left
]

\addplot [color=red, only marks, mark=*, mark options={solid, red}, mark size=0.5pt, forget plot]
  table[row sep=crcr]{%
-11	0\\
-10.8717539919737	0.5\\
-10.7475704729649	1\\
-10.6277023383594	1.5\\
-10.5130111692642	2\\
-10.5	2.05994300004216\\
-10.4046535924391	2.5\\
-10.3036906777655	3\\
-10.2109439158165	3.5\\
-10.1269775861577	4\\
-10.0520780014005	4.5\\
-10	4.89528729129458\\
-9.98623786811337	5\\
-9.92914695213945	5.5\\
-9.88019479159082	6\\
-9.83851981582021	6.5\\
-9.80300409734263	7\\
-9.77227396612907	7.5\\
-9.74468763626286	8\\
-9.71832338475329	8.5\\
-9.69097996705695	9\\
-9.66017732707102	9.5\\
-9.62320595409765	10\\
-9.57722145667378	10.5\\
-9.5193811422337	11\\
-9.5	11.1343580264863\\
-9.44699176485557	11.5\\
-9.35779196481756	12\\
-9.25014209432132	12.5\\
-9.12318793678349	13\\
-9	13.421988834575\\
-8.97695616288028	13.5\\
-8.81222941633321	14\\
-8.63144297278861	14.5\\
-8.5	14.8395754903327\\
-8.43734200789757	15\\
-8.23328518600533	15.5\\
-8.02333403584647	16\\
-8	16.0550151095846\\
-7.81120616985149	16.5\\
-7.60054607432984	17\\
-7.5	17.24353532032\\
-7.39467629998773	17.5\\
-7.19630232667503	18\\
-7.00749821683536	18.5\\
-7	18.5209878610864\\
-6.83041852478708	19\\
-6.66593586303321	19.5\\
-6.51503797811949	20\\
-6.5	20.0546887106591\\
-6.37861235607342	20.5\\
-6.2567646989079	21\\
-6.14970091584155	21.5\\
-6.05750543952703	22\\
-6	22.3711896115959\\
-5.98014536940091	22.5\\
-5.91745443167147	23\\
-5.86917257519805	23.5\\
-5.83501835683328	24\\
-5.81463097760485	24.5\\
-5.80755437126742	25\\
-5.81320824130413	25.5\\
-5.83084401934647	26\\
-5.8594834431802	26.5\\
-5.89783757635868	27\\
-5.94420471064767	27.5\\
-5.99634620823594	28\\
-6	28.0332257822342\\
-6.05135713518743	28.5\\
-6.10556694535134	29\\
-6.15444467679447	29.5\\
-6.19257428900512	30\\
-6.21383477893231	30.5\\
-6.2117649258197	31\\
-6.18013204559359	31.5\\
-6.11363365194396	32\\
-6.00858618836994	32.5\\
-6	32.5301366709799\\
-5.86278784491248	33\\
-5.67793839150889	33.5\\
-5.5	33.9071237188245\\
-5.45826675730668	34\\
-5.20965710767646	34.5\\
-5	34.8926239520709\\
-4.94185231654756	35\\
-4.6631818371773	35.5\\
-4.5	35.7904013490905\\
-4.38243557023841	36\\
-4.10647651862435	36.5\\
-4	36.6989565380421\\
-3.84082445418388	37\\
-3.58887200616264	37.5\\
-3.5	37.6867419991453\\
-3.3532311401258	38\\
-3.13507920337344	38.5\\
-3	38.8361076348781\\
-2.93514969089406	39\\
-2.7536955462215	39.5\\
-2.58981743780698	40\\
-2.5	40.3050572289404\\
-2.44321262751199	40.5\\
-2.31329668845817	41\\
-2.19928977696038	41.5\\
-2.10058221552611	42\\
-2.01654866937787	42.5\\
-2	42.6179087913934\\
-1.94664775392041	43\\
-1.89017189335489	43.5\\
-1.84659980713012	44\\
-1.81549732326216	44.5\\
-1.79648423697844	45\\
-1.78921799653171	45.5\\
-1.7933622388182	46\\
-1.80853967505273	46.5\\
-1.83426536198265	47\\
-1.8698539342475	47.5\\
-1.91429406873498	48\\
-1.96608518779135	48.5\\
-2	48.7979244784127\\
-2.02304053559481	49\\
-2.08199604458686	49.5\\
-2.13861027582879	50\\
-2.18721445482995	50.5\\
-2.22091744983974	51\\
-2.23192411383666	51.5\\
-2.21218865640762	52\\
-2.1544691353136	52.5\\
-2.05329700443858	53\\
-2	53.1829188035646\\
-1.90457138258615	53.5\\
-1.71003356521947	54\\
-1.5	54.453289608935\\
-1.4778676484742	54.5\\
-1.21686800843701	55\\
-1	55.3867719097359\\
-0.935056455683435	55.5\\
-0.636134765767481	56\\
-0.5	56.2282652217151\\
-0.350828076523987	56.5\\
-0.148628639857821	57\\
-0.0699137749396052	57.5\\
-0.0295758193211468	58\\
-0.0102437868164142	58.5\\
11	0\\
10.8717539919737	0.5\\
10.7475704729649	1\\
10.6277023383594	1.5\\
10.5130111692642	2\\
10.5	2.05994300004216\\
10.4046535924391	2.5\\
10.3036906777655	3\\
10.2109439158165	3.5\\
10.1269775861577	4\\
10.0520780014005	4.5\\
10	4.89528729129458\\
9.98623786811337	5\\
9.92914695213945	5.5\\
9.88019479159082	6\\
9.83851981582021	6.5\\
9.80300409734263	7\\
9.77227396612907	7.5\\
9.74468763626286	8\\
9.71832338475329	8.5\\
9.69097996705695	9\\
9.66017732707102	9.5\\
9.62320595409765	10\\
9.57722145667378	10.5\\
9.5193811422337	11\\
9.5	11.1343580264863\\
9.44699176485557	11.5\\
9.35779196481756	12\\
9.25014209432132	12.5\\
9.12318793678349	13\\
9	13.421988834575\\
8.97695616288028	13.5\\
8.81222941633321	14\\
8.63144297278861	14.5\\
8.5	14.8395754903327\\
8.43734200789757	15\\
8.23328518600533	15.5\\
8.02333403584647	16\\
8	16.0550151095846\\
7.81120616985149	16.5\\
7.60054607432984	17\\
7.5	17.24353532032\\
7.39467629998773	17.5\\
7.19630232667503	18\\
7.00749821683536	18.5\\
7	18.5209878610864\\
6.83041852478708	19\\
6.66593586303321	19.5\\
6.51503797811949	20\\
6.5	20.0546887106591\\
6.37861235607342	20.5\\
6.2567646989079	21\\
6.14970091584155	21.5\\
6.05750543952703	22\\
6	22.3711896115959\\
5.98014536940091	22.5\\
5.91745443167147	23\\
5.86917257519805	23.5\\
5.83501835683328	24\\
5.81463097760485	24.5\\
5.80755437126742	25\\
5.81320824130413	25.5\\
5.83084401934647	26\\
5.8594834431802	26.5\\
5.89783757635868	27\\
5.94420471064767	27.5\\
5.99634620823594	28\\
6	28.0332257822342\\
6.05135713518743	28.5\\
6.10556694535134	29\\
6.15444467679447	29.5\\
6.19257428900512	30\\
6.21383477893231	30.5\\
6.2117649258197	31\\
6.18013204559359	31.5\\
6.11363365194396	32\\
6.00858618836994	32.5\\
6	32.5301366709799\\
5.86278784491248	33\\
5.67793839150889	33.5\\
5.5	33.9071237188245\\
5.45826675730668	34\\
5.20965710767646	34.5\\
5	34.8926239520709\\
4.94185231654756	35\\
4.6631818371773	35.5\\
4.5	35.7904013490905\\
4.38243557023841	36\\
4.10647651862435	36.5\\
4	36.6989565380421\\
3.84082445418388	37\\
3.58887200616264	37.5\\
3.5	37.6867419991453\\
3.3532311401258	38\\
3.13507920337344	38.5\\
3	38.8361076348781\\
2.93514969089406	39\\
2.7536955462215	39.5\\
2.58981743780698	40\\
2.5	40.3050572289404\\
2.44321262751199	40.5\\
2.31329668845817	41\\
2.19928977696038	41.5\\
2.10058221552611	42\\
2.01654866937787	42.5\\
2	42.6179087913934\\
1.94664775392041	43\\
1.89017189335489	43.5\\
1.84659980713012	44\\
1.81549732326216	44.5\\
1.79648423697844	45\\
1.78921799653171	45.5\\
1.7933622388182	46\\
1.80853967505273	46.5\\
1.83426536198265	47\\
1.8698539342475	47.5\\
1.91429406873498	48\\
1.96608518779135	48.5\\
2	48.7979244784127\\
2.02304053559481	49\\
2.08199604458686	49.5\\
2.13861027582879	50\\
2.18721445482995	50.5\\
2.22091744983974	51\\
2.23192411383666	51.5\\
2.21218865640762	52\\
2.1544691353136	52.5\\
2.05329700443858	53\\
2	53.1829188035646\\
1.90457138258615	53.5\\
1.71003356521947	54\\
1.5	54.453289608935\\
1.4778676484742	54.5\\
1.21686800843701	55\\
1	55.3867719097359\\
0.935056455683435	55.5\\
0.636134765767481	56\\
0.5	56.2282652217151\\
0.350828076523987	56.5\\
0.148628639857821	57\\
0.0699137749396052	57.5\\
0.0295758193211468	58\\
0.0102437868164142	58.5\\
};

\addplot [color=black, only marks, mark=*, mark options={solid, black}, mark size=0.5pt, forget plot]
  table[row sep=crcr]{%
-11	0\\
-10.7219151272044	1.47718119273835\\
-10.4417661516666	2.9543623854767\\
-10.1595518801349	4.43154357821502\\
-9.88144490951864	5.9087247709534\\
-9.6129049952455	7.38590596369171\\
-9.35621329130449	8.86308715643004\\
-9.1078328797749	10.3402683491684\\
-8.85993048595936	11.8174495419068\\
-8.60130604825603	13.2946307346451\\
-8.32084475855851	14.7718119273834\\
-8.01706293285777	16.2489931201217\\
-7.69304885242828	17.7261743128601\\
-7.34692659901218	19.2033555055986\\
-6.98813668858756	20.6805366983368\\
-6.63382113689705	22.1577178910753\\
-6.31504345158154	23.6348990838136\\
-6.05471772889797	25.1120802765522\\
-5.84441685549626	26.5892614692902\\
-5.68383815187449	28.0664426620286\\
-5.54542356428665	29.5436238547669\\
-5.39321944966144	31.0208050475052\\
-5.19652681544212	32.4979862402434\\
-4.93671098759007	33.9751674329818\\
-4.6067575414306	35.4523486257202\\
-4.20548898842378	36.9295298184587\\
-3.74813807390739	38.4067110111972\\
-3.25272984047917	39.8838922039351\\
-2.77044634645687	41.3610733966736\\
-2.37232037189377	42.838254589412\\
-2.08437928907078	44.3154357821506\\
-1.88610443212537	45.7926169748888\\
-1.77962923863829	47.2697981676272\\
-1.73441059546781	48.7469793603653\\
-1.69902475338832	50.2241605531043\\
-1.63042440964937	51.701341745842\\
-1.46448528825503	53.1785229385805\\
-1.18365711366699	54.6557041313188\\
-0.789219855512229	56.1328853240572\\
-0.328224274220253	57.6100665167953\\
-0.0449904652768976	59.0872477095337\\
11	0\\
10.7219151272044	1.47718119273835\\
10.4417661516666	2.9543623854767\\
10.1595518801349	4.43154357821502\\
9.88144490951864	5.9087247709534\\
9.6129049952455	7.38590596369171\\
9.35621329130449	8.86308715643004\\
9.1078328797749	10.3402683491684\\
8.85993048595936	11.8174495419068\\
8.60130604825603	13.2946307346451\\
8.32084475855851	14.7718119273834\\
8.01706293285777	16.2489931201217\\
7.69304885242828	17.7261743128601\\
7.34692659901218	19.2033555055986\\
6.98813668858756	20.6805366983368\\
6.63382113689705	22.1577178910753\\
6.31504345158154	23.6348990838136\\
6.05471772889797	25.1120802765522\\
5.84441685549626	26.5892614692902\\
5.68383815187449	28.0664426620286\\
5.54542356428665	29.5436238547669\\
5.39321944966144	31.0208050475052\\
5.19652681544212	32.4979862402434\\
4.93671098759007	33.9751674329818\\
4.6067575414306	35.4523486257202\\
4.20548898842378	36.9295298184587\\
3.74813807390739	38.4067110111972\\
3.25272984047917	39.8838922039351\\
2.77044634645687	41.3610733966736\\
2.37232037189377	42.838254589412\\
2.08437928907078	44.3154357821506\\
1.88610443212537	45.7926169748888\\
1.77962923863829	47.2697981676272\\
1.73441059546781	48.7469793603653\\
1.69902475338832	50.2241605531043\\
1.63042440964937	51.701341745842\\
1.46448528825503	53.1785229385805\\
1.18365711366699	54.6557041313188\\
0.789219855512229	56.1328853240572\\
0.328224274220253	57.6100665167953\\
0.0449904652768976	59.0872477095337\\
};
\end{axis}

\begin{axis}[%
width=1.743in,
height=1.777in,
at={(2.971in,0.573in)},
scale only axis,
xmin=-11,
xmax=11,
xtick={-10,   0,  10},
xlabel style={font=\color{white!15!black}},
xlabel={$r$ [mm]},
ymin=0,
ymax=80,
ytick={ 0, 20, 40, 60, 80},
axis background/.style={fill=white},
title={$t/\mathcal{T} = 3/4$},
axis x line*=bottom,
axis y line*=left
]

\addplot [color=red, only marks, mark=*, mark options={solid, red}, mark size=0.5pt, forget plot]
  table[row sep=crcr]{%
-11	0\\
-10.9092218337769	0.5\\
-10.8303656001159	1\\
-10.7595917998683	1.5\\
-10.6953236994441	2\\
-10.6360915181586	2.5\\
-10.5803639813921	3\\
-10.5263847164495	3.5\\
-10.5	3.74336198266301\\
-10.4721644982189	4\\
-10.4155600611791	4.5\\
-10.3543389603901	5\\
-10.2862733641276	5.5\\
-10.2092462050688	6\\
-10.1213711932321	6.5\\
-10.0211334120351	7\\
-10	7.09342934886818\\
-9.90743516459039	7.5\\
-9.77998451609275	8\\
-9.63916667089919	8.5\\
-9.5	8.9547515153071\\
-9.48605142045612	9\\
-9.32217931356089	9.5\\
-9.15006581946477	10\\
-9	10.4223698999099\\
-8.97234054572478	10.5\\
-8.79176354430718	11\\
-8.61121129440829	11.5\\
-8.5	11.8123437955776\\
-8.43339288942595	12\\
-8.26072586694054	12.5\\
-8.09518867564608	13\\
-8	13.3032730446312\\
-7.93866133690628	13.5\\
-7.79256592525547	14\\
-7.65787637079089	14.5\\
-7.53547985869599	15\\
-7.5	15.1614730664843\\
-7.42608742588436	15.5\\
-7.32999316602118	16\\
-7.24739709426455	16.5\\
-7.17839058393425	17\\
-7.12292392372741	17.5\\
-7.08081381332551	18\\
-7.05174422753778	18.5\\
-7.03525852522816	19\\
-7.03073949654765	19.5\\
-7.03737521496889	20\\
-7.05410974672766	20.5\\
-7.07957871051028	21\\
-7.11203069495732	21.5\\
-7.14923759327194	22\\
-7.18841402512209	22.5\\
-7.22617019912478	23\\
-7.25844024706873	23.5\\
-7.28051195761022	24\\
-7.28718903135077	24.5\\
-7.27308219839688	25\\
-7.23303854594888	25.5\\
-7.16266215427051	26\\
-7.05883492306361	26.5\\
-7	26.714198112542\\
-6.91983208920719	27\\
-6.74631005234107	27.5\\
-6.54181266728464	28\\
-6.5	28.0919005100611\\
-6.31045022031309	28.5\\
-6.05992329731471	29\\
-6	29.1147099692661\\
-5.79704713393085	29.5\\
-5.52941744795757	30\\
-5.5	30.0551116741227\\
-5.2634777485581	30.5\\
-5.00417761476387	31\\
-5	31.0083275099405\\
-4.75620864708817	31.5\\
-4.52149385354783	32\\
-4.5	32.0485035766966\\
-4.30276887579617	32.5\\
-4.10039582092947	33\\
-4	33.2695008771242\\
-3.91530066872649	33.5\\
-3.74757373635477	34\\
-3.59676384776342	34.5\\
-3.5	34.8599151157936\\
-3.46270475208678	35\\
-3.34500989251737	35.5\\
-3.24303250148422	36\\
-3.15631179327846	36.5\\
-3.08438357843511	37\\
-3.02679950453158	37.5\\
-3	37.806585564403\\
-2.98313994212896	38\\
-2.95299931387707	38.5\\
-2.93600381169408	39\\
-2.93179601995188	39.5\\
-2.9399886442605	40\\
-2.96011639311091	40.5\\
-2.99156756155946	41\\
-3	41.1007142937553\\
-3.03350102110092	41.5\\
-3.0846930791138	42\\
-3.14337386858846	42.5\\
-3.20702573381636	43\\
-3.27219364412406	43.5\\
-3.3343352218825	44\\
-3.38759805700689	44.5\\
-3.42490401603978	45\\
-3.4383402653891	45.5\\
-3.4199160613734	46\\
-3.36254246137804	46.5\\
-3.26102045453197	47\\
-3.11305211329842	47.5\\
-3	47.796143898159\\
-2.91942754805099	48\\
-2.68579733439341	48.5\\
-2.5	48.8552272247694\\
-2.42252689976261	49\\
-2.14030142523665	49.5\\
-2	49.7431211563338\\
-1.85152429141032	50\\
-1.56437229536075	50.5\\
-1.5	50.6139785206016\\
-1.28306956442738	51\\
-1.01012842319013	51.5\\
-1	51.5191603275619\\
-0.749510630788834	52\\
-0.505402800132281	52.5\\
-0.5	52.5126118368202\\
-0.312983961623806	53\\
-0.184935349846354	53.5\\
-0.111995169105782	54\\
-0.0683694988518499	54.5\\
-0.0412343568380384	55\\
-0.0255258636063983	55.5\\
-0.0165588661235612	56\\
-0.0118465773881786	56.5\\
11	0\\
10.9092218337769	0.5\\
10.8303656001159	1\\
10.7595917998683	1.5\\
10.6953236994441	2\\
10.6360915181586	2.5\\
10.5803639813921	3\\
10.5263847164495	3.5\\
10.5	3.74336198266301\\
10.4721644982189	4\\
10.4155600611791	4.5\\
10.3543389603901	5\\
10.2862733641276	5.5\\
10.2092462050688	6\\
10.1213711932321	6.5\\
10.0211334120351	7\\
10	7.09342934886818\\
9.90743516459039	7.5\\
9.77998451609275	8\\
9.63916667089919	8.5\\
9.5	8.9547515153071\\
9.48605142045612	9\\
9.32217931356089	9.5\\
9.15006581946477	10\\
9	10.4223698999099\\
8.97234054572478	10.5\\
8.79176354430718	11\\
8.61121129440829	11.5\\
8.5	11.8123437955776\\
8.43339288942595	12\\
8.26072586694054	12.5\\
8.09518867564608	13\\
8	13.3032730446312\\
7.93866133690628	13.5\\
7.79256592525547	14\\
7.65787637079089	14.5\\
7.53547985869599	15\\
7.5	15.1614730664843\\
7.42608742588436	15.5\\
7.32999316602118	16\\
7.24739709426455	16.5\\
7.17839058393425	17\\
7.12292392372741	17.5\\
7.08081381332551	18\\
7.05174422753778	18.5\\
7.03525852522816	19\\
7.03073949654765	19.5\\
7.03737521496889	20\\
7.05410974672766	20.5\\
7.07957871051028	21\\
7.11203069495732	21.5\\
7.14923759327194	22\\
7.18841402512209	22.5\\
7.22617019912478	23\\
7.25844024706873	23.5\\
7.28051195761022	24\\
7.28718903135077	24.5\\
7.27308219839688	25\\
7.23303854594888	25.5\\
7.16266215427051	26\\
7.05883492306361	26.5\\
7	26.714198112542\\
6.91983208920719	27\\
6.74631005234107	27.5\\
6.54181266728464	28\\
6.5	28.0919005100611\\
6.31045022031309	28.5\\
6.05992329731471	29\\
6	29.1147099692661\\
5.79704713393085	29.5\\
5.52941744795757	30\\
5.5	30.0551116741227\\
5.2634777485581	30.5\\
5.00417761476387	31\\
5	31.0083275099405\\
4.75620864708817	31.5\\
4.52149385354783	32\\
4.5	32.0485035766966\\
4.30276887579617	32.5\\
4.10039582092947	33\\
4	33.2695008771242\\
3.91530066872649	33.5\\
3.74757373635477	34\\
3.59676384776342	34.5\\
3.5	34.8599151157936\\
3.46270475208678	35\\
3.34500989251737	35.5\\
3.24303250148422	36\\
3.15631179327846	36.5\\
3.08438357843511	37\\
3.02679950453158	37.5\\
3	37.806585564403\\
2.98313994212896	38\\
2.95299931387707	38.5\\
2.93600381169408	39\\
2.93179601995188	39.5\\
2.9399886442605	40\\
2.96011639311091	40.5\\
2.99156756155946	41\\
3	41.1007142937553\\
3.03350102110092	41.5\\
3.0846930791138	42\\
3.14337386858846	42.5\\
3.20702573381636	43\\
3.27219364412406	43.5\\
3.3343352218825	44\\
3.38759805700689	44.5\\
3.42490401603978	45\\
3.4383402653891	45.5\\
3.4199160613734	46\\
3.36254246137804	46.5\\
3.26102045453197	47\\
3.11305211329842	47.5\\
3	47.796143898159\\
2.91942754805099	48\\
2.68579733439341	48.5\\
2.5	48.8552272247694\\
2.42252689976261	49\\
2.14030142523665	49.5\\
2	49.7431211563338\\
1.85152429141032	50\\
1.56437229536075	50.5\\
1.5	50.6139785206016\\
1.28306956442738	51\\
1.01012842319013	51.5\\
1	51.5191603275619\\
0.749510630788834	52\\
0.505402800132281	52.5\\
0.5	52.5126118368202\\
0.312983961623806	53\\
0.184935349846354	53.5\\
0.111995169105782	54\\
0.0683694988518499	54.5\\
0.0412343568380384	55\\
0.0255258636063983	55.5\\
0.0165588661235612	56\\
0.0118465773881786	56.5\\
};

\addplot [color=black, only marks, mark=*, mark options={solid, black}, mark size=0.5pt, forget plot]
  table[row sep=crcr]{%
-11	0\\
-10.7219151272044	1.47718119273835\\
-10.4417661516666	2.9543623854767\\
-10.1595518801349	4.43154357821502\\
-9.88144490951864	5.9087247709534\\
-9.6129049952455	7.38590596369171\\
-9.35621329130449	8.86308715643004\\
-9.1078328797749	10.3402683491684\\
-8.85993048595936	11.8174495419068\\
-8.60130604825603	13.2946307346451\\
-8.32084475855851	14.7718119273834\\
-8.01706293285777	16.2489931201217\\
-7.69304885242828	17.7261743128601\\
-7.34692659901218	19.2033555055986\\
-6.98813668858756	20.6805366983368\\
-6.63382113689705	22.1577178910753\\
-6.31504345158154	23.6348990838136\\
-6.05471772889797	25.1120802765522\\
-5.84441685549626	26.5892614692902\\
-5.68383815187449	28.0664426620286\\
-5.54542356428665	29.5436238547669\\
-5.39321944966144	31.0208050475052\\
-5.19652681544212	32.4979862402434\\
-4.93671098759007	33.9751674329818\\
-4.6067575414306	35.4523486257202\\
-4.20548898842378	36.9295298184587\\
-3.74813807390739	38.4067110111972\\
-3.25272984047917	39.8838922039351\\
-2.77044634645687	41.3610733966736\\
-2.37232037189377	42.838254589412\\
-2.08437928907078	44.3154357821506\\
-1.88610443212537	45.7926169748888\\
-1.77962923863829	47.2697981676272\\
-1.73441059546781	48.7469793603653\\
-1.69902475338832	50.2241605531043\\
-1.63042440964937	51.701341745842\\
-1.46448528825503	53.1785229385805\\
-1.18365711366699	54.6557041313188\\
-0.789219855512229	56.1328853240572\\
-0.328224274220253	57.6100665167953\\
-0.0449904652768976	59.0872477095337\\
11	0\\
10.7219151272044	1.47718119273835\\
10.4417661516666	2.9543623854767\\
10.1595518801349	4.43154357821502\\
9.88144490951864	5.9087247709534\\
9.6129049952455	7.38590596369171\\
9.35621329130449	8.86308715643004\\
9.1078328797749	10.3402683491684\\
8.85993048595936	11.8174495419068\\
8.60130604825603	13.2946307346451\\
8.32084475855851	14.7718119273834\\
8.01706293285777	16.2489931201217\\
7.69304885242828	17.7261743128601\\
7.34692659901218	19.2033555055986\\
6.98813668858756	20.6805366983368\\
6.63382113689705	22.1577178910753\\
6.31504345158154	23.6348990838136\\
6.05471772889797	25.1120802765522\\
5.84441685549626	26.5892614692902\\
5.68383815187449	28.0664426620286\\
5.54542356428665	29.5436238547669\\
5.39321944966144	31.0208050475052\\
5.19652681544212	32.4979862402434\\
4.93671098759007	33.9751674329818\\
4.6067575414306	35.4523486257202\\
4.20548898842378	36.9295298184587\\
3.74813807390739	38.4067110111972\\
3.25272984047917	39.8838922039351\\
2.77044634645687	41.3610733966736\\
2.37232037189377	42.838254589412\\
2.08437928907078	44.3154357821506\\
1.88610443212537	45.7926169748888\\
1.77962923863829	47.2697981676272\\
1.73441059546781	48.7469793603653\\
1.69902475338832	50.2241605531043\\
1.63042440964937	51.701341745842\\
1.46448528825503	53.1785229385805\\
1.18365711366699	54.6557041313188\\
0.789219855512229	56.1328853240572\\
0.328224274220253	57.6100665167953\\
0.0449904652768976	59.0872477095337\\
};
\end{axis}
\end{tikzpicture}%

%% file: figures/0.0H_FDF.tex
%
%
\definecolor{mycolor1}{rgb}{1.00000,0.00000,1.00000}%
\begin{tikzpicture}

\begin{axis}[%
width=1.743in,
height=1.475in,
at={(0.677in,0.452in)},
scale only axis,
xmin=0,
xmax=160,
xlabel style={font=\color{white!15!black}},
xlabel={$f_\trm{p}\ [\trm{Hz}]$},
ymin=0,
ymax=1.05,
ytick={0,0.2,0.4,0.6,0.8,1},
ylabel style={font=\color{white!15!black}},
ylabel={Gain, $|\mathcal{F}|$},
axis background/.style={fill=white},
axis x line*=bottom,
axis y line*=left,
xmajorgrids,
ymajorgrids
]
\addplot [color=red, mark=x, mark options={solid, red}, forget plot]
  table[row sep=crcr]{%
10	1.02941131105684\\
20	0.940094647998133\\
30	0.638744908303734\\
40	0.344431346608321\\
50	0.419673670765671\\
60	0.45311570497177\\
70	0.304667369385656\\
80	0.189803979439128\\
90	0.233225347281582\\
100	0.215840109957262\\
110	0.122874461877971\\
120	0.076937534902424\\
130	0.115731767092459\\
140	0.109647440542482\\
150	0.0594040937318531\\
160	0.0308419373919394\\
};
\addplot [color=blue, mark=o, mark options={solid, blue}, forget plot]
  table[row sep=crcr]{%
10	1.0242028732384\\
20	0.93353644723734\\
30	0.632799908779805\\
40	0.324716826203292\\
50	0.35949546290596\\
60	0.367397681231892\\
70	0.238545889732196\\
80	0.0989414813648997\\
90	0.106357228696183\\
100	0.137573380314119\\
110	0.119994225813202\\
120	0.0687858583312809\\
130	0.0175858735069224\\
140	0.0434869663493662\\
150	0.060967813341704\\
160	0.05617048286342\\
};
\addplot [color=green, mark=square, mark options={solid, green}, forget plot]
  table[row sep=crcr]{%
10	1.02113702453625\\
20	0.918300374828875\\
30	0.612270141240908\\
40	0.24569408798692\\
50	0.188897440713928\\
60	0.223685035561247\\
70	0.196108955008254\\
80	0.132437444315544\\
90	0.0648350786929306\\
100	0.0203148191155744\\
110	0.0377486107046425\\
120	0.0536771018929702\\
130	0.058901876857462\\
140	0.0542997255019554\\
150	0.043861346086104\\
160	0.0315465301128926\\
};
\addplot [color=mycolor1, mark=triangle, mark options={solid, mycolor1}, forget plot]
  table[row sep=crcr]{%
10	1.01734229521557\\
20	0.896059657925328\\
30	0.567824058518126\\
40	0.238584399469385\\
50	0.0978179769500279\\
60	0.107286170108474\\
70	0.113458203779267\\
80	0.106193531550718\\
90	0.093226528832427\\
100	0.0809183776548012\\
110	0.0692918568624441\\
120	0.0550963415515501\\
130	0.0403193476945035\\
140	0.0281965675432923\\
150	0.0235951382536941\\
160	0.0277588980292631\\
};
\end{axis}

\begin{axis}[%
width=1.743in,
height=1.475in,
at={(2.971in,0.452in)},
scale only axis,
xmin=0,
xmax=160,
xlabel style={font=\color{white!15!black}},
xlabel={$f_\trm{p}\ [\trm{Hz}]$},
ymin=-8,
ymax=0,
ylabel style={font=\color{white!15!black}},
ylabel={Phase, $\angle\mathcal{F}/\pi$},
axis background/.style={fill=white},
axis x line*=bottom,
axis y line*=left,
xmajorgrids,
ymajorgrids
]
\addplot [color=red, mark=x, mark options={solid, red}, forget plot]
  table[row sep=crcr]{%
10	-0.18275\\
20	-0.4655\\
30	-0.8355\\
40	-1.401\\
50	-2.085\\
60	-2.549\\
70	-2.99925\\
80	-3.648\\
90	-4.21825\\
100	-4.63\\
110	-5.056\\
120	-5.746\\
130	-6.26325\\
140	-6.623\\
150	-6.98625\\
160	-7.764\\
};
\addplot [color=blue, mark=o, mark options={solid, blue}, forget plot]
  table[row sep=crcr]{%
10	-0.18325\\
20	-0.4655\\
30	-0.837\\
40	-1.375\\
50	-2.075\\
60	-2.501\\
70	-2.85925\\
80	-3.352\\
90	-4.02475\\
100	-4.375\\
110	-4.63525\\
120	-4.897\\
130	-5.4235\\
140	-6.1435\\
150	-6.3825\\
160	-6.58\\
};
\addplot [color=green, mark=square, mark options={solid, green}, forget plot]
  table[row sep=crcr]{%
10	-0.18275\\
20	-0.4645\\
30	-0.822\\
40	-1.264\\
50	-2\\
60	-2.3255\\
70	-2.54425\\
80	-2.752\\
90	-2.9765\\
100	-3.485\\
110	-4.0715\\
120	-4.273\\
130	-4.429\\
140	-4.5845\\
150	-4.71625\\
160	-4.844\\
};
\addplot [color=mycolor1, mark=triangle, mark options={solid, mycolor1}, forget plot]
  table[row sep=crcr]{%
10	-0.18225\\
20	-0.4635\\
30	-0.78375\\
40	-1.087\\
50	-1.655\\
60	-2.1035\\
70	-2.34825\\
80	-2.56\\
90	-2.7425\\
100	-2.9025\\
110	-3.03675\\
120	-3.128\\
130	-3.19925\\
140	-3.19\\
150	-3.0875\\
160	-2.98\\
};
\end{axis}

\begin{axis}[%
width=5.208in,
height=2.083in,
at={(0in,0in)},
scale only axis,
xmin=0,
xmax=1,
ymin=0,
ymax=1,
axis line style={draw=none},
ticks=none,
axis x line*=bottom,
axis y line*=left
]
\end{axis}
\end{tikzpicture}%

%% file: figures/0.2H_FDF.tex
%
%
\definecolor{mycolor1}{rgb}{1.00000,0.00000,1.00000}%
\begin{tikzpicture}

\begin{axis}[%
width=1.743in,
height=1.475in,
at={(0.677in,0.452in)},
scale only axis,
xmin=0,
xmax=160,
xlabel style={font=\color{white!15!black}},
xlabel={$f_\trm{p}\ [\trm{Hz}]$},
ymin=0,
ymax=1.05,
ytick={0,0.2,0.4,0.6,0.8,1},
ylabel style={font=\color{white!15!black}},
ylabel={Gain, $|\mathcal{F}|$},
axis background/.style={fill=white},
axis x line*=bottom,
axis y line*=left,
xmajorgrids,
ymajorgrids
]
\addplot [color=red, mark=x, mark options={solid, red}, forget plot]
  table[row sep=crcr]{%
10	1.01861007040083\\
20	0.974006350475881\\
30	0.772879047746387\\
40	0.463376366534389\\
50	0.315456077960617\\
60	0.409961450248775\\
70	0.420536707053603\\
80	0.300213243622104\\
90	0.173494859861051\\
100	0.1886892874721\\
110	0.216249551075206\\
120	0.174993615253892\\
130	0.0953387059996341\\
140	0.0698423398137903\\
150	0.104664544063087\\
160	0.106768488662847\\
};
\addplot [color=blue, mark=o, mark options={solid, blue}, forget plot]
  table[row sep=crcr]{%
10	1.01799930426541\\
20	0.970656789422832\\
30	0.764675787487307\\
40	0.456958385622696\\
50	0.285291681025521\\
60	0.348314960962519\\
70	0.351037160648438\\
80	0.251185427929173\\
90	0.126699580312332\\
100	0.0842000180075022\\
110	0.123787727464907\\
120	0.133645528027112\\
130	0.107925713977219\\
140	0.0602214108259073\\
150	0.0156466215533986\\
160	0.0398008868613862\\
};
\addplot [color=green, mark=square, mark options={solid, green}, forget plot]
  table[row sep=crcr]{%
10	1.01677642907123\\
20	0.959989941710425\\
30	0.747338742497384\\
40	0.427961865627609\\
50	0.182504554909969\\
60	0.182331596496499\\
70	0.209434472303857\\
80	0.193165541333371\\
90	0.147805827776791\\
100	0.0937447384293601\\
110	0.0453183926942591\\
120	0.0213100497352622\\
130	0.0371336524574692\\
140	0.0484913412433756\\
150	0.0518303123676246\\
160	0.0500258601255584\\
};
\addplot [color=mycolor1, mark=triangle, mark options={solid, mycolor1}, forget plot]
  table[row sep=crcr]{%
10	1.0144979181487\\
20	0.942597437087653\\
30	0.722312430806376\\
40	0.405237403883629\\
50	0.183095845677978\\
60	0.0789756334317557\\
70	0.0861068779958786\\
80	0.102177047267458\\
90	0.108710395570975\\
100	0.10695812500502\\
110	0.100434292319048\\
120	0.0902955301182399\\
130	0.0784737872138079\\
140	0.0635498782580779\\
150	0.0483542822137393\\
160	0.0373882626899618\\
};
\end{axis}

\begin{axis}[%
width=1.743in,
height=1.475in,
at={(2.971in,0.452in)},
scale only axis,
xmin=0,
xmax=160,
xlabel style={font=\color{white!15!black}},
xlabel={$f_\trm{p}\ [\trm{Hz}]$},
ymin=-8,
ymax=0,
ylabel style={font=\color{white!15!black}},
ylabel={Phase, $\angle\mathcal{F}/\pi$},
axis background/.style={fill=white},
axis x line*=bottom,
axis y line*=left,
xmajorgrids,
ymajorgrids
]
\addplot [color=red, mark=x, mark options={solid, red}, forget plot]
  table[row sep=crcr]{%
10	-0.1485\\
20	-0.382\\
30	-0.65475\\
40	-1.019\\
50	-1.64875\\
60	-2.1755\\
70	-2.5425\\
80	-2.924\\
90	-3.449\\
100	-4.03\\
110	-4.4125\\
120	-4.741\\
130	-5.14725\\
140	-5.806\\
150	-6.24\\
160	-6.548\\
};
\addplot [color=blue, mark=o, mark options={solid, blue}, forget plot]
  table[row sep=crcr]{%
10	-0.1485\\
20	-0.382\\
30	-0.6585\\
40	-1.009\\
50	-1.63625\\
60	-2.1545\\
70	-2.48475\\
80	-2.794\\
90	-3.1655\\
100	-3.78\\
110	-4.198\\
120	-4.459\\
130	-4.69225\\
140	-4.9345\\
150	-5.4325\\
160	-6.108\\
};
\addplot [color=green, mark=square, mark options={solid, green}, forget plot]
  table[row sep=crcr]{%
10	-0.14825\\
20	-0.3825\\
30	-0.65775\\
40	-0.957\\
50	-1.485\\
60	-2.0225\\
70	-2.30275\\
80	-2.522\\
90	-2.70425\\
100	-2.8975\\
110	-3.133\\
120	-3.689\\
130	-4.1105\\
140	-4.3115\\
150	-4.46125\\
160	-4.604\\
};
\addplot [color=mycolor1, mark=triangle, mark options={solid, mycolor1}, forget plot]
  table[row sep=crcr]{%
10	-0.14775\\
20	-0.382\\
30	-0.64425\\
40	-0.879\\
50	-1.20875\\
60	-1.6845\\
70	-2.1155\\
80	-2.366\\
90	-2.549\\
100	-2.7075\\
110	-2.83325\\
120	-2.93\\
130	-3.02375\\
140	-3.0815\\
150	-3.11\\
160	-3.076\\
};
\end{axis}

\begin{axis}[%
width=5.208in,
height=2.083in,
at={(0in,0in)},
scale only axis,
xmin=0,
xmax=1,
ymin=0,
ymax=1,
axis line style={draw=none},
ticks=none,
axis x line*=bottom,
axis y line*=left
]
\end{axis}
\end{tikzpicture}%

%% file: figures/0.4H_FDF.tex
%
%
\definecolor{mycolor1}{rgb}{1.00000,0.00000,1.00000}%
\begin{tikzpicture}

\begin{axis}[%
width=1.743in,
height=1.475in,
at={(0.677in,0.452in)},
scale only axis,
xmin=0,
xmax=160,
xlabel style={font=\color{white!15!black}},
xlabel={$f_\trm{p}\ [\trm{Hz}]$},
ymin=0,
ymax=1.05,
ytick={0,0.2,0.4,0.6,0.8,1},
ylabel style={font=\color{white!15!black}},
ylabel={Gain, $|\mathcal{F}|$},
axis background/.style={fill=white},
axis x line*=bottom,
axis y line*=left,
xmajorgrids,
ymajorgrids
]
\addplot [color=red, mark=x, mark options={solid, red}, forget plot]
  table[row sep=crcr]{%
10	1.00884503015048\\
20	0.991440267810975\\
30	0.881581847080266\\
40	0.665992682431606\\
50	0.410130283087906\\
60	0.274359033975696\\
70	0.334256460797896\\
80	0.380870707277427\\
90	0.338376781925046\\
100	0.232559465821111\\
110	0.132798527240795\\
120	0.135140000153604\\
130	0.17073372705351\\
140	0.168489352188931\\
150	0.127544487862489\\
160	0.06730182406627\\
};
\addplot [color=blue, mark=o, mark options={solid, blue}, forget plot]
  table[row sep=crcr]{%
10	1.01038683902644\\
20	0.991825956195802\\
30	0.877838112031254\\
40	0.66152611287983\\
50	0.405594016684301\\
60	0.250478090270755\\
70	0.285019547871311\\
80	0.325002544347207\\
90	0.296580959402192\\
100	0.215664816552723\\
110	0.120890907964185\\
120	0.0674455359471506\\
130	0.094380313153939\\
140	0.119242862888003\\
150	0.115389279168661\\
160	0.0908020080078645\\
};
\addplot [color=green, mark=square, mark options={solid, green}, forget plot]
  table[row sep=crcr]{%
10	1.0100345121587\\
20	0.986238248973978\\
30	0.860079526939741\\
40	0.646104119129676\\
50	0.388959026697606\\
60	0.188002068828403\\
70	0.144243387683928\\
80	0.175989265868349\\
90	0.187147924456348\\
100	0.17203891629031\\
110	0.140551779676827\\
120	0.102295960524797\\
130	0.064521736747822\\
140	0.0312161802449571\\
150	0.0121837160494724\\
160	0.0245003112850804\\
};
\addplot [color=mycolor1, mark=triangle, mark options={solid, mycolor1}, forget plot]
  table[row sep=crcr]{%
10	1.00792954202801\\
20	0.975192574962335\\
30	0.835061256624686\\
40	0.62347145995946\\
50	0.382177772523045\\
60	0.203539842349333\\
70	0.0870113202813033\\
80	0.0486654035120262\\
90	0.0795590737552652\\
100	0.101792802078459\\
110	0.115503979570423\\
120	0.119223801634332\\
130	0.115908070859429\\
140	0.107979518541884\\
150	0.0965665003357457\\
160	0.0858938602397327\\
};
\end{axis}

\begin{axis}[%
width=1.743in,
height=1.475in,
at={(2.971in,0.452in)},
scale only axis,
xmin=0,
xmax=160,
xlabel style={font=\color{white!15!black}},
xlabel={$f_\trm{p}\ [\trm{Hz}]$},
ymin=-8,
ymax=0,
ylabel style={font=\color{white!15!black}},
ylabel={Phase, $\angle\mathcal{F}/\pi$},
axis background/.style={fill=white},
axis x line*=bottom,
axis y line*=left,
xmajorgrids,
ymajorgrids
]
\addplot [color=red, mark=x, mark options={solid, red}, forget plot]
  table[row sep=crcr]{%
10	-0.11225\\
20	-0.293\\
30	-0.48675\\
40	-0.741\\
50	-1.09625\\
60	-1.599\\
70	-2.049\\
80	-2.372\\
90	-2.6525\\
100	-2.965\\
110	-3.3915\\
120	-4\\
130	-4.26975\\
140	-4.546\\
150	-4.79875\\
160	-5.124\\
};
\addplot [color=blue, mark=o, mark options={solid, blue}, forget plot]
  table[row sep=crcr]{%
10	-0.112\\
20	-0.2935\\
30	-0.489\\
40	-0.737\\
50	-1.0825\\
60	-1.566\\
70	-2.01925\\
80	-2.326\\
90	-2.56925\\
100	-2.8225\\
110	-3.11375\\
120	-3.611\\
130	-4.078\\
140	-4.336\\
150	-4.5325\\
160	-4.724\\
};
\addplot [color=green, mark=square, mark options={solid, green}, forget plot]
  table[row sep=crcr]{%
10	-0.11175\\
20	-0.2945\\
30	-0.49125\\
40	-0.718\\
50	-1.00875\\
60	-1.35\\
70	-2\\
80	-2.174\\
90	-2.3825\\
100	-2.5675\\
110	-2.71775\\
120	-2.864\\
130	-3.04\\
140	-3.239\\
150	-3.68375\\
160	-4.172\\
};
\addplot [color=mycolor1, mark=triangle, mark options={solid, mycolor1}, forget plot]
  table[row sep=crcr]{%
10	-0.11125\\
20	-0.2955\\
30	-0.48975\\
40	-0.687\\
50	-0.90625\\
60	-1.0815\\
70	-1.34575\\
80	-2\\
90	-2.30375\\
100	-2.4725\\
110	-2.5885\\
120	-2.696\\
130	-2.78975\\
140	-2.868\\
150	-2.93375\\
160	-2.972\\
};
\end{axis}

\begin{axis}[%
width=5.208in,
height=2.083in,
at={(0in,0in)},
scale only axis,
xmin=0,
xmax=1,
ymin=0,
ymax=1,
axis line style={draw=none},
ticks=none,
axis x line*=bottom,
axis y line*=left
]
\end{axis}
\end{tikzpicture}%

%% file: figures/0.6H_FDF.tex
%
%
\definecolor{mycolor1}{rgb}{1.00000,0.00000,1.00000}%
\begin{tikzpicture}

\begin{axis}[%
width=1.743in,
height=1.475in,
at={(0.677in,0.452in)},
scale only axis,
xmin=0,
xmax=160,
xlabel style={font=\color{white!15!black}},
xlabel={$f_\trm{p}\ [\trm{Hz}]$},
ymin=0,
ymax=1.05,
ytick={0,0.2,0.4,0.6,0.8,1},
ylabel style={font=\color{white!15!black}},
ylabel={Gain, $|\mathcal{F}|$},
axis background/.style={fill=white},
axis x line*=bottom,
axis y line*=left,
xmajorgrids,
ymajorgrids
]
\addplot [color=red, mark=x, mark options={solid, red}, forget plot]
  table[row sep=crcr]{%
10	0.994302667631739\\
20	0.98386833280278\\
30	0.934635531382865\\
40	0.829986228677161\\
50	0.67221333151544\\
60	0.482008887796557\\
70	0.299123745446017\\
80	0.193307453559379\\
90	0.215982717606052\\
100	0.267370852391224\\
110	0.281870175430441\\
120	0.250834466601036\\
130	0.189402830258457\\
140	0.117996771803126\\
150	0.0549882569958301\\
160	0.0611677622830647\\
};
\addplot [color=blue, mark=o, mark options={solid, blue}, forget plot]
  table[row sep=crcr]{%
10	0.98976817682798\\
20	0.981400352816451\\
30	0.932784821651114\\
40	0.827313633919786\\
50	0.66896831408376\\
60	0.479407316734069\\
70	0.297685545684484\\
80	0.179257456274822\\
90	0.184250532818722\\
100	0.226126480679339\\
110	0.246534130448378\\
120	0.230256651355129\\
130	0.188207125603333\\
140	0.133081161876705\\
150	0.0760114210197663\\
160	0.0326954158165835\\
};
\addplot [color=green, mark=square, mark options={solid, green}, forget plot]
  table[row sep=crcr]{%
10	0.998339051632531\\
20	0.98572482014539\\
30	0.927840738094452\\
40	0.810355910384847\\
50	0.653314205587985\\
60	0.474419549189813\\
70	0.301261242255052\\
80	0.164311077078408\\
90	0.0888392487574236\\
100	0.107636346837986\\
110	0.137016242730043\\
120	0.15604188710829\\
130	0.154839624993399\\
140	0.143223171173424\\
150	0.123228121942476\\
160	0.0969387316884971\\
};
\addplot [color=mycolor1, mark=triangle, mark options={solid, mycolor1}, forget plot]
  table[row sep=crcr]{%
10	0.997274710522109\\
20	0.982345343651387\\
30	0.909054730999228\\
40	0.785306417538301\\
50	0.637352743140129\\
60	0.472259235065113\\
70	0.318006350798031\\
80	0.19244546293388\\
90	0.0972307122695819\\
100	0.0299052226697418\\
110	0.0425526887638849\\
120	0.0748175758577054\\
130	0.100595526873815\\
140	0.116367046783454\\
150	0.122755471394671\\
160	0.12317427147719\\
};
\end{axis}

\begin{axis}[%
width=1.743in,
height=1.475in,
at={(2.971in,0.452in)},
scale only axis,
xmin=0,
xmax=160,
xlabel style={font=\color{white!15!black}},
xlabel={$f_\trm{p}\ [\trm{Hz}]$},
ymin=-8,
ymax=0,
ylabel style={font=\color{white!15!black}},
ylabel={Phase, $\angle\mathcal{F}/\pi$},
axis background/.style={fill=white},
axis x line*=bottom,
axis y line*=left,
xmajorgrids,
ymajorgrids
]
\addplot [color=red, mark=x, mark options={solid, red}, forget plot]
  table[row sep=crcr]{%
10	-0.0755\\
20	-0.2065\\
30	-0.34275\\
40	-0.513\\
50	-0.68875\\
60	-0.897\\
70	-1.1585\\
80	-1.55\\
90	-2\\
100	-2.225\\
110	-2.429\\
120	-2.618\\
130	-2.80925\\
140	-3.022\\
150	-3.38\\
160	-4\\
};
\addplot [color=blue, mark=o, mark options={solid, blue}, forget plot]
  table[row sep=crcr]{%
10	-0.076\\
20	-0.2075\\
30	-0.34425\\
40	-0.511\\
50	-0.68375\\
60	-0.882\\
70	-1.12875\\
80	-1.5\\
90	-2\\
100	-2.1875\\
110	-2.38225\\
120	-2.552\\
130	-2.71825\\
140	-2.8855\\
150	-3.095\\
160	-3.544\\
};
\addplot [color=green, mark=square, mark options={solid, green}, forget plot]
  table[row sep=crcr]{%
10	-0.07575\\
20	-0.208\\
30	-0.345\\
40	-0.503\\
50	-0.6625\\
60	-0.8325\\
70	-1.0185\\
80	-1.238\\
90	-1.674\\
100	-2.0625\\
110	-2.26675\\
120	-2.408\\
130	-2.52975\\
140	-2.6545\\
150	-2.7575\\
160	-2.848\\
};
\addplot [color=mycolor1, mark=triangle, mark options={solid, mycolor1}, forget plot]
  table[row sep=crcr]{%
10	-0.07575\\
20	-0.21\\
30	-0.3465\\
40	-0.493\\
50	-0.63125\\
60	-0.771\\
70	-0.89425\\
80	-0.978\\
90	-1.08\\
100	-2.0325\\
110	-2.38775\\
120	-2.438\\
130	-2.49725\\
140	-2.5775\\
150	-2.63\\
160	-2.692\\
};
\end{axis}

\begin{axis}[%
width=5.208in,
height=2.083in,
at={(0in,0in)},
scale only axis,
xmin=0,
xmax=1,
ymin=0,
ymax=1,
axis line style={draw=none},
ticks=none,
axis x line*=bottom,
axis y line*=left
]
\end{axis}
\end{tikzpicture}%

%% file: figures/0.8H_FDF.tex
%
%
\definecolor{mycolor1}{rgb}{1.00000,0.00000,1.00000}%
\begin{tikzpicture}

\begin{axis}[%
width=1.743in,
height=1.475in,
at={(0.677in,0.452in)},
scale only axis,
xmin=0,
xmax=160,
xlabel style={font=\color{white!15!black}},
xlabel={$f_\trm{p}\ [\trm{Hz}]$},
ymin=0,
ymax=1.05,
ytick={0,0.2,0.4,0.6,0.8,1},
ylabel style={font=\color{white!15!black}},
ylabel={Gain, $|\mathcal{F}|$},
axis background/.style={fill=white},
axis x line*=bottom,
axis y line*=left,
xmajorgrids,
ymajorgrids
]
\addplot [color=red, mark=x, mark options={solid, red}, forget plot]
  table[row sep=crcr]{%
10	0.988452579386244\\
20	0.966534101713545\\
30	0.927860930757199\\
40	0.870762641397192\\
50	0.79212432427336\\
60	0.696974395226658\\
70	0.588239910630546\\
80	0.471805579153268\\
90	0.358710881990761\\
100	0.252406827341709\\
110	0.157423130131921\\
120	0.0789767149553965\\
130	0.032628586327905\\
140	0.053247569723483\\
150	0.0886284966705006\\
160	0.111594197539701\\
};
\addplot [color=blue, mark=o, mark options={solid, blue}, forget plot]
  table[row sep=crcr]{%
10	0.996071525338702\\
20	0.974414760545401\\
30	0.932278295942455\\
40	0.868455750369045\\
50	0.783694439210363\\
60	0.686210264308051\\
70	0.579744059909485\\
80	0.468969089760659\\
90	0.360878037329396\\
100	0.258390301607099\\
110	0.167512041602776\\
120	0.0896400736609252\\
130	0.0348847930454911\\
140	0.0366861552480267\\
150	0.0722560274890554\\
160	0.0952123653106067\\
};
\addplot [color=green, mark=square, mark options={solid, green}, forget plot]
  table[row sep=crcr]{%
10	0.987671103947921\\
20	0.964480656879538\\
30	0.919792398694272\\
40	0.850074524806912\\
50	0.763034869128485\\
60	0.667183130190205\\
70	0.567340915134623\\
80	0.465499666118659\\
90	0.366819835207428\\
100	0.27466931373164\\
110	0.192058684585932\\
120	0.121537212854829\\
130	0.0630700868043442\\
140	0.0200897185244235\\
150	0.0310733017177917\\
160	0.0584675238637805\\
};
\addplot [color=mycolor1, mark=triangle, mark options={solid, mycolor1}, forget plot]
  table[row sep=crcr]{%
10	0.984501555416883\\
20	0.960529057641798\\
30	0.908778887267797\\
40	0.829421519032239\\
50	0.738632890703075\\
60	0.646493850826843\\
70	0.554657409750777\\
80	0.463069070796164\\
90	0.374463476795963\\
100	0.292262877369333\\
110	0.21853018618956\\
120	0.154919958467184\\
130	0.101417030469452\\
140	0.0664083751663117\\
150	0.0417701607587945\\
160	0.0300388089999733\\
};
\end{axis}

\begin{axis}[%
width=1.743in,
height=1.475in,
at={(2.971in,0.452in)},
scale only axis,
xmin=0,
xmax=160,
xlabel style={font=\color{white!15!black}},
xlabel={$f_\trm{p}\ [\trm{Hz}]$},
ymin=-8,
ymax=0,
ylabel style={font=\color{white!15!black}},
ylabel={Phase, $\angle\mathcal{F}/\pi$},
axis background/.style={fill=white},
axis x line*=bottom,
axis y line*=left,
xmajorgrids,
ymajorgrids
]
\addplot [color=red, mark=x, mark options={solid, red}, forget plot]
  table[row sep=crcr]{%
10	-0.046\\
20	-0.135\\
30	-0.23325\\
40	-0.324\\
50	-0.4175\\
60	-0.537\\
70	-0.616\\
80	-0.712\\
90	-0.8145\\
100	-0.9125\\
110	-1.023\\
120	-1.158\\
130	-1.5405\\
140	-2.0245\\
150	-2.18\\
160	-2.276\\
};
\addplot [color=blue, mark=o, mark options={solid, blue}, forget plot]
  table[row sep=crcr]{%
10	-0.04575\\
20	-0.1355\\
30	-0.2325\\
40	-0.325\\
50	-0.41625\\
60	-0.5385\\
70	-0.61425\\
80	-0.708\\
90	-0.80325\\
100	-0.8975\\
110	-0.99275\\
120	-1.101\\
130	-1.36175\\
140	-2.0105\\
150	-2.17625\\
160	-2.272\\
};
\addplot [color=green, mark=square, mark options={solid, green}, forget plot]
  table[row sep=crcr]{%
10	-0.04525\\
20	-0.1385\\
30	-0.234\\
40	-0.327\\
50	-0.415\\
60	-0.531\\
70	-0.59675\\
80	-0.682\\
90	-0.75825\\
100	-0.8375\\
110	-0.90475\\
120	-0.966\\
130	-1.027\\
140	-1.9625\\
150	-2.24375\\
160	-2.264\\
};
\addplot [color=mycolor1, mark=triangle, mark options={solid, mycolor1}, forget plot]
  table[row sep=crcr]{%
10	-0.04575\\
20	-0.1425\\
30	-0.2355\\
40	-0.326\\
50	-0.4075\\
60	-0.5175\\
70	-0.57225\\
80	-0.648\\
90	-0.70875\\
100	-0.7775\\
110	-0.825\\
120	-0.867\\
130	-0.90675\\
140	-1.903\\
150	-1.795\\
160	-2.576\\
};
\end{axis}

\begin{axis}[%
width=5.208in,
height=2.083in,
at={(0in,0in)},
scale only axis,
xmin=0,
xmax=1,
ymin=0,
ymax=1,
axis line style={draw=none},
ticks=none,
axis x line*=bottom,
axis y line*=left
]
\end{axis}
\end{tikzpicture}%

%% file: figures/dA_Compare.tex
%
%
\begin{tikzpicture}[scale = 0.7]

\begin{axis}[
width=1.952in,
height=1.488in,
at={(0.758in,2.554in)},
scale only axis,
scaled ticks=false,
tick label style={/pgf/number format/fixed},
ymin=0,
ymax=1.2,
ytick={   0, 0.25,  0.5, 0.75,    1},
xmin=0,
xmax=0.04,
xtick={    0, 0.01,  0.02, 0.03,  0.04},
ylabel style={font=\color{white!15!black}},
ylabel={$z/z_0$},
y tick label style={
        /pgf/number format/.cd,
        fixed,
        fixed zerofill,
        precision=2,
        /tikz/.cd
    },
axis background/.style={fill=white},
title style={font=\bfseries},
title={$t/\mathcal{T} = 1/4$},
axis x line*=bottom,
axis y line*=left
]
\addplot [color=blue, line width = 2, forget plot]
  table[row sep=crcr]{%
0.0337347416087678	0.00834889152834999\\
0.0330259950819979	0.02504667458505\\
0.0325161034679499	0.0371454200403237\\
0.0322159990514038	0.0454943115686737\\
0.0318029368684861	0.0584422406984499\\
0.031321619008993	0.0737792500312109\\
0.0308483590335897	0.0821281415595609\\
0.0304492342351422	0.0918378068118499\\
0.0301242446136505	0.10853558986855\\
0.0297629152176025	0.118357322072784\\
0.0292919064277596	0.126706213601134\\
0.0287112182441218	0.14193115598195\\
0.0280610855490808	0.154420829616029\\
0.0276115434200574	0.162769721144379\\
0.0271561624873266	0.17532672209535\\
0.0266949427508886	0.190545476041876\\
0.0261618069178771	0.198894367570226\\
0.0256339640948803	0.20872228820875\\
0.0251100571361283	0.22542007126545\\
0.0247268763357129	0.235282826216199\\
0.024336562920339	0.243631717744549\\
0.0240141830984224	0.25881563737885\\
0.0234884333027198	0.272125681633851\\
0.0228444846868193	0.280474573162201\\
0.0224015206589462	0.29221120349225\\
0.0219106331104519	0.30890898654895\\
0.0212946129347485	0.317939840778258\\
0.0207895251452072	0.326288732306608\\
0.0202179518799802	0.34230455266235\\
0.0197277519123626	0.356230723946435\\
0.0192438590339387	0.364579615474785\\
0.0188375372405328	0.375700118775749\\
0.0185630420197552	0.392397901832449\\
0.0181011899633807	0.40385200569609\\
0.0173676949506345	0.41220089722444\\
0.0170960102263123	0.425793467945849\\
0.0166553688396287	0.442491251002549\\
0.0162231886524314	0.452629358710277\\
0.0154975285987166	0.460978250238627\\
0.0151449171393676	0.475886817115949\\
0.0147334214204945	0.492584600172649\\
0.0143300441706236	0.502735537021713\\
0.013934785389755	0.511084428550063\\
0.0139391821246824	0.525980166286049\\
0.013554937116906	0.542677949342749\\
0.0128593468752948	0.554313054089009\\
0.0115250226603564	0.562661945617359\\
0.0115193305730525	0.576073515456149\\
0.0111594878510011	0.592771298512849\\
0.0108070338433383	0.607444439539083\\
0.0104619685500639	0.615793331067433\\
0.0098124879968228	0.626166864626249\\
0.0094736584843155	0.642864647682949\\
0.00914100021044294	0.659562430739649\\
0.00913389640017404	0.6704998326894\\
0.00977973579300126	0.67884872421775\\
0.00942729728815949	0.692957996853049\\
0.00911055077363434	0.709655779909749\\
0.0087981840777722	0.726353562966449\\
0.00838972467131693	0.734963362386385\\
0.00819697652862378	0.743312253914735\\
0.00789188065436984	0.759749129079849\\
0.00758964102846196	0.776446912136549\\
0.00728971944004668	0.792116466144049\\
0.00683674580875922	0.800465357672399\\
0.00605677078316019	0.809842478249949\\
0.00575972522021287	0.826540261306649\\
0.00546315066310318	0.843238044363349\\
0.00526716896239371	0.85794604441014\\
0.0049634922548811	0.86629493593849\\
0.00437668430667851	0.876633610476749\\
0.00407733578049775	0.893331393533449\\
0.00377565406748146	0.910029176590149\\
0.00362627649627869	0.923144661418966\\
0.00354310471699692	0.931493552947316\\
0.00316241888108572	0.943424742703549\\
0.00284945652532771	0.960122525760249\\
0.00253127424790369	0.976820308816949\\
0.00235736167525299	0.986965131308835\\
0.00220699386588956	0.995314022837185\\
0.00188039763972027	1.01021587493035\\
0.00155714464786623	1.02691365798705\\
0.00124489087104511	1.04054761365013\\
0.00119938404519008	1.04889650517848\\
0.000879608406033426	1.06030922410045\\
0.000624350427737541	1.07700700715715\\
0.000488918337894669	1.09370479021385\\
0.000240161402589782	1.11040257327055\\
};
\addplot [color=blue, dotted, line width = 1, forget plot]
  table[row sep=crcr]{%
0.012169634845198	0\\
0.012169634845198	1.2\\
};
\addplot [color=red, line width = 2, forget plot]
  table[row sep=crcr]{%
0.0167197189225401	0.00418756167106131\\
0.0165634302482035	0.0125626850131839\\
0.0164208457829924	0.0209378083553065\\
0.0162852292766577	0.0293129316974291\\
0.0161655918599801	0.0376880550395518\\
0.0160464370675304	0.0460631783816744\\
0.0159325431293394	0.0502918611158067\\
0.0158227262777646	0.054479422786868\\
0.015716340838504	0.0628134250659196\\
0.015613223085443	0.0711885484080422\\
0.0155128828078983	0.0795636717501648\\
0.0154153200058702	0.0879387950922874\\
0.0153354924447584	0.09631391843441\\
0.0152408726139441	0.104689041776533\\
0.015148599252059	0.109754926563246\\
0.0150584043750351	0.113942488234307\\
0.0149700089301936	0.121439288460778\\
0.0148831236394525	0.1298144118029\\
0.0147974495342163	0.138189535145023\\
0.0147129866144849	0.146564658487146\\
0.0146297348802583	0.154939781829268\\
0.0145455466822554	0.163314905171391\\
0.0144616143564995	0.171690028513514\\
0.0143776027067781	0.176406565795688\\
0.0142931694887825	0.18059412746675\\
0.0142079667235869	0.188440275197759\\
0.0141216403708889	0.196815398539881\\
0.0140341750379753	0.205190521882004\\
0.0139455707248461	0.213565645224127\\
0.0138662393529876	0.221940768566249\\
0.013774315993736	0.230315891908372\\
0.0136797644067408	0.235214164423235\\
0.0135821964571068	0.239401726094296\\
0.0134818187334502	0.247066138592617\\
0.0133788822060002	0.25544126193474\\
0.0132855857537266	0.263816385276862\\
0.0131880020723649	0.272191508618985\\
0.0130853526347833	0.279098544351809\\
0.0129663973453122	0.28328610602287\\
0.0128427592258626	0.28894175530323\\
0.0127084665602214	0.297316878645353\\
0.012591712656308	0.305692001987475\\
0.0124685084011878	0.314067125329598\\
0.0123388537948609	0.318324665523\\
0.0121908470116961	0.322512227194061\\
0.0120557012238997	0.330817372013843\\
0.0119138897209279	0.339192495355966\\
0.0117652810698883	0.346392260588277\\
0.0116098752707811	0.350579822259338\\
0.0114117208297071	0.355942742040211\\
0.011249512775182	0.364317865382334\\
0.0110807296957427	0.372692988724456\\
0.0109053715913892	0.377105336931744\\
0.0107234384621214	0.381292898602805\\
0.010542493502402	0.389443235408701\\
0.010356260394184	0.397818358750824\\
0.010165698744543	0.402797246703499\\
0.00997080855347891	0.40698480837456\\
0.00984166892141067	0.414568605435069\\
0.00964367576083106	0.422943728777192\\
0.00944333959613777	0.428174669903421\\
0.00924066042733079	0.432362231574482\\
0.00906502435426901	0.439693975461437\\
0.00886268910357102	0.44806909880356\\
0.00866023623095945	0.453633167309326\\
0.00845766573643429	0.457820728980388\\
0.00828480351417446	0.464819345487805\\
0.00808554138457786	0.473194468829927\\
0.00780641962225276	0.479534649471827\\
0.00744743822719914	0.483722211142889\\
0.00728286499256468	0.489944715514173\\
0.00709271355578903	0.498319838856295\\
0.00690659087347208	0.506252888178625\\
0.00672449694561383	0.510440449849686\\
0.00654643177221429	0.515070085540541\\
0.00636942649616551	0.523445208882663\\
0.00619736151327151	0.531820332224786\\
0.0061127276342769	0.538453838954022\\
0.00611552485918168	0.542641400625083\\
0.00593148539483818	0.548570578909031\\
0.00577506593358642	0.556945702251154\\
0.00562397153820194	0.565320825593276\\
0.00547893147991813	0.572552897756333\\
0.00534066157958695	0.576740459427394\\
0.00518230480013743	0.582071072277521\\
0.00502843359589375	0.590446195619644\\
0.00489823385939848	0.598821318961767\\
0.00477382271615821	0.607196442303889\\
0.0046558454266472	0.613692678374919\\
0.00452106973571042	0.61788024004598\\
0.00439168072686229	0.623946688988135\\
0.00426694912886943	0.632321812330257\\
0.00414615912087989	0.64069693567238\\
0.00404495875718312	0.649072059014502\\
0.00394939485134051	0.657447182356625\\
0.00384004245537908	0.663525025731212\\
0.00373529389213638	0.667712587402273\\
0.00363499749890427	0.67419742904087\\
0.00353900766013493	0.682572552382993\\
0.00344718460711967	0.690947675725116\\
0.00335952833985849	0.699322799067238\\
0.00320066360220678	0.707697922409361\\
0.00311677613657151	0.716073045751483\\
0.00303666836367175	0.724448169093606\\
0.0029602197970851	0.72970005146127\\
0.00288731356176113	0.733887613132331\\
0.00281783619651167	0.741198415777851\\
0.00275167748540506	0.749573539119974\\
0.00268873043605671	0.757948662462096\\
0.00262934146702286	0.766323785804219\\
0.00264922193213051	0.774698909146342\\
0.00258686408439372	0.783074032488464\\
0.00252773484039931	0.791449155830587\\
0.00242000999536522	0.799824279172709\\
0.00217134691651853	0.808199402514832\\
0.00212563778924531	0.816574525856955\\
0.00208249672963443	0.824949649199077\\
0.00204184196520024	0.832458930906138\\
0.00200359104269651	0.836646492577199\\
0.00196765945805504	0.841699895883323\\
0.00193395909517354	0.850075019225445\\
0.00190239642967719	0.858450142567568\\
0.00192459521648775	0.86682526590969\\
0.00209269084290986	0.875200389251813\\
0.00206257327414162	0.883575512593936\\
0.00203439553494421	0.891950635936058\\
0.00200798644309299	0.900325759278181\\
0.00198314955889163	0.908700882620303\\
0.00195965791217205	0.917076005962426\\
0.00193724806727231	0.925451129304549\\
0.00191561345745067	0.933826252646671\\
0.00189439697511115	0.942201375988794\\
0.00187318295313182	0.950576499330916\\
0.00185148878562288	0.958951622673039\\
0.00182875642835096	0.967326746015162\\
0.00180434418282479	0.975701869357284\\
0.00177751940225538	0.984076992699407\\
0.0017474521835965	0.992452116041529\\
0.00171321211761911	1.00082723938365\\
0.00167488314457251	1.00920236272577\\
0.00162165960717672	1.0175774860679\\
0.00157852976831344	1.02595260941002\\
0.00152811274231453	1.03432773275214\\
0.00146916745848395	1.03891954088635\\
0.00140032675893287	1.04310710255741\\
0.00132009081575461	1.05107797943639\\
0.00118265529460026	1.05945310277851\\
0.00105492391528287	1.06782822612063\\
0.000954042517835465	1.07620334946276\\
0.000849481739450846	1.08457847280488\\
0.000718750360612231	1.09065305063514\\
0.000585519884148967	1.0948406123062\\
0.000428004204216563	1.10132871948912\\
0.000339814663194173	1.10970384283125\\
0.000240617114817605	1.11807896617337\\
0.000107640827128204	1.12645408951549\\
};
\addplot [color=red, dotted, line width = 1, forget plot]
  table[row sep=crcr]{%
0.00709281050666122	0\\
0.00709281050666122	1.2\\
};
\end{axis}

\begin{axis}[%
width=1.952in,
height=1.488in,
at={(3.327in,2.554in)},
scale only axis,
scaled ticks=false,
tick label style={/pgf/number format/fixed},
ymin=0,
ymax=1.2,
ytick={   0, 0.25,  0.5, 0.75,    1},
y tick label style={
        /pgf/number format/.cd,
        fixed,
        fixed zerofill,
        precision=2,
        /tikz/.cd
    },
xmin=0,
xmax=0.04,
xtick={    0, 0.01,  0.02, 0.03,  0.04},
axis background/.style={fill=white},
title style={font=\bfseries},
title={$t/\mathcal{T} = 1/2$},
axis x line*=bottom,
axis y line*=left
]
\addplot [color=blue, line width = 2, forget plot]
  table[row sep=crcr]{%
0.0330043995063949	0.00834889152834999\\
0.0323765581720489	0.02504667458505\\
0.0319377370467753	0.0416984675175258\\
0.0316111624432173	0.0500473590458758\\
0.0312753726613057	0.0584422406984499\\
0.0308987568185293	0.0751400237551499\\
0.0305368110334445	0.0908591731400597\\
0.0301946826011822	0.0992080646684097\\
0.0299222305710834	0.10853558986855\\
0.0295465941804365	0.12523337292525\\
0.0291644105752628	0.138632735711176\\
0.0287756797555623	0.146981627239526\\
0.0284329697803277	0.15862893903865\\
0.0280862741891297	0.17532672209535\\
0.0277878822911654	0.185129658690051\\
0.0274246506344473	0.193478550218401\\
0.0268955727439324	0.20872228820875\\
0.0261507895702799	0.22216818649487\\
0.0257653719563582	0.23051707802322\\
0.0253660080164454	0.24211785432215\\
0.0249526977505415	0.258358625995326\\
0.0244728730996538	0.266707517523676\\
0.023992812255858	0.27551342043555\\
0.0235125152191541	0.29221120349225\\
0.0231451254415295	0.302250633710436\\
0.0228525626319705	0.310599525238786\\
0.0226348267904769	0.32560676960565\\
0.0219709704892803	0.337373026315544\\
0.0215075980660315	0.345721917843894\\
0.0209869117993861	0.359002335719049\\
0.0204089116893442	0.372268076259606\\
0.0199312670620653	0.380616967787956\\
0.0192853438513695	0.392397901832449\\
0.0184711420572568	0.407116153607468\\
0.0179850845645912	0.415465045135818\\
0.0175622317552781	0.425793467945849\\
0.0172025836293175	0.442091612038762\\
0.0164029298305066	0.450440503567112\\
0.0159639870526723	0.459189034059249\\
0.0154221309752057	0.475886817115949\\
0.0145534140798715	0.485725184580942\\
0.0141655547544208	0.494074076109292\\
0.0136335750722933	0.509282383229349\\
0.0126001389146815	0.521491229781953\\
0.0121323406257082	0.529840121310303\\
0.0116726639841149	0.542677949342749\\
0.0114510268941758	0.557905718997208\\
0.010652071626766	0.566254610525558\\
0.0103621573237616	0.576073515456149\\
0.0101076919091655	0.592771298512849\\
0.00934959106140662	0.603520657514402\\
0.00916737089355816	0.611869549042752\\
0.00923825535567595	0.626166864626249\\
0.00843761817679781	0.641810795220433\\
0.00750768002464576	0.650159686748783\\
0.0071159919773263	0.659562430739649\\
0.00701013318284232	0.676260213796349\\
0.00625248496326676	0.689706035882772\\
0.00565255202158809	0.698054927411122\\
0.00554751876190226	0.709655779909749\\
0.00512931018359606	0.726353562966449\\
0.00453839688017486	0.739029895075374\\
0.00419562968417629	0.747378786603724\\
0.00432552603591551	0.759749129079849\\
0.00393903727650484	0.776446912136549\\
0.00353839255796413	0.790014964534392\\
0.00320707523614199	0.798363856062742\\
0.00319428715533353	0.809842478249949\\
0.00283492065783219	0.826540261306649\\
0.00248347072303439	0.842637462590695\\
0.00223946544683414	0.850986354119045\\
0.0021398070821947	0.859935827420049\\
0.00180889876698044	0.876633610476749\\
0.00149508888987029	0.893331393533449\\
0.00122699657306665	0.905434314787166\\
0.00124274764027673	0.913783206315516\\
0.000963710871903779	0.926726959646849\\
0.000723049522185732	0.943424742703549\\
0.000560498454365315	0.960122525760249\\
0.000411861822154024	0.976820308816949\\
0.00023687538545195	0.993518091873649\\
0.00013415609209871	1.01021587493035\\
};
\addplot [color=blue, dotted, line width = 1, forget plot]
  table[row sep=crcr]{%
0.012169634845198	0\\
0.012169634845198	1.2\\
};
\addplot [color=red, line width = 2, forget plot]
  table[row sep=crcr]{%
0.0168671576964176	0.00418756167106131\\
0.0166530337806817	0.0125626850131839\\
0.0164678728417404	0.0209378083553065\\
0.0163151488862365	0.0293129316974291\\
0.0161777710960996	0.036310365170937\\
0.0160218846376168	0.0404979268419983\\
0.0158747535549231	0.0460631783816744\\
0.0157318434935398	0.054438301723797\\
0.0155154891940068	0.0628134250659196\\
0.0153964850994095	0.0711885484080422\\
0.0152819485887018	0.0795636717501648\\
0.0151517152829092	0.0839823334681593\\
0.0150256246275426	0.0881698951392206\\
0.0149036621584623	0.09631391843441\\
0.0147853112857197	0.104689041776533\\
0.014670049660413	0.113064165118655\\
0.0145727500874151	0.121439288460778\\
0.0145562509201892	0.1298144118029\\
0.0144501757814753	0.136104992766111\\
0.0143478027219606	0.140292554437173\\
0.0142489901572755	0.146564658487146\\
0.014153580705851	0.154939781829268\\
0.0140614024757307	0.163314905171391\\
0.0139724554669147	0.171690028513514\\
0.0138867396794031	0.180065151855636\\
0.0138004559165015	0.188440275197759\\
0.0137168106747034	0.196815398539881\\
0.0136355850258851	0.203234950969259\\
0.0135565481568663	0.20742251264032\\
0.0134794566734873	0.213565645224127\\
0.013404054276897	0.221940768566249\\
0.0133300715442589	0.230315891908372\\
0.0132575180937845	0.238691015250494\\
0.0131863939254738	0.247066138592617\\
0.0131056254311484	0.25544126193474\\
0.0130341509270309	0.263816385276862\\
0.0129628811200181	0.272191508618985\\
0.0128914797476966	0.279829428696919\\
0.0128195983006444	0.28401699036798\\
0.0127468769793991	0.28894175530323\\
0.0126729456793499	0.297316878645353\\
0.012597424835439	0.305692001987475\\
0.0125203144476663	0.314067125329598\\
0.0124416145160318	0.322442248671721\\
0.0123617361411618	0.330817372013843\\
0.0122790588794977	0.339192495355966\\
0.0121931576408858	0.346248234771842\\
0.0121035973846496	0.350435796442903\\
0.0120099306313668	0.355942742040211\\
0.0119124765250278	0.364317865382334\\
0.011811605012479	0.372692988724456\\
0.0117186816285037	0.381068112066579\\
0.0116197280417235	0.389443235408701\\
0.0115046789886525	0.39467193049539\\
0.0113840450682508	0.398859492166451\\
0.0112584230166249	0.406193482092947\\
0.0111426111526369	0.414568605435069\\
0.0110197150874988	0.422943728777192\\
0.0108230292672691	0.429024577281026\\
0.0105525536919479	0.433212138952087\\
0.0104142438680996	0.439693975461437\\
0.0102676364219856	0.44806909880356\\
0.0101125042488202	0.45620211712857\\
0.00994884734860343	0.460389678799631\\
0.00976063960713212	0.464819345487805\\
0.00958825006194019	0.473194468829927\\
0.00933866425110943	0.48156959217205\\
0.00907858772858133	0.485918818651334\\
0.00904286870744377	0.490106380322396\\
0.00884652282153437	0.498319838856295\\
0.00864399177597011	0.506398402630517\\
0.00843650595477017	0.510585964301579\\
0.00825347146845904	0.515070085540541\\
0.00808035150889583	0.523445208882663\\
0.0078626339478303	0.530704814882206\\
0.00771094446321849	0.534892376553267\\
0.00765734285029765	0.540195455566909\\
0.00743525630324784	0.548570578909031\\
0.00721234814688365	0.555025249016918\\
0.00698861838120509	0.559212810687979\\
0.00679706844363131	0.565320825593276\\
0.00657683779271232	0.573695948935399\\
0.00635848935079215	0.579680826633163\\
0.00614202311787081	0.583868388304225\\
0.00595975580229314	0.590446195619644\\
0.00574959818926693	0.598821318961767\\
0.00554327487152421	0.605002304344065\\
0.00534078584906499	0.609189866015126\\
0.00517259695439522	0.615571565646012\\
0.00497791192991233	0.623946688988135\\
0.00478909038170046	0.631361831859385\\
0.00460613230975963	0.635549393530446\\
0.00442903771408983	0.64069693567238\\
0.00425248408213442	0.649072059014502\\
0.00408216588827173	0.657447182356625\\
0.0039189377589108	0.663458186774746\\
0.00376279969405163	0.667645748445807\\
0.00358328586118826	0.67419742904087\\
0.00343151049307338	0.682572552382993\\
0.00328574547996361	0.690947675725116\\
0.00314675882192209	0.697772979392962\\
0.00301528427462157	0.701960541064023\\
0.00277167613077568	0.707697922409361\\
0.00248006335033437	0.716073045751483\\
0.00235681855013876	0.724448169093606\\
0.00223956105812331	0.732823292435729\\
0.00212889977855429	0.739644101614773\\
0.00200209870539158	0.743831663285834\\
0.00188054008876129	0.749573539119974\\
0.00173769810694776	0.757948662462096\\
0.00149153312954946	0.766323785804219\\
0.00148923268502014	0.774698909146342\\
0.00154508923917303	0.783074032488464\\
0.0014423578063175	0.790792186323398\\
0.00134388571078361	0.79497974799446\\
0.00124951637052141	0.799824279172709\\
0.00115910536580425	0.808199402514832\\
0.00107251991438902	0.816574525856955\\
0.00101593278292392	0.824949649199077\\
0.00100726430788861	0.8333247725412\\
0.000910493352489811	0.841699895883323\\
0.000835440519221688	0.850075019225445\\
0.000764561222436852	0.858450142567568\\
0.00069793463716524	0.86350957191126\\
0.000635648301134096	0.867697133582321\\
0.000577790782827327	0.875200389251813\\
0.000524440967983587	0.883575512593936\\
0.000475660108307403	0.891950635936058\\
0.000495276208107175	0.900325759278181\\
0.00046518089709254	0.908700882620303\\
0.000421837471792016	0.917076005962426\\
0.00038264368893146	0.925451129304549\\
0.000347356105229923	0.933826252646671\\
0.000315706384926046	0.942201375988794\\
0.000287409911723971	0.950576499330916\\
0.000262177953284929	0.958951622673039\\
0.000239726918158402	0.967326746015162\\
0.000219784834539052	0.975701869357284\\
0.000202095487831578	0.984076992699407\\
0.000186420296578594	0.992452116041529\\
0.000172538637342464	1.00082723938365\\
0.000160247338961496	1.00920236272577\\
0.000149359443944185	1.0175774860679\\
0.000139702797909324	1.02595260941002\\
0.000131118225557444	1.03432773275214\\
0.000122428348036027	1.04270285609427\\
0.000115082210965502	1.05107797943639\\
0.000108799340692459	1.05945310277851\\
0.000103326982066567	1.06782822612063\\
};
\addplot [color=red, dotted, line width = 1, forget plot]
  table[row sep=crcr]{%
0.00709281050666122	0\\
0.00709281050666122	1.2\\
};
\end{axis}

\begin{axis}[%
width=1.952in,
height=1.488in,
at={(0.758in,0.481in)},
scale only axis,
scaled ticks=false,
tick label style={/pgf/number format/fixed},
ymin=0,
ymax=1.2,
ytick={   0, 0.25,  0.5, 0.75,    1},
xlabel style={font=\color{white!15!black}},
xlabel={$\Delta A/A_0$},
xmin=0,
xmax=0.04,
xtick={    0, 0.01,  0.02, 0.03,  0.04},
ylabel style={font=\color{white!15!black}},
ylabel={$z/z_0$},
y tick label style={
        /pgf/number format/.cd,
        fixed,
        fixed zerofill,
        precision=2,
        /tikz/.cd
    },
axis background/.style={fill=white},
title style={font=\bfseries},
title={$t/\mathcal{T} = 1$},
axis x line*=bottom,
axis y line*=left
]
\addplot [color=blue, line width = 2, forget plot]
  table[row sep=crcr]{%
0.0340112489401303	0.00834889152834999\\
0.033042837119525	0.02504667458505\\
0.0323899967316276	0.0365706821520739\\
0.0319360807531188	0.0449195736804239\\
0.0315111306241071	0.0584422406984499\\
0.0309579947200365	0.0738932053798423\\
0.0304390690649909	0.0822420969081923\\
0.030031757430436	0.0918378068118499\\
0.0296629194965413	0.10853558986855\\
0.029157237800691	0.120441306661545\\
0.02837141606089	0.128790198189895\\
0.0279361255643456	0.14193115598195\\
0.0272645715116849	0.15862893903865\\
0.026414311281381	0.167934245279023\\
0.0260131032382186	0.176283136807373\\
0.0256232465755025	0.19202450515205\\
0.0249093404740539	0.208175887683234\\
0.0239445252537034	0.216524779211584\\
0.023568836847526	0.22542007126545\\
0.0234831128088209	0.24211785432215\\
0.02305796921571	0.258006658124522\\
0.0228283184042538	0.266355549652872\\
0.0227746306106051	0.27551342043555\\
0.0223684754079764	0.29221120349225\\
0.0219713255262496	0.30890898654895\\
0.0218411780131375	0.317516314204513\\
0.0219780328686401	0.325865205732863\\
0.0215978593065846	0.34230455266235\\
0.0212253875672417	0.359002335719049\\
0.0208593302852822	0.369995891389257\\
0.0204996874607062	0.378344782917607\\
0.0202183978472139	0.392397901832449\\
0.0198657993773364	0.409095684889149\\
0.0195186346227947	0.423577525233122\\
0.0191769035835888	0.431926416761472\\
0.0188406062597187	0.442491251002549\\
0.0185025216289481	0.459189034059249\\
0.0181680198378254	0.475886817115949\\
0.0178371008863506	0.486236770912871\\
0.0175097647745237	0.494585662441221\\
0.0171805212818075	0.509282383229349\\
0.0168528544736072	0.525980166286049\\
0.016526053497018	0.540687708796003\\
0.01620011835204	0.549036600324353\\
0.0155686880604973	0.559375732399449\\
0.0152413419674254	0.576073515456149\\
0.0149124648609011	0.592771298512849\\
0.0145820567409244	0.602890618056528\\
0.0142501176074953	0.611239509584878\\
0.0139155097118968	0.626166864626249\\
0.0135765400694596	0.642864647682949\\
0.0132334050381807	0.655419703802533\\
0.0128861046180602	0.663768595330883\\
0.0128880057126003	0.676260213796349\\
0.0125292895485004	0.692957996853049\\
0.0121633619779906	0.706115432197296\\
0.011790223001071	0.714464323725646\\
0.0114594289232739	0.726353562966449\\
0.011056953832154	0.743051346023149\\
0.0106313407764477	0.754748703138286\\
0.0102787481864723	0.763097594666636\\
0.00990964142751515	0.776446912136549\\
0.00947701457424999	0.793044958849572\\
0.00878532097411245	0.801393850377922\\
0.00838668142299312	0.809842478249949\\
0.00765117977808098	0.826540261306649\\
0.00684452822546287	0.83783672482534\\
0.00642282259150135	0.84618561635369\\
0.0057753983831512	0.859935827420049\\
0.00499438811295947	0.872794053987546\\
0.00451884195541605	0.881142945515896\\
0.0038626117085727	0.893331393533449\\
0.0034732483023258	0.906548543083295\\
0.00295733279409314	0.914897434611645\\
0.00264804705046089	0.926726959646849\\
0.00242181525013488	0.939532324806973\\
0.00188496743177676	0.947881216335323\\
0.00159233354809982	0.960122525760249\\
0.00118890354890409	0.971923704070454\\
0.00101298356919407	0.980272595598804\\
0.000634983639089897	0.993518091873649\\
0.000302601977139999	1.01021587493035\\
};
\addplot [color=blue, dotted, line width = 1, forget plot]
  table[row sep=crcr]{%
0.012169634845198	0\\
0.012169634845198	1.2\\
};
\addplot [color=red, line width = 2, forget plot]
  table[row sep=crcr]{%
0.0166227014991592	0.00418756167106131\\
0.016528097159637	0.0125626850131839\\
0.0164248825473606	0.0209378083553065\\
0.0163182527807389	0.0293129316974291\\
0.0162213394595617	0.0376880550395518\\
0.0161251679205819	0.0460631783816744\\
0.0160251836617698	0.052256426687353\\
0.0159224380102575	0.0564439883584143\\
0.0158170785848619	0.0628134250659196\\
0.0157093358569236	0.0711885484080422\\
0.0156121873053675	0.0795636717501648\\
0.0155112415764142	0.0879387950922874\\
0.015405880091206	0.0953751401047919\\
0.0152843223755545	0.0995627017758532\\
0.015158862083752	0.104689041776533\\
0.0150299537125053	0.113064165118655\\
0.0149133824496386	0.121439288460778\\
0.0147918663645404	0.1298144118029\\
0.0146647602018643	0.135150025341609\\
0.0145193733164677	0.13933758701267\\
0.0143700970621297	0.146564658487146\\
0.0142356206349872	0.154939781829268\\
0.0140967568264214	0.163314905171391\\
0.0139530147039536	0.1687934626012\\
0.0138043942675839	0.172981024272261\\
0.013635600311258	0.180065151855636\\
0.013484851946726	0.188440275197759\\
0.0133307954801917	0.196815398539881\\
0.013173430911655	0.201451414385302\\
0.0130127582411159	0.205638976056363\\
0.0128533555737047	0.213565645224127\\
0.0126920279483833	0.221940768566249\\
0.0125292902900025	0.229616260500001\\
0.0123651425985624	0.233803822171062\\
0.0121995848740629	0.238691015250494\\
0.0120355866375111	0.247066138592617\\
0.0118717590073173	0.25544126193474\\
0.0117083888277852	0.262226530418593\\
0.0115454760989148	0.266414092089654\\
0.0113603720982968	0.272191508618985\\
0.0111999123837553	0.280566631961107\\
0.0110414454291486	0.28894175530323\\
0.0108854352603764	0.295629581326902\\
0.0106339588989711	0.299817142997964\\
0.0102631096427873	0.305692001987475\\
0.0101126406881898	0.314067125329598\\
0.00996540098726667	0.322442248671721\\
0.0098219461213884	0.330446090112366\\
0.00968284209246708	0.334633651783427\\
0.00952427432251545	0.339192495355966\\
0.00936887034530288	0.347567618698088\\
0.00923629317813932	0.355942742040211\\
0.00920609647369483	0.364317865382334\\
0.00927886160247938	0.37176142114705\\
0.00913880106987409	0.375948982818112\\
0.00900335288635742	0.381068112066579\\
0.00887196147055875	0.389443235408701\\
0.00874406082056604	0.397818358750824\\
0.00863625680459362	0.406193482092947\\
0.00853363665643182	0.414568605435069\\
0.00841661597236332	0.421036211310989\\
0.00830409634768337	0.42522377298205\\
0.00819601814758302	0.431318852119315\\
0.00809231489356932	0.439693975461437\\
0.00794802076077361	0.44806909880356\\
0.00776313574919589	0.456444222145682\\
0.00768342620765959	0.464819345487805\\
0.00759238159332619	0.473194468829927\\
0.00750537844524141	0.48134568699813\\
0.00742232559726408	0.485533248669191\\
0.00734312970991066	0.489944715514173\\
0.00726769520169638	0.498319838856295\\
0.00719592379890704	0.506694962198418\\
0.00717260726259936	0.515070085540541\\
0.0071981314997223	0.523445208882663\\
0.00712710163096503	0.531820332224786\\
0.00705977861328495	0.540195455566909\\
0.00699569867384601	0.548570578909031\\
0.00693441033929706	0.556945702251154\\
0.00688240008684333	0.565320825593276\\
0.00683328401654375	0.573695948935399\\
0.00678693369021823	0.579488933615056\\
0.0067432126047926	0.583676495286117\\
0.00670197466472378	0.590446195619644\\
0.00666306249150805	0.598821318961767\\
0.0066263055480977	0.607196442303889\\
0.00659185824647817	0.615571565646012\\
0.00655984028880781	0.623946688988135\\
0.00652340747241391	0.632321812330257\\
0.0064885560437692	0.64069693567238\\
0.00645499250693608	0.649072059014502\\
0.00642239247388728	0.657447182356625\\
0.00639039722973664	0.665822305698748\\
0.00635861025920731	0.67419742904087\\
0.00632659386321888	0.682572552382993\\
0.00629386600304998	0.690947675725116\\
0.00625989751895011	0.699322799067238\\
0.0062241098659448	0.707697922409361\\
0.00618646454328611	0.716073045751483\\
0.00614697036628485	0.724448169093606\\
0.00610904616925846	0.732823292435729\\
0.00606737601727955	0.741198415777851\\
0.00602123745735883	0.746770879907504\\
0.00596986184976519	0.750958441578565\\
0.00591243387583991	0.757948662462096\\
0.00584808981151632	0.766323785804219\\
0.00575094232852952	0.774698909146342\\
0.0056209914268795	0.783074032488464\\
0.00554466116895187	0.791449155830587\\
0.00545383625006246	0.799824279172709\\
0.00535205053622975	0.80802342760899\\
0.00521562710470164	0.812210989280051\\
0.00504558654900409	0.816574525856955\\
0.00492874843313366	0.824949649199077\\
0.0047984755306386	0.8333247725412\\
0.00464334030263596	0.841518821582672\\
0.00446334274912573	0.845706383253733\\
0.00430308977628602	0.850075019225445\\
0.00412657302594641	0.858450142567568\\
0.00393569347049605	0.866491331513995\\
0.00375503290994436	0.870678893185057\\
0.00361030460138789	0.875200389251813\\
0.00330929934192845	0.883575512593936\\
0.00289887907155007	0.889571052560573\\
0.00271970896951033	0.893758614231635\\
0.00254729389920942	0.900325759278181\\
0.00225060289208363	0.907541566671652\\
0.00201529587729427	0.911729128342713\\
0.00176995587096276	0.917076005962426\\
0.00150813434441932	0.925112525538079\\
0.00122628985257785	0.929300087209141\\
0.00103621394544885	0.933826252646671\\
0.000924672093616008	0.942201375988794\\
0.000724023146246458	0.946716811818782\\
0.000379743700475979	0.950904373489843\\
0.000168827723384486	0.958951622673039\\
};
\addplot [color=red, dotted, line width = 1, forget plot]
  table[row sep=crcr]{%
0.00709281050666122	0\\
0.00709281050666122	1.2\\
};
\end{axis}

\begin{axis}[%
width=1.952in,
height=1.488in,
at={(3.327in,0.481in)},
scale only axis,
scaled ticks=false,
tick label style={/pgf/number format/fixed},
ymin=0,
ymax=1.2,
ytick={   0, 0.25,  0.5, 0.75,    1},
xlabel style={font=\color{white!15!black}},
xlabel={$\Delta A/A_0$},
xmin=0,
xmax=0.04,
xtick={    0, 0.01,  0.02, 0.03,  0.04},
y tick label style={
        /pgf/number format/.cd,
        fixed,
        fixed zerofill,
        precision=2,
        /tikz/.cd
    },
axis background/.style={fill=white},
title style={font=\bfseries},
title={$t/\mathcal{T} = 3/4$},
axis x line*=bottom,
axis y line*=left
]
\addplot [color=blue, line width = 2, forget plot]
  table[row sep=crcr]{%
0.0333031289110138	0.00834889152834999\\
0.0315202037421756	0.02504667458505\\
0.0302474971242968	0.0409716777048783\\
0.0303281512054888	0.0493205692332283\\
0.0301173428956896	0.0584422406984499\\
0.0296721001376024	0.0751400237551499\\
0.0292762152254429	0.0918378068118499\\
0.0292163138339249	0.100408054528163\\
0.0294923959630484	0.108756946056513\\
0.0291119403613543	0.12523337292525\\
0.028737912736479	0.14193115598195\\
0.0283686749605785	0.152152853565623\\
0.0280042270336528	0.160501745093973\\
0.0277281535124639	0.17532672209535\\
0.0273672593674898	0.19202450515205\\
0.0270090899211131	0.204062540171673\\
0.0266536451733337	0.212411431700023\\
0.0263515093930986	0.22542007126545\\
0.025996577225345	0.24211785432215\\
0.0256422445791712	0.255591004094003\\
0.0252885114545773	0.263939895622353\\
0.0249850881028066	0.27551342043555\\
0.0246283298047667	0.29221120349225\\
0.024269833750356	0.306215017581993\\
0.0239095999395744	0.314563909110343\\
0.0235975237017375	0.32560676960565\\
0.0232302316944654	0.34230455266235\\
0.0228585606411772	0.355549279598415\\
0.0224825105418728	0.363898171126765\\
0.0221533540013027	0.375700118775749\\
0.0214958694852277	0.392397901832449\\
0.02056125254679	0.403418050616388\\
0.0202168346561117	0.411766942144738\\
0.0198610598574984	0.425793467945849\\
0.0194440328216345	0.441478182891494\\
0.0190185426219529	0.449827074419844\\
0.0186417790464576	0.459189034059249\\
0.0183137420951488	0.475886817115949\\
0.017983159163276	0.486532101555733\\
0.0178511929108298	0.494880993084083\\
0.0179178433378101	0.509282383229349\\
0.017315778974095	0.522098062575435\\
0.0168847400919598	0.530446954103785\\
0.0164098217777389	0.542677949342749\\
0.0158910240314325	0.556745150962699\\
0.0154292982931651	0.565094042491049\\
0.0149232670203447	0.576073515456149\\
0.0143729302129712	0.590684085028204\\
0.0138816078597453	0.599032976556554\\
0.0134492999606671	0.609469081569549\\
0.0130760065157366	0.624119471376252\\
0.0125589706307667	0.632468362904602\\
0.0121130972194229	0.642864647682949\\
0.0117383862817053	0.657241954675768\\
0.0112021289813808	0.665590846204118\\
0.0102985455269308	0.676260213796349\\
0.00919338883537222	0.690228483906614\\
0.00864630483982737	0.698577375434964\\
0.00780549979133666	0.709655779909749\\
0.00686375266397368	0.723245098681629\\
0.00631500299383161	0.731593990209979\\
0.00551914548736185	0.743051346023149\\
0.00460670851534423	0.756449674883884\\
0.00406514665296525	0.764798566412234\\
0.00374757956148972	0.776446912136549\\
0.00352938481043021	0.789989428713299\\
0.00300054294281921	0.798338320241649\\
0.00272125410856893	0.809842478249949\\
0.00247920476782474	0.823889092426241\\
0.00197269983371493	0.832237983954591\\
0.00169245203886332	0.843238044363349\\
0.00152811698349891	0.858311828188355\\
0.00108536459380488	0.866660719716705\\
0.000921851199620199	0.876633610476749\\
0.000478772831899551	0.893331393533449\\
0.000328343886629537	0.910029176590149\\
0.000107074862816585	0.926726959646849\\
};
\addplot [color=blue, dotted, line width = 1, forget plot]
  table[row sep=crcr]{%
0.012169634845198	0\\
0.012169634845198	1.2\\
};
\addplot [color=red, line width = 2, forget plot]
  table[row sep=crcr]{%
0.0168914667129619	0.00418756167106131\\
0.0166980530233348	0.0125626850131839\\
0.0165147692389947	0.0209378083553065\\
0.0163585043505739	0.0293129316974291\\
0.016209204097258	0.0344691070600118\\
0.0160444230084395	0.0386566687310731\\
0.015880929434934	0.0460631783816744\\
0.0157382090293104	0.054438301723797\\
0.0155950058113931	0.0628134250659196\\
0.0154515095222613	0.0691747716159512\\
0.0153079600009477	0.0733623332870125\\
0.0151452620841747	0.0795636717501648\\
0.0149778249936363	0.0879387950922874\\
0.0148372867745281	0.09631391843441\\
0.0146984422411204	0.104689041776533\\
0.0145616903433202	0.108942697106919\\
0.0143185849417138	0.113130258777981\\
0.0139895880279604	0.121439288460778\\
0.0138575616614684	0.1298144118029\\
0.0137283268450411	0.138189535145023\\
0.0136023667165719	0.146323153849661\\
0.0134855189500907	0.150510715520722\\
0.0133467103522712	0.154939781829268\\
0.0132111461088428	0.163314905171391\\
0.0130784272698987	0.171690028513514\\
0.0130534939603897	0.180065151855636\\
0.0131206830705118	0.188440275197759\\
0.0129974469428454	0.195186604584035\\
0.0128782237881668	0.199374166255096\\
0.012763022936008	0.205190521882004\\
0.0126518365955535	0.213565645224127\\
0.012544164532753	0.221940768566249\\
0.0124400067476063	0.230315891908372\\
0.0123529382300578	0.238691015250494\\
0.0122535831576734	0.247066138592617\\
0.0121580419079699	0.253658315300632\\
0.0120662320442654	0.257845876971693\\
0.0119780591792677	0.263816385276862\\
0.0118934170621464	0.272191508618985\\
0.0118121870842763	0.280566631961107\\
0.0117347140757279	0.28894175530323\\
0.0116468935734674	0.297316878645353\\
0.0115351505875507	0.305692001987475\\
0.011463316527238	0.314067125329598\\
0.0113942942170013	0.322442248671721\\
0.0113279030844216	0.330769287695427\\
0.0112639492970122	0.334956849366488\\
0.0112022251601917	0.339192495355966\\
0.0111425084269459	0.347567618698088\\
0.0110845616452218	0.355942742040211\\
0.0110425576409384	0.364317865382334\\
0.0110165233564757	0.372692988724456\\
0.0109547894372463	0.381068112066579\\
0.0108939523482608	0.389443235408701\\
0.0108336584755848	0.397818358750824\\
0.0107736969863852	0.406193482092947\\
0.0107138893586481	0.414568605435069\\
0.0106602718872891	0.422943728777192\\
0.0106059008566006	0.431318852119315\\
0.0105503768888091	0.43645650955017\\
0.0104932823938119	0.440644071221232\\
0.010434182347594	0.44806909880356\\
0.0103726254454932	0.456444222145682\\
0.0103081452925589	0.464819345487805\\
0.0102274730711692	0.473194468829927\\
0.0101306087813243	0.48156959217205\\
0.0100588252460322	0.489944715514173\\
0.00998261178215805	0.498319838856295\\
0.00990142606357996	0.506646743964106\\
0.00981470273643474	0.510834305635167\\
0.00972185009599683	0.515070085540541\\
0.00962338288973328	0.523445208882663\\
0.00951987999256184	0.531820332224786\\
0.00943577208786444	0.540195455566909\\
0.00935612492585136	0.548570578909031\\
0.00923203850171579	0.553642205867969\\
0.00909606165872164	0.55782976753903\\
0.00895197566853926	0.565320825593276\\
0.00879675397336129	0.573695948935399\\
0.00863144027172915	0.582071072277521\\
0.00837506572407279	0.586268786971105\\
0.00820935852036118	0.590456348642166\\
0.00793039190765243	0.598821318961767\\
0.00755310013573628	0.607196442303889\\
0.00736837584943429	0.611876872094239\\
0.00717478215422644	0.6160644337653\\
0.00696941111945233	0.623946688988135\\
0.00675624530450556	0.631865811289897\\
0.00656476971345703	0.636053372960958\\
0.00644380981887207	0.64069693567238\\
0.00621163743078172	0.649072059014502\\
0.00608087862216669	0.655312473033663\\
0.00604605276961392	0.659500034704724\\
0.00581771404443231	0.665822305698748\\
0.00560109751870289	0.67419742904087\\
0.0053875656826733	0.678484676584552\\
0.00517169457381504	0.682672238255613\\
0.00492399918805723	0.690947675725116\\
0.00462210871972241	0.697525651812677\\
0.00441198871296407	0.701713213483739\\
0.00417887266246265	0.707697922409361\\
0.00394069549737579	0.716073045751483\\
0.00369288791190802	0.721102293420829\\
0.00343654193595854	0.72528985509189\\
0.00317851321208986	0.732823292435729\\
0.00293091767232482	0.741130383412622\\
0.00269189796443648	0.745317945083683\\
0.00240960780680114	0.749573539119974\\
0.00211566669057311	0.757948662462096\\
0.00189810918803925	0.766323785804219\\
0.00168300657785492	0.770556889492147\\
0.00147595375322221	0.774744451163208\\
0.00126625702853468	0.783074032488464\\
0.00108079029582186	0.791449155830587\\
0.000904583719152067	0.797107695412395\\
0.000742675105495484	0.801295257083457\\
0.000704392670299192	0.808199402514832\\
0.00061451296858403	0.816574525856955\\
0.000481675776576169	0.824949649199077\\
0.000334762227178825	0.82966945903365\\
0.000238505249209966	0.833857020704711\\
0.000162154162797566	0.841699895883323\\
0.000112388973378325	0.850075019225445\\
};
\addplot [color=red, dotted, line width = 1, forget plot]
  table[row sep=crcr]{%
0.00709281050666122	0\\
0.00709281050666122	1.2\\
};
\end{axis}

\begin{axis}[%
width=5.208in,
height=5.208in,
at={(0in,0in)},
scale only axis,
xmin=0,
xmax=1,
ymin=0,
ymax=1,
axis line style={draw=none},
ticks=none,
axis x line*=bottom,
axis y line*=left
]
\end{axis}

\end{tikzpicture}%

%% file: figures/A_Transient.tex
\begin{tikzpicture}[scale = 0.7]
\pgfplotsset{
    scale only axis,
    scaled ticks=false,
    tick label style={/pgf/number format/fixed},
    xmin=0, xmax=1,
    xtick={0,0.5,1},
    ymajorgrids = true,
    legend pos=south east,
}

\begin{axis}[
  axis y line*=left,
  ymin=-80, ymax=80,
  xlabel=$t/\mathcal{T}$,
  ylabel={$A'(t)\ [\trm{mm}^2]$},
]
\addplot[smooth,red,line width = 2]
  table{
    0   48.1534
    0.0050   48.1722
    0.0100   48.1684
    0.0150   48.1488
    0.0200   48.1102
    0.0250   48.0485
    0.0300   47.9672
    0.0350   47.8701
    0.0400   47.7523
    0.0450   47.6116
    0.0500   47.4544
    0.0550   47.2788
    0.0600   47.0802
    0.0650   46.8608
    0.0700   46.6222
    0.0750   46.3603
    0.0800   46.0772
    0.0850   45.7771
    0.0900   45.4560
    0.0950   45.1107
    0.1000   44.7460
    0.1050   44.3605
    0.1100   43.9498
    0.1150   43.5159
    0.1200   43.0623
    0.1250   42.5847
    0.1300   42.0804
    0.1350   41.5558
    0.1400   41.0074
    0.1450   40.4318
    0.1500   39.8319
    0.1550   39.2078
    0.1600   38.5548
    0.1650   37.8770
    0.1700   37.1774
    0.1750   36.4501
    0.1800   35.6937
    0.1850   34.9139
    0.1900   34.1070
    0.1950   33.2689
    0.2000   32.4044
    0.2050   31.5120
    0.2100   30.5884
    0.2150   29.6351
    0.2200   28.6543
    0.2250   27.6414
    0.2300   26.5957
    0.2350   25.5191
    0.2400   24.4098
    0.2450   23.2684
    0.2500   22.0957
    0.2550   20.8922
    0.2600   19.6564
    0.2650   18.3881
    0.2700   17.0865
    0.2750   15.7531
    0.2800   14.3879
    0.2850   12.9899
    0.2900   11.5573
    0.2950   10.0926
    0.3000    8.5954
    0.3050    7.0635
    0.3100    5.4958
    0.3150    3.8926
    0.3200    2.2542
    0.3250    0.5791
    0.3300   -1.1323
    0.3350   -2.8800
    0.3400   -4.6651
    0.3450   -6.4889
    0.3500   -8.3508
    0.3550  -10.2503
    0.3600  -12.1869
    0.3650  -14.1602
    0.3700  -16.1689
    0.3750  -18.2108
    0.3800  -20.2833
    0.3850  -22.3809
    0.3900  -24.4978
    0.3950  -26.6244
    0.4000  -28.7455
    0.4050  -30.8376
    0.4100  -32.8746
    0.4150  -34.8400
    0.4200  -36.7201
    0.4250  -38.5002
    0.4300  -40.1633
    0.4350  -41.6930
    0.4400  -43.0768
    0.4450  -44.3053
    0.4500  -45.3711
    0.4550  -46.2669
    0.4600  -46.9881
    0.4650  -47.5351
    0.4700  -47.9113
    0.4750  -48.1215
    0.4800  -48.1704
    0.4850  -48.0640
    0.4900  -47.8116
    0.4950  -47.4236
    0.5000  -46.9092
    0.5050  -46.2770
    0.5100  -45.5358
    0.5150  -44.6951
    0.5200  -43.7640
    0.5250  -42.7513
    0.5300  -41.6642
    0.5350  -40.5097
    0.5400  -39.2946
    0.5450  -38.0259
    0.5500  -36.7092
    0.5550  -35.3497
    0.5600  -33.9525
    0.5650  -32.5223
    0.5700  -31.0628
    0.5750  -29.5787
    0.5800  -28.0737
    0.5850  -26.5511
    0.5900  -25.0142
    0.5950  -23.4660
    0.6000  -21.9096
    0.6050  -20.3475
    0.6100  -18.7818
    0.6150  -17.2144
    0.6200  -15.6476
    0.6250  -14.0840
    0.6300  -12.5247
    0.6350  -10.9710
    0.6400   -9.4246
    0.6450   -7.8873
    0.6500   -6.3606
    0.6550   -4.8451
    0.6600   -3.3422
    0.6650   -1.8535
    0.6700   -0.3797
    0.6750    1.0785
    0.6800    2.5201
    0.6850    3.9440
    0.6900    5.3495
    0.6950    6.7359
    0.7000    8.1027
    0.7050    9.4488
    0.7100   10.7739
    0.7150   12.0776
    0.7200   13.3599
    0.7250   14.6200
    0.7300   15.8574
    0.7350   17.0719
    0.7400   18.2635
    0.7450   19.4313
    0.7500   20.5748
    0.7550   21.6942
    0.7600   22.7906
    0.7650   23.8631
    0.7700   24.9112
    0.7750   25.9343
    0.7800   26.9325
    0.7850   27.9063
    0.7900   28.8564
    0.7950   29.7816
    0.8000   30.6826
    0.8050   31.5582
    0.8100   32.4092
    0.8150   33.2362
    0.8200   34.0398
    0.8250   34.8192
    0.8300   35.5737
    0.8350   36.3040
    0.8400   37.0116
    0.8450   37.6962
    0.8500   38.3576
    0.8550   38.9943
    0.8600   39.6085
    0.8650   40.2001
    0.8700   40.7681
    0.8750   41.3120
    0.8800   41.8337
    0.8850   42.3351
    0.8900   42.8148
    0.8950   43.2715
    0.9000   43.7065
    0.9050   44.1215
    0.9100   44.5153
    0.9150   44.8866
    0.9200   45.2375
    0.9250   45.5690
    0.9300   45.8800
    0.9350   46.1687
    0.9400   46.4384
    0.9450   46.6892
    0.9500   46.9202
    0.9550   47.1295
    0.9600   47.3216
    0.9650   47.4955
    0.9700   47.6486
    0.9750   47.7796
    0.9800   47.8930
    0.9850   47.9869
    0.9900   48.0596
    0.9950   48.1147
    1.0000   48.1535
}; \label{plot_one}

\addplot[smooth,blue,line width = 2]
  table{
    0   64.5286
    0.0050   64.5746
    0.0100   64.5634
    0.0150   64.4950
    0.0200   64.3685
    0.0250   64.1830
    0.0300   63.9374
    0.0350   63.6329
    0.0400   63.2704
    0.0450   62.8505
    0.0500   62.3727
    0.0550   61.8363
    0.0600   61.2401
    0.0650   60.5832
    0.0700   59.8687
    0.0750   59.0981
    0.0800   58.2715
    0.0850   57.3886
    0.0900   56.4486
    0.0950   55.4510
    0.1000   54.3954
    0.1050   53.2842
    0.1100   52.1195
    0.1150   50.9022
    0.1200   49.6327
    0.1250   48.3111
    0.1300   46.9375
    0.1350   45.5123
    0.1400   44.0360
    0.1450   42.5097
    0.1500   40.9364
    0.1550   39.3174
    0.1600   37.6539
    0.1650   35.9467
    0.1700   34.1966
    0.1750   32.4047
    0.1800   30.5728
    0.1850   28.7030
    0.1900   26.7963
    0.1950   24.8535
    0.2000   22.8760
    0.2050   20.8662
    0.2100   18.8283
    0.2150   16.7642
    0.2200   14.6738
    0.2250   12.5583
    0.2300   10.4213
    0.2350    8.2674
    0.2400    6.0997
    0.2450    3.9201
    0.2500    1.7306
    0.2550   -0.4662
    0.2600   -2.6664
    0.2650   -4.8670
    0.2700   -7.0665
    0.2750   -9.2638
    0.2800  -11.4572
    0.2850  -13.6432
    0.2900  -15.8171
    0.2950  -17.9741
    0.3000  -20.1107
    0.3050  -22.2245
    0.3100  -24.3128
    0.3150  -26.3708
    0.3200  -28.3970
    0.3250  -30.3895
    0.3300  -32.3461
    0.3350  -34.2651
    0.3400  -36.1426
    0.3450  -37.9752
    0.3500  -39.7621
    0.3550  -41.5003
    0.3600  -43.1874
    0.3650  -44.8206
    0.3700  -46.3972
    0.3750  -47.9159
    0.3800  -49.3756
    0.3850  -50.7754
    0.3900  -52.1137
    0.3950  -53.3886
    0.4000  -54.5983
    0.4050  -55.7423
    0.4100  -56.8200
    0.4150  -57.8304
    0.4200  -58.7723
    0.4250  -59.6443
    0.4300  -60.4461
    0.4350  -61.1773
    0.4400  -61.8374
    0.4450  -62.4264
    0.4500  -62.9439
    0.4550  -63.3902
    0.4600  -63.7651
    0.4650  -64.0687
    0.4700  -64.3013
    0.4750  -64.4629
    0.4800  -64.5543
    0.4850  -64.5760
    0.4900  -64.5281
    0.4950  -64.4115
    0.5000  -64.2270
    0.5050  -63.9749
    0.5100  -63.6565
    0.5150  -63.2726
    0.5200  -62.8239
    0.5250  -62.3116
    0.5300  -61.7365
    0.5350  -61.1002
    0.5400  -60.4031
    0.5450  -59.6469
    0.5500  -58.8327
    0.5550  -57.9616
    0.5600  -57.0350
    0.5650  -56.0542
    0.5700  -55.0205
    0.5750  -53.9356
    0.5800  -52.8008
    0.5850  -51.6175
    0.5900  -50.3869
    0.5950  -49.1107
    0.6000  -47.7905
    0.6050  -46.4277
    0.6100  -45.0238
    0.6150  -43.5806
    0.6200  -42.0995
    0.6250  -40.5821
    0.6300  -39.0301
    0.6350  -37.4449
    0.6400  -35.8282
    0.6450  -34.1817
    0.6500  -32.5070
    0.6550  -30.8059
    0.6600  -29.0800
    0.6650  -27.3307
    0.6700  -25.5596
    0.6750  -23.7683
    0.6800  -21.9587
    0.6850  -20.1323
    0.6900  -18.2907
    0.6950  -16.4359
    0.7000  -14.5694
    0.7050  -12.6929
    0.7100  -10.8079
    0.7150   -8.9158
    0.7200   -7.0185
    0.7250   -5.1174
    0.7300   -3.2144
    0.7350   -1.3110
    0.7400    0.5910
    0.7450    2.4900
    0.7500    4.3847
    0.7550    6.2735
    0.7600    8.1546
    0.7650   10.0274
    0.7700   11.8902
    0.7750   13.7415
    0.7800   15.5797
    0.7850   17.4035
    0.7900   19.2114
    0.7950   21.0022
    0.8000   22.7742
    0.8050   24.5258
    0.8100   26.2551
    0.8150   27.9600
    0.8200   29.6410
    0.8250   31.2968
    0.8300   32.9259
    0.8350   34.5273
    0.8400   36.0993
    0.8450   37.6400
    0.8500   39.1479
    0.8550   40.6214
    0.8600   42.0586
    0.8650   43.4611
    0.8700   44.8281
    0.8750   46.1589
    0.8800   47.4520
    0.8850   48.7058
    0.8900   49.9188
    0.8950   51.0894
    0.9000   52.2175
    0.9050   53.3040
    0.9100   54.3476
    0.9150   55.3471
    0.9200   56.3003
    0.9250   57.2051
    0.9300   58.0595
    0.9350   58.8632
    0.9400   59.6179
    0.9450   60.3224
    0.9500   60.9759
    0.9550   61.5776
    0.9600   62.1258
    0.9650   62.6183
    0.9700   63.0568
    0.9750   63.4417
    0.9800   63.7718
    0.9850   64.0463
    0.9900   64.2648
    0.9950   64.4263
    1.0000   64.5288
}; \label{plot_two}

\end{axis}

\begin{axis}[
  axis y line*=right,
  axis x line=none,
  ymin=-0.08, ymax=0.08,
  ylabel=$A'(t)/A_0$
]
\addlegendimage{/pgfplots/refstyle=plot_one}\addlegendentry{0\% H $[\trm{mm}^2]$}
\addlegendimage{/pgfplots/refstyle=plot_two}\addlegendentry{60\% H $[\trm{mm}^2]$}
\addplot[dashed,red,line width = 2]
  table{
    0    0.0230
    0.0050    0.0230
    0.0100    0.0230
    0.0150    0.0230
    0.0200    0.0230
    0.0250    0.0230
    0.0300    0.0229
    0.0350    0.0229
    0.0400    0.0228
    0.0450    0.0228
    0.0500    0.0227
    0.0550    0.0226
    0.0600    0.0225
    0.0650    0.0224
    0.0700    0.0223
    0.0750    0.0222
    0.0800    0.0220
    0.0850    0.0219
    0.0900    0.0217
    0.0950    0.0216
    0.1000    0.0214
    0.1050    0.0212
    0.1100    0.0210
    0.1150    0.0208
    0.1200    0.0206
    0.1250    0.0204
    0.1300    0.0201
    0.1350    0.0199
    0.1400    0.0196
    0.1450    0.0193
    0.1500    0.0190
    0.1550    0.0187
    0.1600    0.0184
    0.1650    0.0181
    0.1700    0.0178
    0.1750    0.0174
    0.1800    0.0171
    0.1850    0.0167
    0.1900    0.0163
    0.1950    0.0159
    0.2000    0.0155
    0.2050    0.0151
    0.2100    0.0146
    0.2150    0.0142
    0.2200    0.0137
    0.2250    0.0132
    0.2300    0.0127
    0.2350    0.0122
    0.2400    0.0117
    0.2450    0.0111
    0.2500    0.0106
    0.2550    0.0100
    0.2600    0.0094
    0.2650    0.0088
    0.2700    0.0082
    0.2750    0.0075
    0.2800    0.0069
    0.2850    0.0062
    0.2900    0.0055
    0.2950    0.0048
    0.3000    0.0041
    0.3050    0.0034
    0.3100    0.0026
    0.3150    0.0019
    0.3200    0.0011
    0.3250    0.0003
    0.3300   -0.0005
    0.3350   -0.0014
    0.3400   -0.0022
    0.3450   -0.0031
    0.3500   -0.0040
    0.3550   -0.0049
    0.3600   -0.0058
    0.3650   -0.0068
    0.3700   -0.0077
    0.3750   -0.0087
    0.3800   -0.0097
    0.3850   -0.0107
    0.3900   -0.0117
    0.3950   -0.0127
    0.4000   -0.0137
    0.4050   -0.0147
    0.4100   -0.0157
    0.4150   -0.0167
    0.4200   -0.0176
    0.4250   -0.0184
    0.4300   -0.0192
    0.4350   -0.0199
    0.4400   -0.0206
    0.4450   -0.0212
    0.4500   -0.0217
    0.4550   -0.0221
    0.4600   -0.0225
    0.4650   -0.0227
    0.4700   -0.0229
    0.4750   -0.0230
    0.4800   -0.0230
    0.4850   -0.0230
    0.4900   -0.0229
    0.4950   -0.0227
    0.5000   -0.0224
    0.5050   -0.0221
    0.5100   -0.0218
    0.5150   -0.0214
    0.5200   -0.0209
    0.5250   -0.0204
    0.5300   -0.0199
    0.5350   -0.0194
    0.5400   -0.0188
    0.5450   -0.0182
    0.5500   -0.0175
    0.5550   -0.0169
    0.5600   -0.0162
    0.5650   -0.0155
    0.5700   -0.0148
    0.5750   -0.0141
    0.5800   -0.0134
    0.5850   -0.0127
    0.5900   -0.0120
    0.5950   -0.0112
    0.6000   -0.0105
    0.6050   -0.0097
    0.6100   -0.0090
    0.6150   -0.0082
    0.6200   -0.0075
    0.6250   -0.0067
    0.6300   -0.0060
    0.6350   -0.0052
    0.6400   -0.0045
    0.6450   -0.0038
    0.6500   -0.0030
    0.6550   -0.0023
    0.6600   -0.0016
    0.6650   -0.0009
    0.6700   -0.0002
    0.6750    0.0005
    0.6800    0.0012
    0.6850    0.0019
    0.6900    0.0026
    0.6950    0.0032
    0.7000    0.0039
    0.7050    0.0045
    0.7100    0.0051
    0.7150    0.0058
    0.7200    0.0064
    0.7250    0.0070
    0.7300    0.0076
    0.7350    0.0082
    0.7400    0.0087
    0.7450    0.0093
    0.7500    0.0098
    0.7550    0.0104
    0.7600    0.0109
    0.7650    0.0114
    0.7700    0.0119
    0.7750    0.0124
    0.7800    0.0129
    0.7850    0.0133
    0.7900    0.0138
    0.7950    0.0142
    0.8000    0.0147
    0.8050    0.0151
    0.8100    0.0155
    0.8150    0.0159
    0.8200    0.0163
    0.8250    0.0166
    0.8300    0.0170
    0.8350    0.0174
    0.8400    0.0177
    0.8450    0.0180
    0.8500    0.0183
    0.8550    0.0186
    0.8600    0.0189
    0.8650    0.0192
    0.8700    0.0195
    0.8750    0.0197
    0.8800    0.0200
    0.8850    0.0202
    0.8900    0.0205
    0.8950    0.0207
    0.9000    0.0209
    0.9050    0.0211
    0.9100    0.0213
    0.9150    0.0215
    0.9200    0.0216
    0.9250    0.0218
    0.9300    0.0219
    0.9350    0.0221
    0.9400    0.0222
    0.9450    0.0223
    0.9500    0.0224
    0.9550    0.0225
    0.9600    0.0226
    0.9650    0.0227
    0.9700    0.0228
    0.9750    0.0228
    0.9800    0.0229
    0.9850    0.0229
    0.9900    0.0230
    0.9950    0.0230
    1.0000    0.0230
}; \addlegendentry{0\% H $[\varnothing]$}

\addplot[dashed,blue,line width = 2]
  table{
    0    0.0588
    0.0050    0.0588
    0.0100    0.0588
    0.0150    0.0588
    0.0200    0.0586
    0.0250    0.0585
    0.0300    0.0583
    0.0350    0.0580
    0.0400    0.0576
    0.0450    0.0573
    0.0500    0.0568
    0.0550    0.0563
    0.0600    0.0558
    0.0650    0.0552
    0.0700    0.0545
    0.0750    0.0538
    0.0800    0.0531
    0.0850    0.0523
    0.0900    0.0514
    0.0950    0.0505
    0.1000    0.0496
    0.1050    0.0485
    0.1100    0.0475
    0.1150    0.0464
    0.1200    0.0452
    0.1250    0.0440
    0.1300    0.0428
    0.1350    0.0415
    0.1400    0.0401
    0.1450    0.0387
    0.1500    0.0373
    0.1550    0.0358
    0.1600    0.0343
    0.1650    0.0328
    0.1700    0.0312
    0.1750    0.0295
    0.1800    0.0279
    0.1850    0.0262
    0.1900    0.0244
    0.1950    0.0226
    0.2000    0.0208
    0.2050    0.0190
    0.2100    0.0172
    0.2150    0.0153
    0.2200    0.0134
    0.2250    0.0114
    0.2300    0.0095
    0.2350    0.0075
    0.2400    0.0056
    0.2450    0.0036
    0.2500    0.0016
    0.2550   -0.0004
    0.2600   -0.0024
    0.2650   -0.0044
    0.2700   -0.0064
    0.2750   -0.0084
    0.2800   -0.0104
    0.2850   -0.0124
    0.2900   -0.0144
    0.2950   -0.0164
    0.3000   -0.0183
    0.3050   -0.0202
    0.3100   -0.0222
    0.3150   -0.0240
    0.3200   -0.0259
    0.3250   -0.0277
    0.3300   -0.0295
    0.3350   -0.0312
    0.3400   -0.0329
    0.3450   -0.0346
    0.3500   -0.0362
    0.3550   -0.0378
    0.3600   -0.0393
    0.3650   -0.0408
    0.3700   -0.0423
    0.3750   -0.0437
    0.3800   -0.0450
    0.3850   -0.0463
    0.3900   -0.0475
    0.3950   -0.0486
    0.4000   -0.0497
    0.4050   -0.0508
    0.4100   -0.0518
    0.4150   -0.0527
    0.4200   -0.0535
    0.4250   -0.0543
    0.4300   -0.0551
    0.4350   -0.0557
    0.4400   -0.0563
    0.4450   -0.0569
    0.4500   -0.0573
    0.4550   -0.0578
    0.4600   -0.0581
    0.4650   -0.0584
    0.4700   -0.0586
    0.4750   -0.0587
    0.4800   -0.0588
    0.4850   -0.0588
    0.4900   -0.0588
    0.4950   -0.0587
    0.5000   -0.0585
    0.5050   -0.0583
    0.5100   -0.0580
    0.5150   -0.0576
    0.5200   -0.0572
    0.5250   -0.0568
    0.5300   -0.0562
    0.5350   -0.0557
    0.5400   -0.0550
    0.5450   -0.0543
    0.5500   -0.0536
    0.5550   -0.0528
    0.5600   -0.0520
    0.5650   -0.0511
    0.5700   -0.0501
    0.5750   -0.0491
    0.5800   -0.0481
    0.5850   -0.0470
    0.5900   -0.0459
    0.5950   -0.0447
    0.6000   -0.0435
    0.6050   -0.0423
    0.6100   -0.0410
    0.6150   -0.0397
    0.6200   -0.0384
    0.6250   -0.0370
    0.6300   -0.0356
    0.6350   -0.0341
    0.6400   -0.0326
    0.6450   -0.0311
    0.6500   -0.0296
    0.6550   -0.0281
    0.6600   -0.0265
    0.6650   -0.0249
    0.6700   -0.0233
    0.6750   -0.0217
    0.6800   -0.0200
    0.6850   -0.0183
    0.6900   -0.0167
    0.6950   -0.0150
    0.7000   -0.0133
    0.7050   -0.0116
    0.7100   -0.0098
    0.7150   -0.0081
    0.7200   -0.0064
    0.7250   -0.0047
    0.7300   -0.0029
    0.7350   -0.0012
    0.7400    0.0005
    0.7450    0.0023
    0.7500    0.0040
    0.7550    0.0057
    0.7600    0.0074
    0.7650    0.0091
    0.7700    0.0108
    0.7750    0.0125
    0.7800    0.0142
    0.7850    0.0159
    0.7900    0.0175
    0.7950    0.0191
    0.8000    0.0207
    0.8050    0.0223
    0.8100    0.0239
    0.8150    0.0255
    0.8200    0.0270
    0.8250    0.0285
    0.8300    0.0300
    0.8350    0.0315
    0.8400    0.0329
    0.8450    0.0343
    0.8500    0.0357
    0.8550    0.0370
    0.8600    0.0383
    0.8650    0.0396
    0.8700    0.0408
    0.8750    0.0421
    0.8800    0.0432
    0.8850    0.0444
    0.8900    0.0455
    0.8950    0.0465
    0.9000    0.0476
    0.9050    0.0486
    0.9100    0.0495
    0.9150    0.0504
    0.9200    0.0513
    0.9250    0.0521
    0.9300    0.0529
    0.9350    0.0536
    0.9400    0.0543
    0.9450    0.0550
    0.9500    0.0556
    0.9550    0.0561
    0.9600    0.0566
    0.9650    0.0571
    0.9700    0.0575
    0.9750    0.0578
    0.9800    0.0581
    0.9850    0.0584
    0.9900    0.0586
    0.9950    0.0587
    1.0000    0.0588
}; \addlegendentry{60\% H $[\varnothing]$}
\end{axis}
\end{tikzpicture}

%% file: figures/L_Compare.tex
%
%
\definecolor{mycolor1}{rgb}{1.00000,0.00000,1.00000}%
\begin{tikzpicture}[scale = 0.7]

\begin{axis}[%
width=2.789in,
height=2.865in,
at={(1.083in,0.507in)},
scale only axis,
xmin=0,
xmax=10,
xlabel style={font=\color{white!15!black}},
xlabel={Flame Strouhal Number, $\trm{St}$},
ymin=0,
ymax=1.05,
ylabel style={font=\color{white!15!black}},
ylabel={Gain, $|\mathcal{F}|$},
axis background/.style={fill=white},
axis x line*=bottom,
axis y line*=left,
xmajorgrids,
ymajorgrids,
legend style={legend cell align=left, align=left, draw=white!15!black}
]
\addplot [color=red]
  table[row sep=crcr]{%
0.570	1.01892750750576\\
1.140	0.735477530736867\\
1.710	0.358422149812802\\
2.280	0.415476927721999\\
2.850	0.206086335799495\\
3.421	0.0487465270633696\\
3.991	0.0964792490882687\\
4.561	0.0836996725860327\\
5.131	0.0417107688830808\\
5.701	0.0228667608206968\\
6.271	0.0347771573669871\\
6.841	0.043016807479264\\
7.411	0.0382808386499727\\
7.981	0.0286753529842654\\
8.551	0.0215262540287896\\
9.122	0.0222657386490709\\
};
\addlegendentry{$\etah = 0$}

\addplot [color=blue]
  table[row sep=crcr]{%
0.49	1.02424538631104\\
0.98	0.843514073475044\\
1.47	0.436510302462882\\
1.96	0.403813951060945\\
2.45	0.357739895579472\\
2.94	0.146110007299195\\
3.43	0.0637593146233138\\
3.92	0.107341805745971\\
4.41	0.0898204651807299\\
4.90	0.0526285499428657\\
5.38	0.019686498055664\\
5.87	0.0343343202508193\\
6.36	0.0467538189102381\\
6.85	0.0445040437510826\\
7.34	0.0367255227475522\\
7.83	0.0275363195335774\\
};
\addlegendentry{$\etah = 0.2$}

\addplot [color=green]
  table[row sep=crcr]{%
0.40	1.02472533841226\\
0.79	0.942233676727431\\
1.19	0.641830125929634\\
1.59	0.363604112801394\\
1.99	0.415123152003438\\
2.38	0.361493693267123\\
2.78	0.202372200781795\\
3.18	0.0753545788388971\\
3.58	0.096790448175528\\
3.97	0.125151396473492\\
4.37	0.110958397133312\\
4.77	0.0673268132685087\\
5.16	0.0223704642140757\\
5.56	0.0275260007821294\\
5.96	0.0492552954541181\\
6.36	0.0545847727982158\\
};
\addlegendentry{$\etah = 0.4$}

\addplot [color=mycolor1]
  table[row sep=crcr]{%
0.29	1.01710878063702\\
0.58	1.00506334840609\\
0.88	0.871479993118961\\
1.17	0.620354955948675\\
1.46	0.386863585375601\\
1.75	0.382461259720538\\
2.05	0.420869450592229\\
2.34	0.368926011545435\\
2.63	0.252131788071497\\
2.92	0.131942691668336\\
3.22	0.0802613248214179\\
3.51	0.116988524775057\\
3.80	0.138766278948429\\
4.09	0.128635331000246\\
4.39	0.101370747564939\\
4.68	0.0729192576434464\\
};
\addlegendentry{$\etah = 0.6$}

\addplot [color=red, dashed, forget plot]
  table[row sep=crcr]{%
0.570	1.01815621575349\\
1.140	0.731495523373038\\
1.710	0.351844564233205\\
2.280	0.414583793181057\\
2.850	0.212096960792554\\
3.421	0.0564979134045574\\
3.991	0.103444793702093\\
4.561	0.084190189875077\\
5.131	0.0365222547796611\\
5.701	0.0248170457308074\\
6.271	0.0432758330905917\\
6.841	0.0448897820273523\\
7.411	0.0327986250370344\\
7.981	0.0210155636062483\\
8.551	0.0209789257414359\\
9.122	0.0266539851404247\\
};

\addplot [color=blue, dashed, forget plot]
  table[row sep=crcr]{%
0.49	1.02362744717293\\
0.98	0.837444469569008\\
1.47	0.425463894730452\\
1.96	0.393204537670457\\
2.45	0.362108168864812\\
2.94	0.154924126559985\\
3.43	0.0738907037456632\\
3.92	0.116783261669543\\
4.41	0.0925203041969287\\
4.90	0.0442759255825055\\
5.38	0.0146247629379599\\
5.87	0.0438235251741725\\
6.36	0.0501487488083786\\
6.85	0.0424030717546319\\
7.34	0.0271852992358466\\
7.83	0.0179381554530133\\
};

\addplot [color=green, dashed, forget plot]
  table[row sep=crcr]{%
0.40	1.02415968428569\\
0.79	0.933800519369679\\
1.19	0.625062154385397\\
1.59	0.339912258283278\\
1.99	0.398446547408303\\
2.38	0.363359652713249\\
2.78	0.202012740741286\\
3.18	0.0772615216270755\\
3.58	0.113307209891324\\
3.97	0.130899965334506\\
4.37	0.0982244799149119\\
4.77	0.0467475764751989\\
5.16	0.012127295406604\\
5.56	0.0423800976992101\\
5.96	0.058218356852067\\
6.36	0.0533100165072467\\
};

\addplot [color=mycolor1, dashed, forget plot]
  table[row sep=crcr]{%
0.29	1.01809843537832\\
0.58	0.997196804286943\\
0.88	0.849213319472081\\
1.17	0.585567459240842\\
1.46	0.341463412359483\\
1.75	0.33483556482902\\
2.05	0.388028473784246\\
2.34	0.343027305705011\\
2.63	0.229704945285647\\
2.92	0.114451410734757\\
3.22	0.0869902533838957\\
3.51	0.123694347439685\\
3.80	0.136871720919839\\
4.09	0.120457314752884\\
4.39	0.083544401953001\\
4.68	0.0381895715119324\\
};

\end{axis}

\begin{axis}[%
width=2.789in,
height=2.865in,
at={(4.753in,0.507in)},
scale only axis,
xmin=0,
xmax=10,
xlabel style={font=\color{white!15!black}},
xlabel={Flame Strouhal Number, $\trm{St}$},
ymin=-6,
ymax=0,
ylabel style={font=\color{white!15!black}},
ylabel={Normalised Phase Shift, $\angle F/\pi$},
axis background/.style={fill=white},
axis x line*=bottom,
axis y line*=left,
xmajorgrids,
ymajorgrids,
legend style={legend cell align=left, align=left, draw=white!15!black}
]
\addplot [color=red]
  table[row sep=crcr]{%
0.570	-0.294\\
1.140	-0.733\\
1.710	-1.51125\\
2.280	-2.337\\
2.850	-2.8925\\
3.421	-4\\
3.991	-4.3325\\
4.561	-4.568\\
5.131	-4.684\\
5.701	-4.48\\
6.271	-4.38775\\
6.841	-4.477\\
7.411	-4.56225\\
7.981	-4.6125\\
8.551	-4.5325\\
9.122	-4.448\\
};
\addlegendentry{$\etah = 0$}

\addplot [color=blue]
  table[row sep=crcr]{%
0.49	-0.24275\\
0.98	-0.6105\\
1.47	-1.17375\\
1.96	-2\\
2.45	-2.5575\\
2.94	-3.023\\
3.43	-4\\
3.92	-4.328\\
4.41	-4.5715\\
4.90	-4.75\\
5.38	-4.5995\\
5.87	-4.36\\
6.36	-4.4355\\
6.85	-4.5285\\
7.34	-4.5775\\
7.83	-4.564\\
};
\addlegendentry{$\etah = 0.2$}

\addplot [color=green]
  table[row sep=crcr]{%
0.40	-0.18575\\
0.79	-0.476\\
1.19	-0.87225\\
1.59	-1.455\\
1.99	-2.14\\
2.38	-2.5775\\
2.78	-2.945\\
3.18	-3.482\\
3.58	-4.117\\
3.97	-4.41\\
4.37	-4.60775\\
4.77	-4.768\\
5.16	-4.8515\\
5.56	-4.203\\
5.96	-4.29625\\
6.36	-4.416\\
};
\addlegendentry{$\etah = 0.4$}

\addplot [color=mycolor1]
  table[row sep=crcr]{%
0.29	-0.1265\\
0.58	-0.3305\\
0.88	-0.56475\\
1.17	-0.866\\
1.46	-1.36125\\
1.75	-2\\
2.05	-2.2835\\
2.34	-2.57\\
2.63	-2.82125\\
2.92	-3.1275\\
3.22	-3.65\\
3.51	-4.027\\
3.80	-4.24375\\
4.09	-4.434\\
4.39	-4.58875\\
4.68	-4.764\\
};
\addlegendentry{$\etah = 0.6$}

\addplot [color=red, dashed, forget plot]
  table[row sep=crcr]{%
0.570	-0.29475\\
1.140	-0.7335\\
1.710	-1.51125\\
2.280	-2.335\\
2.850	-2.9\\
3.421	-4\\
3.991	-4.3465\\
4.561	-4.6\\
5.131	-4.71775\\
5.701	-4.3975\\
6.271	-4.396\\
6.841	-4.525\\
7.411	-4.60775\\
7.981	-4.581\\
8.551	-4.44625\\
9.122	-4.428\\
};

\addplot [color=blue, dashed, forget plot]
  table[row sep=crcr]{%
0.49	-0.244\\
0.98	-0.612\\
1.47	-1.173\\
1.96	-2\\
2.45	-2.555\\
2.94	-3.0395\\
3.43	-4\\
3.92	-4.356\\
4.41	-4.612\\
4.90	-4.7975\\
5.38	-4.352\\
5.87	-4.339\\
6.36	-4.47775\\
6.85	-4.609\\
7.34	-4.64875\\
7.83	-4.544\\
};

\addplot [color=green, dashed, forget plot]
  table[row sep=crcr]{%
0.40	-0.18775\\
0.79	-0.4785\\
1.19	-0.87225\\
1.59	-1.454\\
1.99	-2.13875\\
2.38	-2.573\\
2.78	-2.9555\\
3.18	-3.57\\
3.58	-4.16425\\
3.97	-4.46\\
4.37	-4.6875\\
4.77	-4.891\\
5.16	-5\\
5.56	-4.245\\
5.96	-4.40875\\
6.36	-4.564\\
};

\addplot [color=mycolor1, dashed, forget plot]
  table[row sep=crcr]{%
0.29	-0.12775\\
0.58	-0.3345\\
0.88	-0.5685\\
1.17	-0.866\\
1.46	-1.355\\
1.75	-2\\
2.05	-2.2905\\
2.34	-2.594\\
2.63	-2.87975\\
2.92	-3.25\\
3.22	-3.8205\\
3.51	-4.18\\
3.80	-4.41925\\
4.09	-4.6335\\
4.39	-4.825\\
4.68	-5.06\\
};

\end{axis}
\end{tikzpicture}%

%% file: figures/GainSensitivity.tex
\begin{tikzpicture}[scale=0.8]
\begin{axis}[
    title={Flame Transfer Function Test Cases},
    ylabel={Gain, $|\mathcal{F}|$},
    xmin=0, xmax=160,
    ymin=0, ymax=1.05,
    xtick={0,50,100,150},
    ytick={0,0.2,0.4,0.6,0.8,1},
    y tick label style={
        /pgf/number format/.cd,
        fixed,
        fixed zerofill,
        precision=1,
        /tikz/.cd},
    legend pos=north east,
    legend cell align={left},
    ymajorgrids=true,
    xmajorgrids=true,
    grid style=dashed,
]

\addplot[
    color=black,
    mark=x,
    ]
    table {
    10.0000    1.0189
   20.0000    0.7355
   30.0000    0.3584
   40.0000    0.4155
   50.0000    0.2061
   60.0000    0.0488
   70.0000    0.0965
   80.0000    0.0837
   90.0000    0.0417
  100.0000    0.0229
  110.0000    0.0348
  120.0000    0.0430
  130.0000    0.0382
  140.0000    0.0286
  150.0000    0.0215
  160.0000    0.0223
    };
    \addlegendentry{$\etah=0$ Ref.}

\addplot[
    color=black,
    mark=x,
    dashed,
    ]
    table {
    10.0000    1.0171
   20.0000    1.0050
   30.0000    0.8714
   40.0000    0.6201
   50.0000    0.3867
   60.0000    0.3825
   70.0000    0.4209
   80.0000    0.3688
   90.0000    0.2519
  100.0000    0.1317
  110.0000    0.0803
  120.0000    0.1171
  130.0000    0.1388
  140.0000    0.1286
  150.0000    0.1013
  160.0000    0.0729
    };
\addlegendentry{$\etah=0.6$ Ref.}

\addplot[
    color=red,
    mark=o,
    ]
    table {
    10.0000    1.0187
   20.0000    0.7355
   30.0000    0.3585
   40.0000    0.4154
   50.0000    0.2057
   60.0000    0.0483
   70.0000    0.0957
   80.0000    0.0836
   90.0000    0.0423
  100.0000    0.0228
  110.0000    0.0346
  120.0000    0.0427
  130.0000    0.0384
  140.0000    0.0285
  150.0000    0.0220
  160.0000    0.0223
    };
\addlegendentry{$\ta{u}_1$ Modified}

\addplot[
    color=blue,
    mark=square,
    ]
    table {
    10.0000    1.0196
   20.0000    1.0023
   30.0000    0.8593
   40.0000    0.5995
   50.0000    0.3587
   60.0000    0.3527
   70.0000    0.4001
   80.0000    0.3490
   90.0000    0.2363
  100.0000    0.1219
  110.0000    0.0838
  120.0000    0.1164
  130.0000    0.1352
  140.0000    0.1267
  150.0000    0.0963
  160.0000    0.0558
    };
\addlegendentry{$s_\trm{L0}$ Modified}

\addplot[
    color=green,
    mark=triangle,
    ]
    table {
    10.0000    1.0196
   20.0000    0.7414
   30.0000    0.3688
   40.0000    0.4142
   50.0000    0.1978
   60.0000    0.0491
   70.0000    0.0936
   80.0000    0.0852
   90.0000    0.0496
  100.0000    0.0263
  110.0000    0.0397
  120.0000    0.0446
  130.0000    0.0419
  140.0000    0.0340
  150.0000    0.0249
  160.0000    0.0270
    };
\addlegendentry{$\mathcal{L}$ Modified}

\end{axis}
\end{tikzpicture}

%% file: figures/PhaseSensitivity.tex
\begin{tikzpicture}[scale=0.8]
\begin{axis}[
    xlabel={Perturbation Frequency, $f_\trm{p}\ [\trm{Hz}]$},
    ylabel={Normalised Phase Shift, $\angle\mathcal{F}/\pi$},
    xmin=0, xmax=160,
    ymin=-5, ymax=0,
    xtick={0,50,100,150},
    legend pos=north east,
    legend cell align={left},
    ymajorgrids=true,
    xmajorgrids=true,
    grid style=dashed,
]

\addplot[
    color=black,
    mark=x,
    ]
    table {
   10.0000   -0.2940
   20.0000   -0.7330
   30.0000   -1.5112
   40.0000   -2.3370
   50.0000   -2.8937
   60.0000   -4.0000
   70.0000   -4.3325
   80.0000   -4.5700
   90.0000   -4.6840
  100.0000   -4.4800
  110.0000   -4.3878
  120.0000   -4.4770
  130.0000   -4.5623
  140.0000   -4.6125
  150.0000   -4.5325
  160.0000   -4.4480
    };
    \addlegendentry{$\etah=0$ Ref.}

\addplot[
    color=black,
    mark=x,
    dashed,
    ]
    table {
    10.0000   -0.1265
   20.0000   -0.3305
   30.0000   -0.5647
   40.0000   -0.8670
   50.0000   -1.3613
   60.0000   -2.0000
   70.0000   -2.2835
   80.0000   -2.5700
   90.0000   -2.8213
  100.0000   -3.1275
  110.0000   -3.6500
  120.0000   -4.0270
  130.0000   -4.2438
  140.0000   -4.4340
  150.0000   -4.5888
  160.0000   -4.7640
    };
\addlegendentry{$\etah=0.6$ Ref.}

\addplot[
    color=red,
    mark=o,
    ]
    table {
    10.0000   -0.2628
   20.0000   -0.7470
   30.0000   -1.3478
   40.0000   -2.3520
   50.0000   -2.8913
   60.0000   -4.0000
   70.0000   -4.3308
   80.0000   -4.5740
   90.0000   -4.6705
  100.0000   -4.4825
  110.0000   -4.3878
  120.0000   -4.4680
  130.0000   -4.5623
  140.0000   -4.6125
  150.0000   -4.5288
  160.0000   -4.4520
    };
\addlegendentry{$\ta{u}_1$ Modified}

\addplot[
    color=blue,
    mark=square,
    ]
    table {
    10.0000   -0.0845
   20.0000   -0.3425
   30.0000   -0.5040
   40.0000   -0.8820
   50.0000   -1.3562
   60.0000   -2.0000
   70.0000   -2.2940
   80.0000   -2.5960
   90.0000   -2.8550
  100.0000   -3.2100
  110.0000   -3.7545
  120.0000   -4.1290
  130.0000   -4.3705
  140.0000   -4.5705
  150.0000   -4.7350
  160.0000   -4.9200
    };
\addlegendentry{$s_\trm{L0}$ Modified}

\addplot[
    color=green,
    mark=triangle,
    ]
    table {
    10.0000   -0.2618
   20.0000   -0.7465
   30.0000   -1.3470
   40.0000   -2.3560
   50.0000   -2.8825
   60.0000   -3.6980
   70.0000   -4.3133
   80.0000   -4.5760
   90.0000   -4.6885
  100.0000   -4.5275
  110.0000   -4.4070
  120.0000   -4.4590
  130.0000   -4.5038
  140.0000   -4.5390
  150.0000   -4.5137
  160.0000   -4.5120
    };
\addlegendentry{$\mathcal{L}$ Modified}

\end{axis}
\end{tikzpicture}